\documentclass[aps]{revtex4}
\usepackage{eurosym}
\usepackage{amsfonts}
\usepackage{amsmath}
\usepackage{amssymb,epsf}
\usepackage{color}
\usepackage{graphicx}
\usepackage{epstopdf}
\usepackage{float}
\usepackage{caption}
\usepackage{subfig}

\begin{document}

\title{Thermodynamics and Optical Properties of Phantom AdS Black Holes in
Massive Gravity}
\author{Kh. Jafarzade$^{1}$\footnote{email address: khadijeh.jafarzade@gmail.com}, B. Eslam Panah$^{1,2,3}$\footnote{email address: eslampanah@umz.ac.ir}, and M. E. Rodrigues$^{4,5}$ \footnote{email address: esialg@gmail.com}}
\affiliation{$^{1}$ Department of Theoretical Physics, Faculty of Science, University of
Mazandaran, P. O. Box 47416-95447, Babolsar, Iran}
\affiliation{$^{2}$ ICRANet-Mazandaran, University of Mazandaran, P. O. Box 47416-95447
Babolsar, Iran}
\affiliation{$^{3}$ ICRANet, Piazza della Repubblica 10, I-65122 Pescara, Italy}
\affiliation{$^{4}$ Faculdade de Ci\^{e}ncias Exatas e Tecnologia, Universidade Federal
do Par\'{a}\\
Campus Universit\'{a}rio de Abaetetuba, 68440-000, Abaetetuba, Par\'{a}, Brazil}
\affiliation{$^{5}$ Faculdade de F\'{\i}sica, Programa de P\'{o}s-Gradua\c{c}\~{a}o em F%
\'{\i}sica}

\begin{abstract}
Motivated by high interest in Lorentz invariant massive gravity models known
as dRGT massive gravity, we present an exact phantom black hole solution in
this theory of gravity and discuss the thermodynamic structure of the black
hole in the canonical ensemble. Calculating the conserved and thermodynamic
quantities, we check the validity of the first law of thermodynamics and the
Smarr relation in the extended phase space. In addition, we investigate both
the local and global stability of these black holes and show how massive
parameters affect the regions of stability. We extend our study to
investigate the optical features of the black holes such as the shadow
geometrical shape, energy emission rate, and deflection angle. Also, we
discuss how these optical quantities are affected by massive coefficients.
Finally, we consider a massive scalar perturbation minimally coupled to the
background geometry of the black hole and examine the quasinormal modes
(QNMs) by employing the WKB approximation.
\end{abstract}

\maketitle

\address{$^{1}$  Department of Theoretical Physics, Faculty of Science, University of Mazandaran, P. O. Box 47415-416, Babolsar, Iran\\
$^{2}$ ICRANet-Mazandaran, University of Mazandaran, P. O. Box 47415-416, Babolsar, Iran\\
$^{3}$ ICRANet, Piazza della Repubblica 10, I-65122 Pescara, Italy}

\section{Introduction}

Modified theories of gravity can explain some phenomena both from the
perspective of cosmology and from the perspective of the astrophysical
field, which general relativity was unable to describe. For example, the
observational evidence of the current acceleration of our universe, dark
energy, and dark matter problems. In addition, they almost inevitably
require a spin$-2$ particle at their foundation. To include these issues,
massive gravity is a fascinating theory of gravity that may include all of
them. In this theory, the graviton includes mass which can describe, for
example, the recent accelerated expansion at large distances. Historically, Pauli and Fierz introduced a linear theory of massive gravity in 1939 \cite{FP}, to include massive graviton. However, this linear theory of massive gravity suffers from a fundamental law because in
the massless limit of the graviton, it does not satisfy the observational
evidence of the solar system. In this regard, Vainshtein extended it to include a non-linear case for removing this
problem \cite{Vainshtein}. However, it was indicated the existence of the
ghost in this non-linear theory becomes another problematic factor which is
known as Boulware and Deser (BD) ghost \cite{BD}. The new massive gravity in
three-dimensional spacetime is one of these attempts to remove BD-ghost in
massive gravity \cite{Bergshoe2009}. Another commendable effort to remove
this ghost is by introducing a reference metric which is known as dRGT
massive gravity \cite{dRGT1,dRGT2}. In addition, this theory of massive
gravity removes BD-ghost in arbitrary dimensional of spacetime \cite{Do104003,Do044022}. On the other hand, the black hole
solutions may provide important tests of both the theoretical and
phenomenological viability of dRGT massive gravity. In this regard, the
black hole solutions in dRGT massive gravity are obtained extensively in the
literature \cite%
{BHmass1,BHmass2,BHmass3,BHmass4,BHmass5,BHmass6,BHmass7,BHmass8,BHmass9,BHmass10,BHmass11,BHmass12,BHmass13,BHmass14,BHmass15}%
. The presence of additional terms in the black hole solutions can justify
dark matter and dark energy in massive gravity. It was indicated that the
massive graviton could play the role of the cosmological constant in cosmic
distances \cite{Comelli2012,Langlois2012,Kobayashi2012,Felice2013}. In
addition, the end state of Hawking evaporation led to a black hole remnant,
which could help to ameliorate the information paradox \cite{BHmass14}.
Also, in Ref. \cite{Hinterbichler2012} multi-gravity theory has been
introduced to solve the BD-ghost in multi-dimensional spacetime for
interacting spin-2 fields by applying the vielbein formulation. In another
fascinating effort, it was indicated that by using a dynamic reference
metric (instead of a static reference metric) the BD-ghost could be
neglected in a non-linear bimetric theory for a massless spin-2 field \cite%
{Hassan2012}. Vegh introduced another branch of dRGT massive gravity by
applying holographic principles and a singular reference metric \cite{Vegh}.
This theory was a ghost-free theory of massive gravity. By considering this
theory of massive gravity, many works were done. For example, black hole
solutions and their thermodynamic properties are investigated in Refs. \cite%
{VBHmass1,VBHmass2,VBHmass3,VBHmass4,VBHmass5,VBHmass6,Maetal2020,VBHmass7,VBHmass8,VBHmass9}%
. It was shown that there is a correspondence between black hole solutions
of conformal and this theory of massive gravity \cite{Panah2019}. From an
astrophysical point of view, the effect of massive graviton on the structure
of neutron stars, white dwarfs, and dark energy stars has been studied in
Refs. \cite{NeutronStar,Whitedwarf,DarkEnergyStar}. The results indicated
that the maximum mass of these compact objects could be more than three
times of solar mass. In addition, by considering Vegh's massive theory of
gravity, the mass of the graviton could generate a range of new phase
transitions for topological black holes \cite{Hendi2017}. Therefore, these
issues motivated us to consider this theory of massive gravity.

In physics in general, we have some systems that present a peculiar
characteristic, the energy density of the system is negative. We can cite
some examples here \cite{Visser}: a) the so-called usual Casimir effect, the
case between two neutral plates, and the topological case; b) the case of
the squeezed vacuum; c) the case of the Hartle-Hawking vacuum; and d) the
case of a phantom field responsible for the accelerated expansion of the
universe \cite{Caldwell}. In gravitation, we also have similar cases. An
example that draws our attention, is that of the quasi-charged bridge, built
ad hoc by Einstein and Rosen in 1935 \cite{Einstein}. This example
introduces a charge of the system that appears with a sign opposite to that
of the Reissner-Nordstr\"{o}m solution, $q^2\rightarrow -q^2$. Thus, we
have, in the action of the system, the first introduction of a spin-1
phantom field, with negative energy density. We also have the possibility of
zero spin phantom fields, such as the anti-Fisher solution \cite{Bronnikov}.
The most general solutions of coupling scalar and Maxwell phantom fields are
called Einstein-anti-Maxwell-anti-Dilaton \cite{Gerard}. We have solutions
with the cosmological constant \cite{Jardim2012} and regular \cite{Fabris}.
We have generalizations for phantom fields, using the Sigma model method,
phantom solutions with rotation \cite{Gerard2}. Furthermore, we have the
studies of the light path \cite{Azreg}; gravitational collapse \cite{Na};
gravitational lenses \cite{Gy}; global monopole \cite{Chen}; quasi-black
holes \cite{Bronnikov2}; dyon black holes \cite{Abishev}; wormholes \cite%
{Huang}; spherical accretion \cite{Azreg2}; topological black holes in $f(R)$
\cite{Rodrigues}; and finally black bounces \cite{Silva}.

Thermodynamic aspects of black holes have been among the most interesting
subjects for many years. The studies on the black hole as a thermodynamic
system date back to the work of Hawking and Bekenstein \cite{Bekenstein:1974}
who demonstrated that geometrical quantities such as surface gravity and
horizon area correspond to the thermodynamic quantities such as temperature
and entropy. The analogy between the geometrical properties of the black
holes and thermodynamic variables provides deep insight into understanding
the relations between the physical properties of gravity and classical
thermodynamics. Among the thermodynamical properties of black holes, phase
transition and thermal stability have been widely studied in the literature 
\cite{Davies:197,Hawking:1983,Chamblin:1999}. Thermodynamic phase transition
is one of the most interesting phenomena in black hole physics, which can
provide insight into the underlying structure of spacetime geometry. Phase
transition also plays an important role in elementary particle \cite%
{Kleinert:085001}, condensed matter \cite{Greer:1947}, usual thermodynamics 
\cite{Callen:1985}, cosmology \cite{Layzer:1991}, black holes \cite{Zou:2017}%
, and other branches of physics. In recent years, considerable attention has
arisen to investigate the thermodynamic phase transition of black holes in
anti-de-Sitter (AdS) spacetime. This is mainly motivated by AdS/CFT duality,
which states that there exists a correspondence between gravity in an $%
(n+1)- $dimensional AdS spacetime and the conformal field theory living on
the boundary of its n-dimensional spacetime \cite{Witten:1998}. Besides,
studying the thermodynamic phase transition of black holes in such a
spacetime provides a deep insight to understand the quantum nature of
gravity, which is one of the most important theoretical subjects in physical
communities. A significant interest has arisen to study the phase transition
of AdS black holes in an extended phase space in which the cosmological
constant can be regarded as a variable \cite{Kubiznak:1207}. In this
viewpoint, the cosmological constant is identified as thermodynamic
pressure, and the mass of the black hole is interpreted as the enthalpy \cite%
{Kastor:195011}. The study on phase transition in this phase space has
disclosed some interesting phenomena, such as Van der Waals liquid-vapor
phase transition \cite{Kubiznak:1207,Hendi2013,Cai2013,Xu2015,Singh2020,EslamPanah2020}, zeroth-order
phase transition \cite{Doneva2010}, reentrant phase transition \cite%
{Kubiznak:1211}, triple critical point \cite{Wei:044057}. There are several
methods to study the phase transition of black holes. The first method is
based on studying heat capacity. It was pointed out that divergencies of the
heat capacity represent phase transition points. Another approach is based
on studying the van der Waals-like behavior of black holes in the extended
phase space. In general, stability criteria can be categorized into two
classes; dynamical stability and thermodynamical one. Thermodynamic
stability of the black hole is of great importance because the instability
of black holes puts restrictions on the validation of the physical
solutions. One approach to investigating thermodynamic stability is through
calculating heat capacity in the canonical ensemble. The sign of heat
capacity represents the stability/instability of the black holes. Systems
with positive (negative) heat capacity are denoted to be in thermally stable
(unstable) states. Studying dynamical stability can be done by analyzing
quasinormal modes (QNMs). In the literature, we will investigate both cases
associated with phantom solutions \cite{Rodrigues3,Rodrigues4}.

The breakthrough discovery of the reconstruction of the event-horizon-scale
images of the supermassive black holes by the Event Horizon Telescope (EHT)
project shed light on the physics of black holes and provided us better
insight into understanding these mysterious celestial objects. The
experimental results reported by EHT \cite{Akiyama:2019L1} not only directly
prove the existence of black holes, but also allow us to directly observe
the shadows of black holes. Shadow is the image formed by null geodesics in
the strong gravity regime. The formation of a black hole shadow depends on
the angular momentum of photons. For low values of the angular momentum ($L$%
), ingoing photons fall into the black hole and form a dark area for a
distant observer, whereas, for larger values of $L$, they will be deflected
by the gravitational potential of the black hole. An interesting phenomenon
is related to the photons with critical angular momentum. In this case, the
photons orbit the black hole due to their non-vanishing transverse velocity
and surround the dark interior, which are, respectively, called the photon
ring and the black hole shadow. The shape and size of the shadow are
determined by the black hole parameters, i.e., mass, angular momentum, and
electric charge \cite{Vries:1999123}, as well as spacetime properties \cite%
{Johannsen:446,Cunha:024039} and the position of the observer. The image of
a black hole gives us important information concerning jets and matter
dynamics around black holes, and can be a useful tool for comparing
alternative theories of gravity with general relativity. Moreover, the black
hole shadow can be used to extract information about the deviations in the
spacetime geometry \cite{Pedro:5042,Wei:08030,Belhaj:215004}. These
deviations might be due to some parameters from different alternative
theories of gravity \cite{Khodadi:104050,Uniyal:101178}, or the
astrophysical environment in which the black hole is immersed \cite%
{Afrin:101195,Konoplya:933}. The history of the theoretical analysis of
black hole shadow began with the seminal paper of Synge \cite{Synge:131},
who calculated the angular radius of the shadow of Schwarzschild black hole
and showed the boundary of the shadow is a perfect circle for a non-rotating
black hole. Then, Bardeen et al. \cite{Bardeen:347} analyzed the shadow of
the Kerr black hole and argued that the black hole's angular momentum causes
the deformation of its shadow. Studying shadows has been extended to black
holes in the modified theory of gravity \cite{Amarilla:124045,Kumar:100881},
to black holes with higher or extra dimensions \cite%
{Papnoi:024073,Amir:39978}, and black holes surrounded by plasma \cite%
{Atamurotov:084005}.

Black holes are interacting with the matter and radiations in the
surroundings. The gravitational waves are the response of black holes to
these perturbations, which were recently detected by LIGO Scientific
Collaboration and Virgo Collaboration \cite{LIGO:241102}. The issue of the
perturbations of black holes was first considered by Regge and Wheeler, who
explored the stability of Schwarzschild black holes \cite{Wheeler:1063}.
Studying black hole perturbations has been applied not only in the framework
of GR \cite{Kokkota:1219,Cardoso:163001,Horowitz:024027}, but also in
alternative gravity theories \cite{Kim:990,Kim:763}. The signal of the
gravitational waves is dominated by the fundamental mode of the so-called
QNMs of the black holes. QNMs are characteristic frequencies that encode the
information on how black holes relax after the perturbation. From an
astrophysical point of view, QNMs play an increasingly essential role in
contemporary gravitational waves astronomy \cite{LIGO:061102,LIGO:161101},
because they can be regarded as characteristic "sound" of black holes \cite%
{Nollert:1999} which serves as the basis of black holes spectroscopy. QNMs
provide us with valuable information about the properties of black holes,
such as their mass, angular momentum, and the nature of surrounding
spacetime. QNM frequencies not only are an important factor in determining
the parameters of black holes but also play a significant role in
determining their space-time stability under perturbation. Studying QNMs can
deepen our insight into the structure and evolution of black holes, and
their role in astrophysical phenomena. Empirical detection of QNMs would not
only provide us an opportunity to test general relativity and the validity
of the famous "no-hair" theorem of black holes \cite%
{Krishnan:787,Cardoso:064030,Giesler:111102} but also allow us to constrain
modified gravity theories \cite{Liu:124011,Franciolini:12702,Gonzalez:2021}
and examine strong cosmic censorship conjecture \cite%
{Cardoso:2018,Hod:941,Hod:2018}. Given the significance of QNMs introduced
above, it will therefore be interesting to study QNMs when a new black hole
solution is obtained.

Recently, it has been suggested that the real part of QNMs in
the eikonal limit can be connected to the shadow radius of a black hole.
Cardoso et al. \cite{Cardoso;2009} showed that the real part of the QNMs
corresponds to the angular velocity of the unstable null geodesic while the
imaginary part of the QNMs is related to the Lyapunov exponent that
determines the instability timescale of the orbits. Such a connection can,
alternatively, provide a physical picture at the semi-classical level, where
the gravitational waves can be understood as a phenomenon in which a
massless particle propagates along an outmost and unstable orbit of null
geodesics and spreads transversely out to infinity \cite{Guo;084057}. The
significance of this issue is in general to reach the relationship between
gravitational waves and shadows of black holes, the two amazing achievements
in the new century. The study connecting the real part of the QNMs and the
shadow radius has been explored for the static \cite%
{Jusufi;084055,Melgar;13596} and the rotating spacetime \cite{Jusufi;124063}
and has been also applied to different black holes \cite%
{Lan;115539,Wei;115103,Singh2022sU,Myrzakulov2023}. 

This work aims to obtain, through the equations of motion, a phantom AdS
black hole solution, coupled with dRGT massive gravity; to study the
horizons of the causal structure, the singularity, the thermodynamic system,
and the main optical properties of the solution.

The paper is divided as follows: in section \ref{sec2}, we define the action
of the theory and establish the equations of motion; in section \ref{sec3},
we impose the symmetry of the metric and the material content, studying the
possible horizons and singularities; in section \ref{sec4}, we study the
thermodynamics of the system, establishing the first law, the local and
global stability; in section \ref{sec5}, we present the main optical
properties of our solution; we study the QNMs in section \ref{sec6}; and we
end by presenting the final considerations in section \ref{sec7}.

\section{Field Equations}

\label{sec2}

The $d$-dimensional form of Lagrangian of dRGT massive gravity is given by 
\cite{dRGT1,dRGT2}%
\begin{equation}
\mathcal{L}_{massive}=m_{g}^{2}\sum_{i}^{d}c_{i}\mathcal{U}_{i}(g,f),
\label{massiveLag}
\end{equation}%
where $m_{g}$ is the graviton mass parameter, $c_{i}$'s are arbitrary
massive couplings and $\mathcal{U}_{i}$'s are graviton's self-interaction
potentials constructed from the building blocks $\mathcal{K}%
_{\,\,\,\nu }^{\mu }=\delta _{\nu }^{\mu }-\sqrt{g^{\mu \sigma
}f_{ab}\partial _{\sigma }\phi ^{a}\partial _{\nu }\phi ^{b}}$ (where $%
g_{\mu \nu }$ is the dynamical (physical) metric, and $f_{\mu \nu }$ is the
auxiliary reference metric, needed to define the mass term for gravitons as
well as $\phi ^{a}$'s are scalar fields well-known as St\"{u}ckelberg fields), which can be written as 
\begin{equation}
\mathcal{U}_{i}=\sum_{y=1}^{i}\left( -1\right) ^{y+1}\frac{\left( i-1\right)
!}{\left( i-y\right) !}\mathcal{U}_{i-y}\left[ \mathcal{K}^{y}\right] ,
\end{equation}%
where $\mathcal{U}_{i-y}=1$, when $i=y$. The square root in $%
\mathcal{K}$ means $\left( \mathcal{K}\right) _{\lambda }^{\mu }\left( 
\mathcal{K}\right) _{\nu }^{\lambda }=\mathcal{K}_{\nu }^{\mu }$ and the
rectangular brackets denote traces, $[\mathcal{K}]=\mathcal{K}_{\mu }^{\mu }$%
. The explicit form of $\mathcal{U}_{i}$'s can be written as follows 
\begin{eqnarray}
\mathcal{U}_{1} &=&\left[ \mathcal{K}\right] ,\;\;\;\;\;\mathcal{U}_{2}=%
\left[ \mathcal{K}\right] ^{2}-\left[ \mathcal{K}^{2}\right] ,\;\;\;\;\;%
\mathcal{U}_{3}=\left[ \mathcal{K}\right] ^{3}-3\left[ \mathcal{K}\right] %
\left[ \mathcal{K}^{2}\right] +2\left[ \mathcal{K}^{3}\right] ,  \notag \\
&&\mathcal{U}_{4}=\left[ \mathcal{K}\right] ^{4}-6\left[ \mathcal{K}^{2}%
\right] \left[ \mathcal{K}\right] ^{2}+8\left[ \mathcal{K}^{3}\right] \left[ 
\mathcal{K}\right] +3\left[ \mathcal{K}^{2}\right] ^{2}-6\left[ \mathcal{K}%
^{4}\right] .
\end{eqnarray}%

The action of dRGT massive gravity with phantom Maxwell field on the $4-$%
dimensional spacetime and in the presence of the cosmological constant can
be written as 
\begin{equation}
\mathcal{I}=\int \sqrt{-g}\left[ \mathcal{R}-2\Lambda -2\eta F_{\mu \nu
}F^{\mu \nu }-\mathcal{L}_{massive}\right] d^{4}x,  \label{Action}
\end{equation}%
in which we used geometric units in this work. Also, $\mathcal{R}$ and $%
\Lambda $ are the scalar curvature and the cosmological constant,
respectively. In addition, $F_{\mu \nu }=\partial _{\mu }A_{\nu
}-\partial_{\nu }A_{\mu }$ is the Faraday tensor with $A_{\mu }$ as the
gauge potential. It is notable that, the first term characterizes the
Einstein-Hilbert action, and the second term is related to the cosmological
constant, which can behave as dS ($\Lambda >0$) or AdS ($\Lambda <0$). The
third term is the coupling with the Maxwell field when $\eta =1$, or a
phantom field of spin$-1$, when $\eta =-1$. In other words, the constant $%
\eta $ indicates the nature of the electromagnetic field. Indeed, we obtain
the classical Einstein-Maxwell theory when $\eta =1$, whereas for $\eta =-1$%
, the Maxwell field is phantom (this phantom is because the energy density
of the field of spin$-1$ is negative (see Ref. \cite{Jardim2012}, for more
details)). The last term in the above action is related to the Lagrangian of
dRGT massive gravity (Eq. (\ref{massiveLag})).

Taking the action (\ref{Action}) into account and using the variational
principle, we obtain the field equations corresponding to the gravitation
and gauge fields as%
\begin{eqnarray}
G_{\mu \nu }-\Lambda g_{\mu \nu }-m_{g}^{2}\chi _{\mu \nu } &=&2\eta \left( 
\frac{g_{\mu \nu }}{4}F_{\mu \nu }F^{\mu \nu }-F_{\mu \rho }F_{\nu }^{\rho
}\right) ,  \label{Field
equation} \\
&&  \notag \\
\partial _{\mu }\left( \sqrt{-g}F^{\mu \nu }\right) &=&0,
\label{Maxwell equation}
\end{eqnarray}%
where $\chi _{\mu \nu }$ is given by 
\begin{equation}
\chi _{\mu \nu }=-\sum_{i=1}^{d-2}\frac{c_{i}}{2}\left[ \mathcal{U}%
_{i}g_{\mu \nu }+\sum_{y=1}^{i}\left( -1\right) ^{y}\frac{i!}{\left(
i-y\right) !}\mathcal{U}_{i-y} \mathcal{K}_{\mu \nu }^{y} \right] .  \notag
\end{equation}

The field equations turn to general relativity's field equations when the
graviton mass is zero ($m_{g}=0$), i.e., $G_{\mu \nu }-\Lambda g_{\mu\nu
}=2\eta \left( \frac{g_{\mu \nu }}{4}F_{\mu \nu }F^{\mu \nu }-F_{\mu\rho
}F_{\nu }^{\rho }\right) $ \cite{Jardim2012}.

\section{Phantom black hole solutions}

\label{sec3}

Since we are interested in static phantom charged black hole solutions in
massive gravity, we consider the metric of $4$-dimensional spacetime with
the following form 
\begin{equation}
ds^{2}=\psi \left( r\right) dt^{2}-\frac{dr^{2}}{\psi (r)}-r^{2}\left(
d\theta ^{2}+\sin ^{2}\theta d\varphi ^{2}\right) ,  \label{metric}
\end{equation}%
in which $\psi (r)$ is the metric function of our black holes. In addition,
the conventions adopted here are with the description of the metric whose
signature is $(+,-,-,-)$.

Our main motivation is to obtain exact phantom black holes in the context of
massive gravity. In this regard, the appropriate ansatz for the reference
metric is as \cite{VBHmass1,VBHmass2} 
\begin{equation}
f_{\mu \nu }=diag\left( 0,0,-c^{2},-c^{2}\sin ^{2}\theta \right).  \label{RM1}
\end{equation}%

Before deriving the exact solution for AdS phantom black holes,
we would like to explain the importance of choosing such a reference metric.
As we know, to construct the non-linear massive gravity, introducing an
auxiliary non-dynamical reference metric is necessary. However, the
reference metric breaks the diffeomorphism invariance (also known as general
covariance or coordinate invariance) of general relativity. The breakdown of
diffeomorphism invariance implies that the two helicities $\pm 2$ of the
massless graviton would be joined by four degrees to give six degrees of
freedom \cite{BD}. The sixth degree of freedom corresponds to BD ghost. One
way to solve this problem is introducing a degenerate reference metric as $%
f_{\mu \nu }=diag(0,0,1,1)$ which was first defined by Vegh \cite{Vegh}.
Such a choice of reference metric depends only on the spatial components;
therefore, the diffeomorphism invariance (general reference) is preserved in
the $t$ and $r$ coordinates but is broken in the spatial dimensions. One can
also imagine a more general reference metric that does not respect
diffeomorphism invariance in the $r-direction$. For instance, to preserve
rotational invariance on the sphere and general time parametrization
invariance, one can choose $f_{\mu \nu }=diag(0,1,c^{2},c^{2}sin^{2}\theta )$
as a natural ansatz. One can also consider a different generalization of $%
f_{\mu \nu }$, with $sin^{2}\theta f_{\theta \theta }=f_{\varphi \varphi
}=F(r)$, and all other components to be zero \cite{Adams91}. This can lead
to an ability to add arbitrary polynomial terms in $r$ to the emblackening
factor. The massive gravity of this choice of reference metric has important
applications in gauge/gravity duality. Massive gravities on AdS with a
degenerate reference metric can be viewed as effective dual-field theories
of different phases of condensed matter systems with broken translational
symmetry such as solids, (perfect) fluids, and liquid crystals \cite%
{Alberte114,Alberte171,Alberte129}. In fact, AdS black holes in massive
gravity with a singular (degenerate) reference metric are proving useful in
building holographic models for normal conductors that are close to
realistic ones, with finite direct-current (DC) conductivity \cite%
{Blake88,Alberte91}, the desired property for normal conductors that is
absent in massless gravities. However, such an investigation is not the
purpose of this paper.

It is notable that for $d-$dimensional spacetime, the reference
metric is $f_{\mu \nu }=diag\left( 0,0,c^{2}h_{ij}\right) $ \cite{VBHmass1},
in which $h_{ij}$ determines by $d-$dimensional dynamical metric (i.e. $%
ds^{2}=\psi \left( r\right) dt^{2}-\psi
^{-1}(r)dr^{2}-r^{2}h_{ij}dx^{i}dx^{j}$, $i,j=1,2,3,...,n$). In this case,
the explicit functional forms of $U_{i}$'s are given by \cite{VBHmass1} 
\begin{eqnarray}
\mathcal{U}_{1} &=&\frac{\left( d-2\right) c}{r},  \notag \\
\mathcal{U}_{2} &=&\frac{\left( d-2\right) \left( d-3\right) c^{2}}{r^{2}}, 
\notag \\
\mathcal{U}_{3} &=&\frac{\left( d-2\right) \left( d-3\right) \left(
d-4\right) c^{3}}{r^{3}},  \notag \\
\mathcal{U}_{4} &=&\frac{\left( d-2\right) \left( d-3\right) \left(
d-4\right) \left( d-5\right) c^{4}}{r^{4}}.
\end{eqnarray}

Using the mentioned information and ansatz (\ref{RM1}), one can extract the
explicit functional forms of $\mathcal{U}_{i}$ 's, Since we are interested
in $4-$dimensional solutions (i.e., $d=4$), the only non-zero terms of $%
\mathcal{U}_{i}$ are $\mathcal{U}_{1}$ and $\mathcal{U}_{2}$\ while the
quartic terms all vanish. By taking this fact into account, one can find the
following action governing our black holes of interest 
\begin{eqnarray}
\mathcal{U}_{1} &=&\frac{2c}{r},~~~\&~~~\mathcal{U}_{2}=\frac{2c^{2}}{r^{2}},
\notag \\
&&  \notag \\
\mathcal{U}_{i} &=&0,\text{ \ \ when \ \ }i>2  \label{U1}
\end{eqnarray}

To have a radial electric field, we consider the following gauge potential 
\begin{equation}
A_{\mu }=h(r)\delta _{\mu }^{t},  \label{gauge
potential}
\end{equation}%
where by using the metric (\ref{metric}) with the Maxwell field equation (%
\ref{Maxwell equation}), one can find the following differential equation%
\begin{equation}
h^{\prime }(r)+rh^{\prime \prime }(r)=0,  \label{heq}
\end{equation}%
in which the prime and double prime are representing the first and second
derivatives with respect to $r$, respectively. It is a matter of calculation
to solve Eq. (\ref{heq}), yielding 
\begin{equation}
h(r)=-\frac{q}{r},  \label{h(r)}
\end{equation}%
where $q$ is an integration constant related to the electric charge. It is
worthwhile to mention that the corresponding electromagnetic field tensor is 
$F_{tr}=\frac{q}{r^{2}}$.

In order to obtain the metric function, $f(r)$, we use Eq. (\ref{metric})
with Eq. (\ref{Field equation}), and obtain the following differential
equations 
\begin{eqnarray}
e_{tt} &=&e_{rr}=\Lambda r^{2}+\left( \psi ^{\prime }(r)-C_{1}\right) r+\psi
\left( r\right) -1-C_{2}+\frac{\eta q^{2}}{r^{2}},  \label{eqENMax1} \\
&&  \notag \\
e_{\theta \theta } &=&e_{\varphi \varphi }=\left( \psi ^{\prime \prime
}(r)+2\Lambda \right) r^{2}+\left( 2\psi ^{\prime }(r)-C_{1}\right) r-\frac{%
2\eta q^{2}}{r^{2}},  \label{eqENMax2}
\end{eqnarray}%
where $C_{1}=m_{g}^{2}cc_{1}$, and $C_{2}=m_{g}^{2}c^{2}c_{2}$. Also, $%
e_{tt} $, $e_{rr}$, $e_{\theta \theta }$ and $e_{\varphi \varphi }$
correspond to $tt$, $rr$, $\theta \theta $ and $\varphi \varphi $\
components of Eq. (\ref{Field equation}), respectively. Considering the
equations (\ref{eqENMax1}) and (\ref{eqENMax2}), we can obtain the metric
function in the following form%
\begin{equation}
\psi (r)=1-\frac{2m_{0}}{r}-\frac{\Lambda r^{2}}{3}+\frac{\eta q^{2}}{r^{2}}+%
\frac{C_{1}r}{2}+C_{2},  \label{f(r)ENMax}
\end{equation}%
where $m_{0}$ is an integration constant related to the total mass of the
black hole. It is worthwhile to mention that the resulting metric function (%
\ref{f(r)ENMax}) satisfies all the components of the field equation (\ref%
{Field equation}), simultaneously. Whereas the constant term $C_{2}$ appears
due to the existence of massive gravity (which acts as a modified term to
Einstein's gravity) and it compares with $1$, so this term cannot be more
than $1$ (i.e., $\left\vert C_{2}\right\vert <1$). In addition, in the
absence of massive parameter ($C_{1}=C_{2}=0$), the metric function Eq. (\ref%
{f(r)ENMax}) reduces to 
\begin{equation}
\psi (r)=1-\frac{2m_{0}}{r}-\frac{\Lambda r^{2}}{3}+\frac{\eta q^{2}}{r^{2}}.
\end{equation}

Our next step is the examination of the geometrical structure of solutions.
First, we should look for the existence of essential singularity(ies). The
Ricci and Kretschmann scalars of the solutions are, respectively, 
\begin{eqnarray}
R &=&-4\Lambda +\frac{3C_{1}}{r}+\frac{2C_{2}}{r^{2}}, \\
&&  \notag \\
R_{\alpha \beta \gamma \delta }R^{\alpha \beta \gamma \delta } &=&\frac{%
8\Lambda ^{2}}{3}-\frac{4C_{1}\Lambda }{r}+\frac{2C_{1}^{2}-\frac{8\Lambda
C_{2}}{3}}{r^{2}}+\frac{4C_{1}C_{2}}{r^{3}}-\frac{16C_{2}m_{0}+4C_{1}\eta
q^{2}}{r^{5}}  \notag \\
&&+\frac{48m_{0}^{2}+8C_{2}\eta q^{2}}{r^{6}}+\frac{4C_{2}^{2}}{r^{4}}-\frac{%
96m_{0}\eta q^{2}}{r^{7}}+\frac{56\eta ^{2}q^{4}}{r^{8}}.
\end{eqnarray}

These relations confirm that there is an essential curvature singularity at $%
r=0$. For the limit of $r\longrightarrow \infty $, the Ricci and Kretschmann
scalars yield the values $4\Lambda $ and $\frac{8\Lambda ^{2}}{3}$,
respectively, which show that for $\Lambda >0$ ($\Lambda <0$), the
asymptotical behavior of the solution is (A)dS.

\begin{figure}[tbph]
\centering
\includegraphics[width=0.4\linewidth]{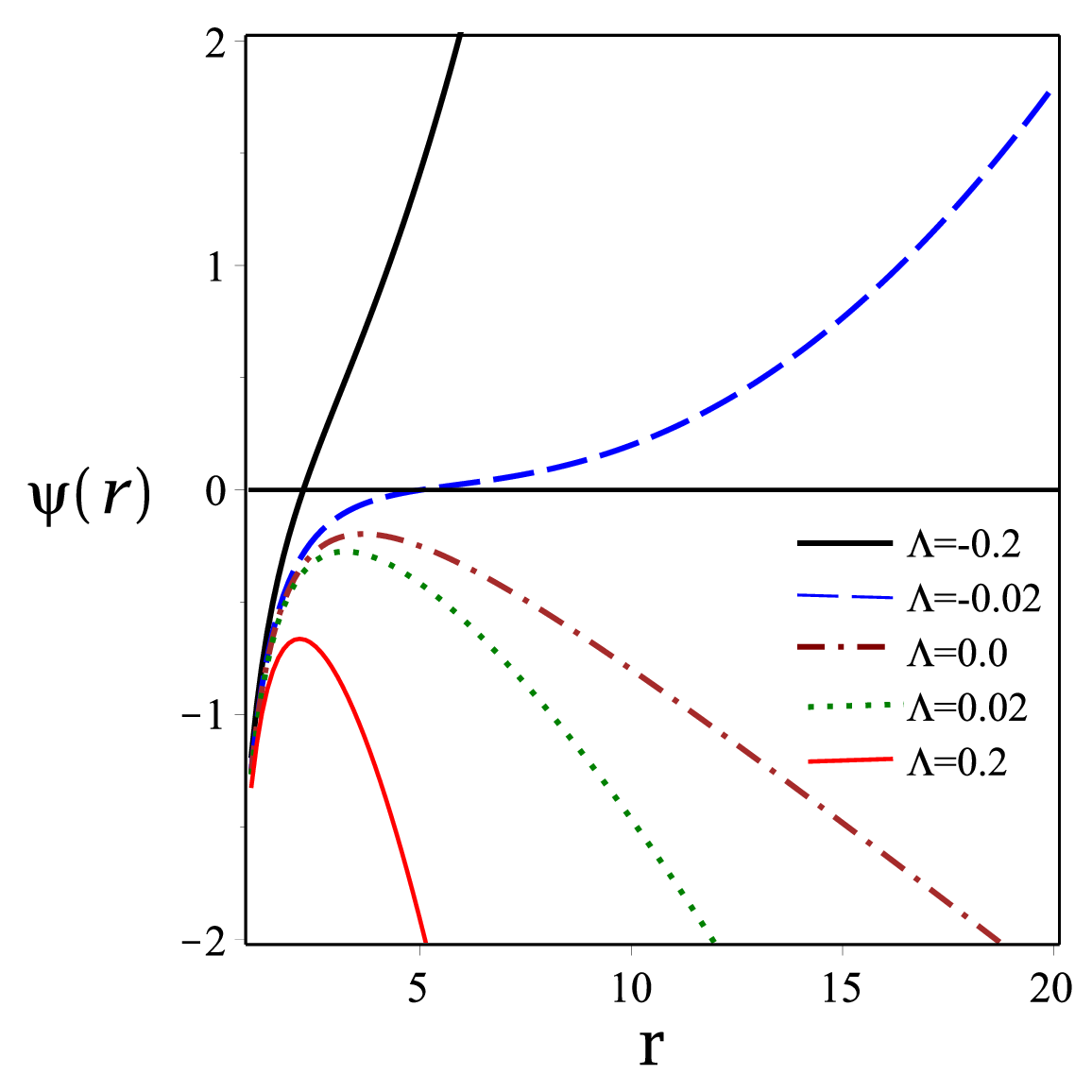} \newline
\caption{The metric function $\protect\psi(r)$ versus $r$ for $q=0.1$, $%
m_{0}=1$, $C_{1}=-0.3$, $C_{2}=-0.1$, $\protect\eta =-1$ and different
values of the cosmological constant.}
\label{Fig1a}
\end{figure}
\begin{figure*}[tbph]
\centering
\subfloat[$C_{2}=-0.069$ and $ \Lambda=-0.02
$]{\includegraphics[width=0.32\textwidth]{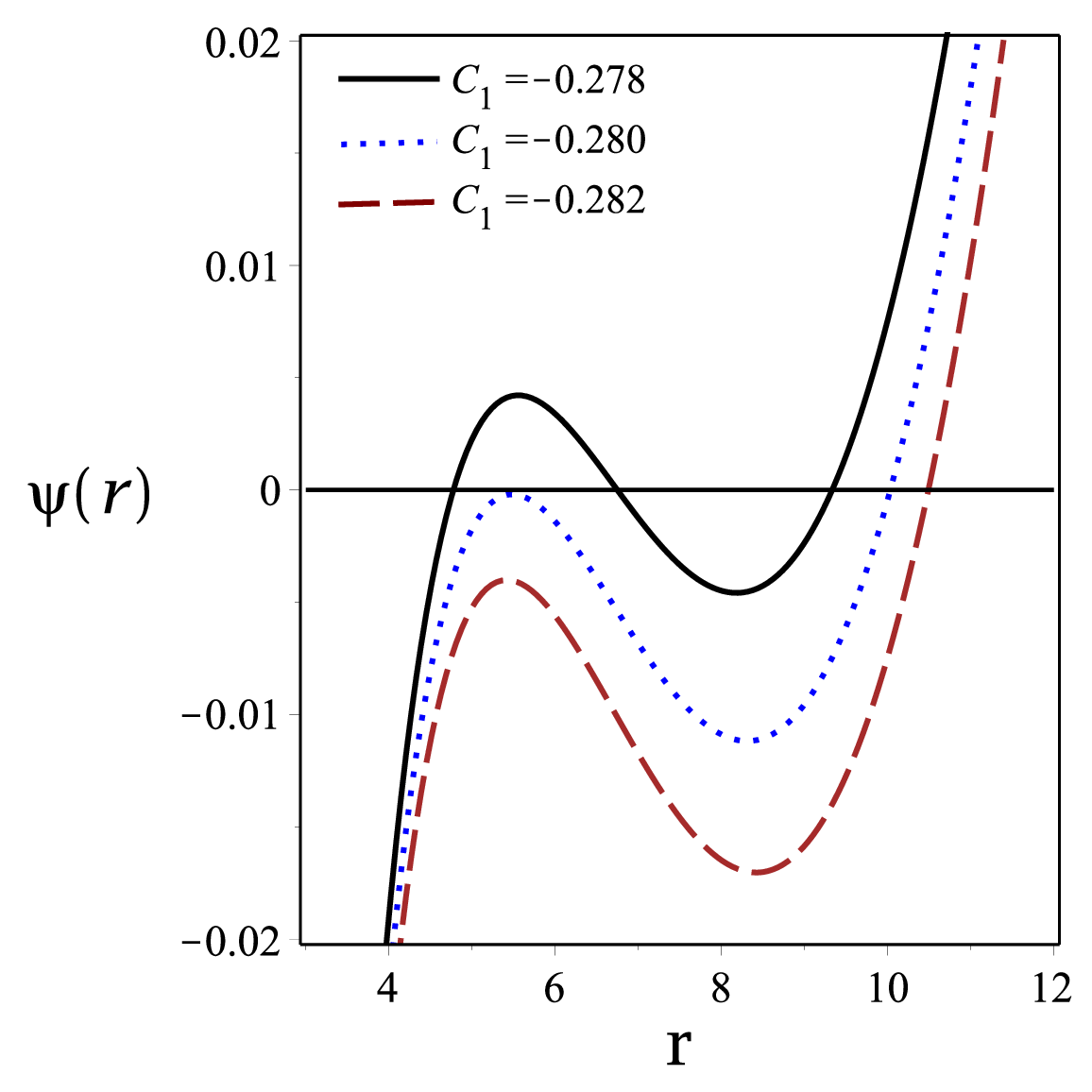}} 
\subfloat[$C_{2}=0.98$
and $ \Lambda=-0.2 $]{\includegraphics[width=0.335\textwidth]{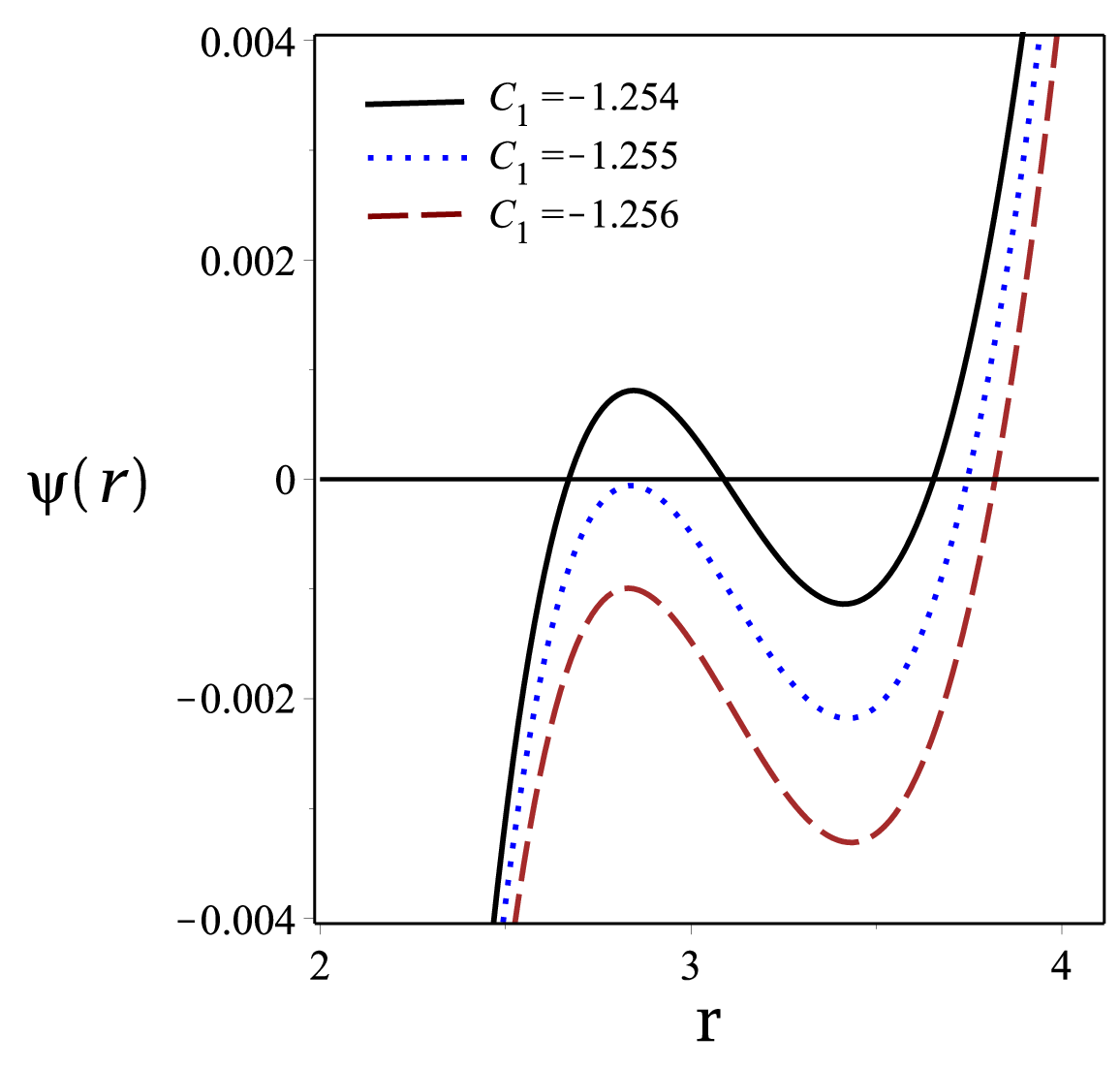}} 
\newline
\caption{The metric function $\protect\psi(r)$ versus $r$ for different
values of $C_{1}$ and $\protect\eta=-1$.}
\label{Fig11b}
\end{figure*}

In Ref. \cite{Panah2019} a correspondence between black hole solutions of
conformal and massive theories of gravity is found which imposed some
constraints on the parameters of massive gravity, which are;

\textit{Constraint (i)}: $C_{1}<0$ and $C_{2}<0$.

\textit{Constraint (ii)}: $C_{1}>0$ and $C_{2}<0$.

Also, these constraints are compatible from the astrophysical point of view
for the study of neutron stars, white dwarfs, and dark energy stars \cite%
{NeutronStar,Whitedwarf,DarkEnergyStar}. Applying \textit{Constraint (i)},
i.e., $C_{1}<0$ and $C_{2}<0$, and special negative values of $C_{1}$ and $%
C_{2}$, we plot the Figure. \ref{Fig1a}. Our results show that there is no
root for $\Lambda >0$ and $\Lambda =0$. Indeed, for special negative values
of parameters $C_{1}$ and $C_{2}$, phantom dS and flat black holes do not
have any root, but phantom AdS black holes (i.e., $\Lambda <0$) have one
root or an event horizon (see the continuous and dashed lines in Fig. \ref%
{Fig1a}).

Another interesting result is related to the existence of multi-horizons for
phantom AdS black holes. In other words, by considering negative values for
the cosmological constant and other parameters, we can find black holes with
one root or an event horizon (see the dashed line in Fig. \ref{Fig11b}(a));
two roots, an inner root, and an event horizon (see the dotted line in Fig. %
\ref{Fig11b}(a)), and multi-horizons which include two inner roots and an
event horizon (see the continuous line in Fig. \ref{Fig11b}(a)). In Fig. \ref{Fig11b}(b), we considered $\Lambda=-0.2$ and studied the
existence of multi-horizons for the mentioned black holes. One can see that
such behavior is observed for $C_{2}$ very close to 1, which is not
acceptable according to what was already mentioned. Therefore, the existence
of multi-horizons for phantom AdS black holes is only visible for very small
values of $\vert \Lambda \vert$.

\begin{figure}[tbph]
\centering
\includegraphics[width=0.3\linewidth]{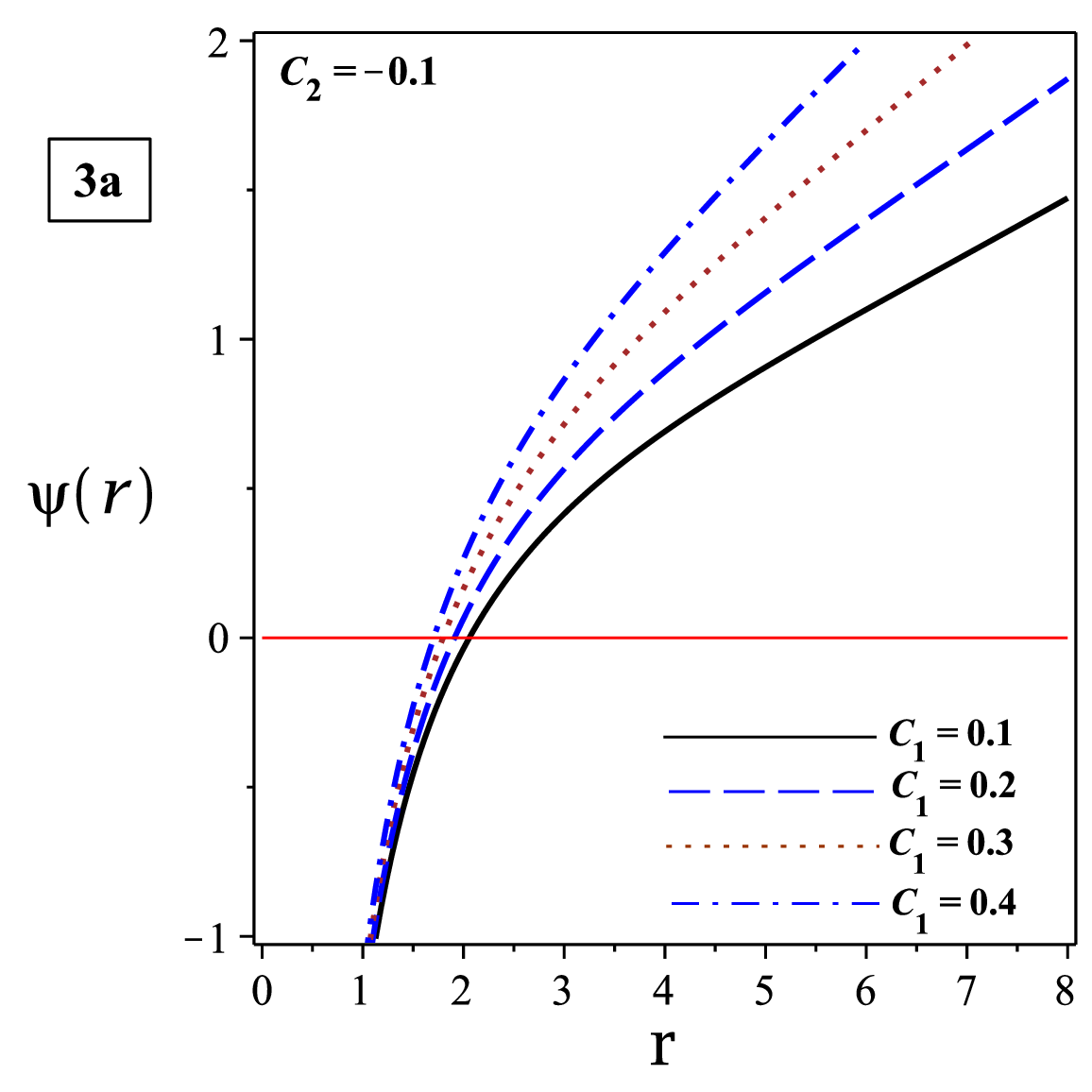} \includegraphics[width=0.3%
\linewidth]{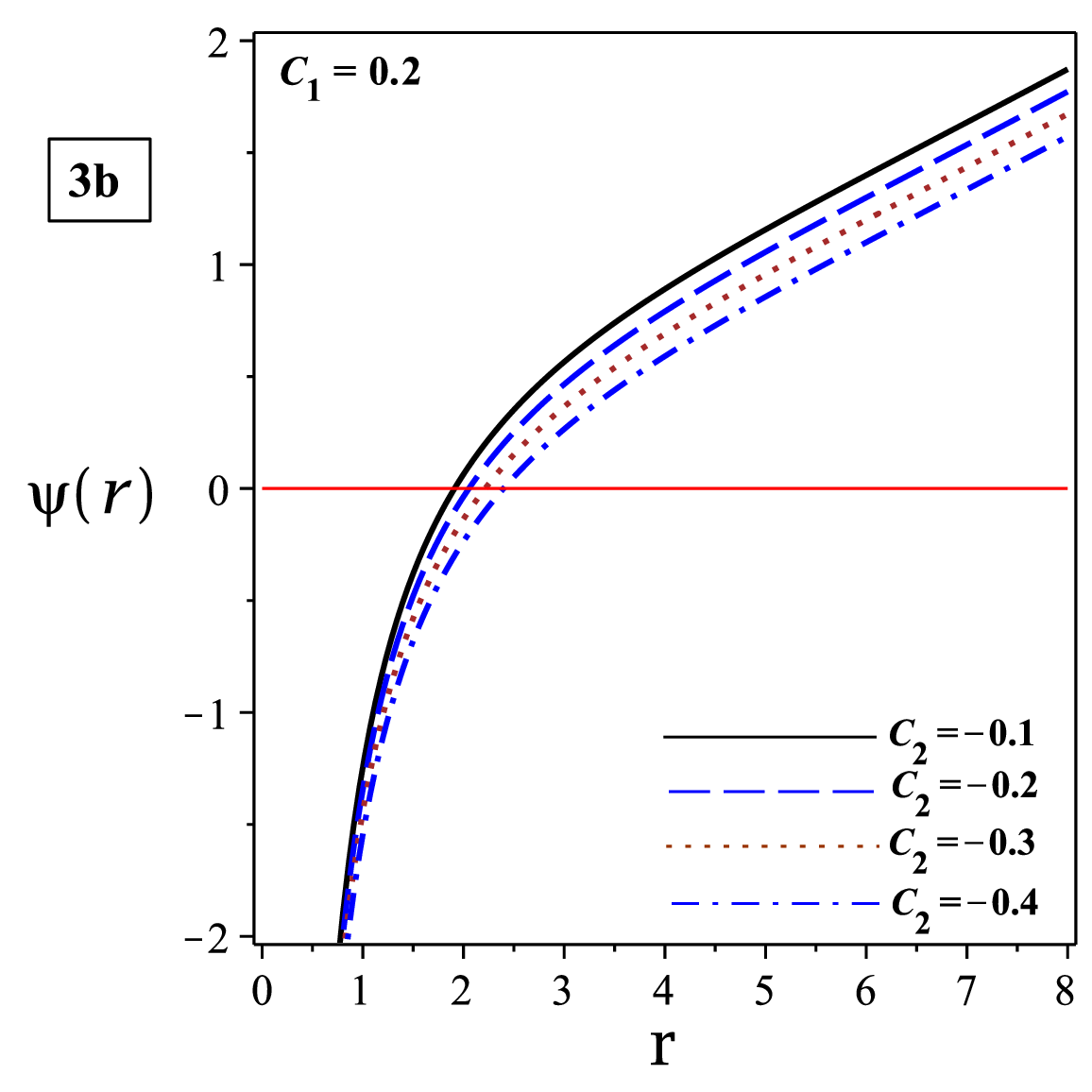} \newline
\includegraphics[width=0.3\linewidth]{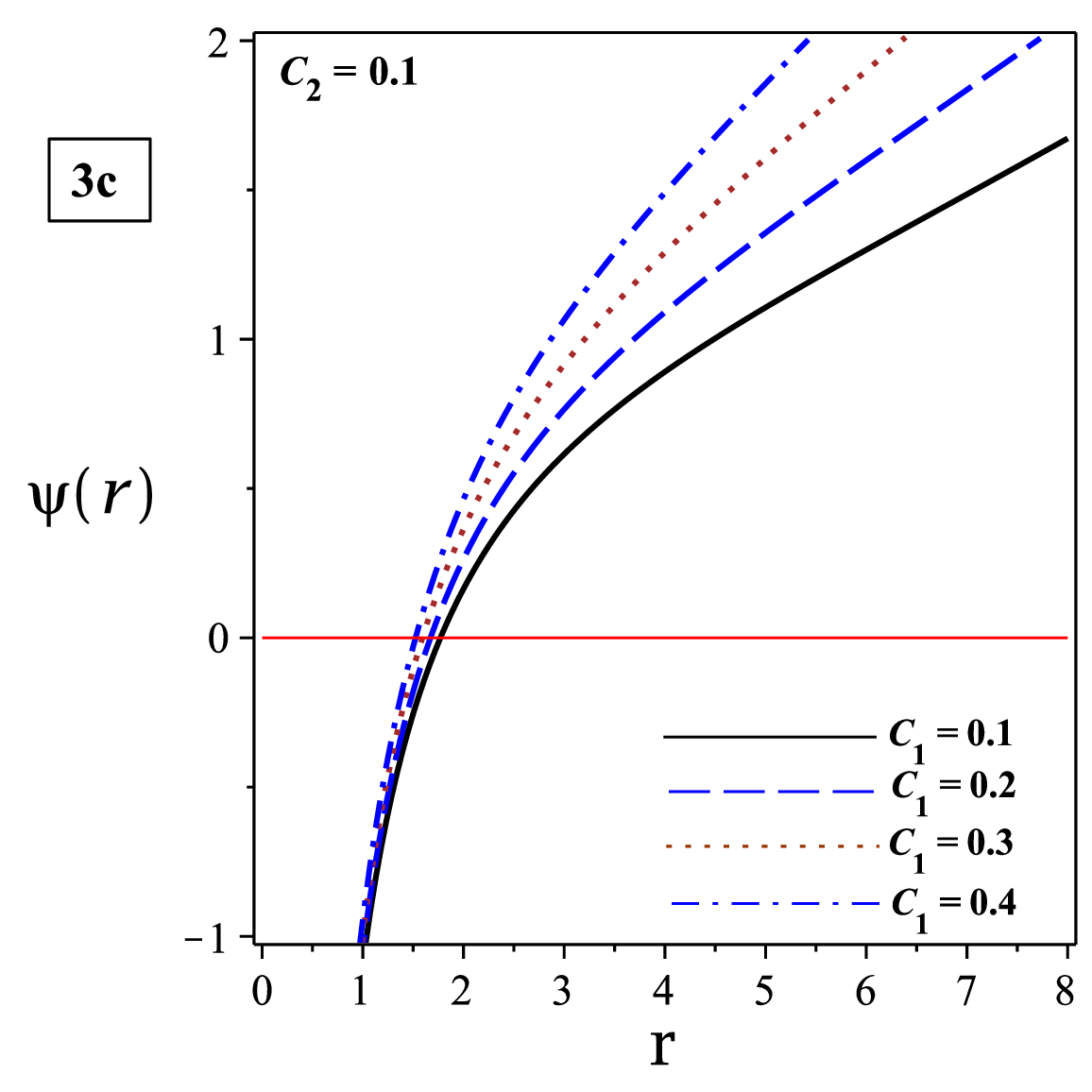} \includegraphics[width=0.3%
\linewidth]{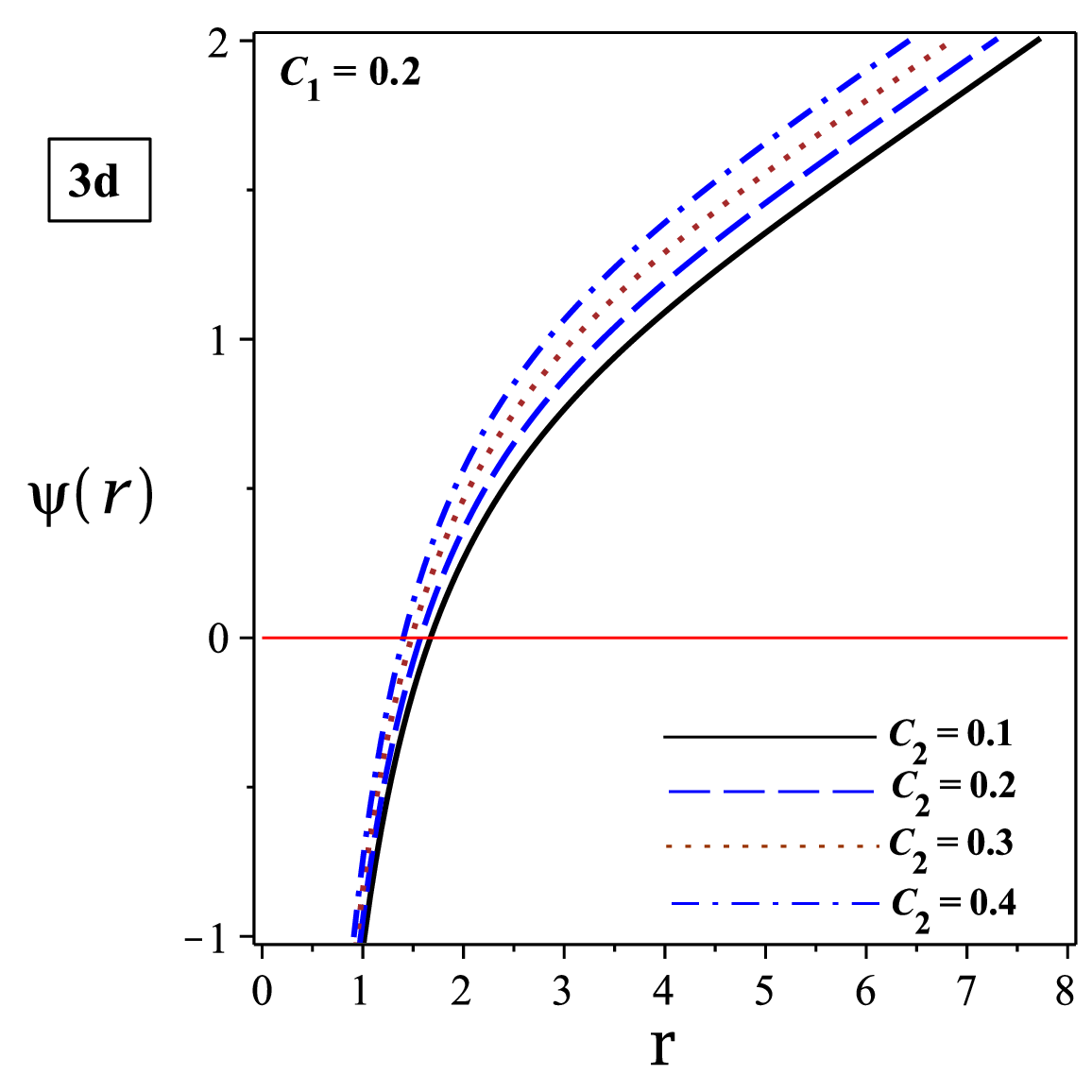} \newline
\includegraphics[width=0.3\linewidth]{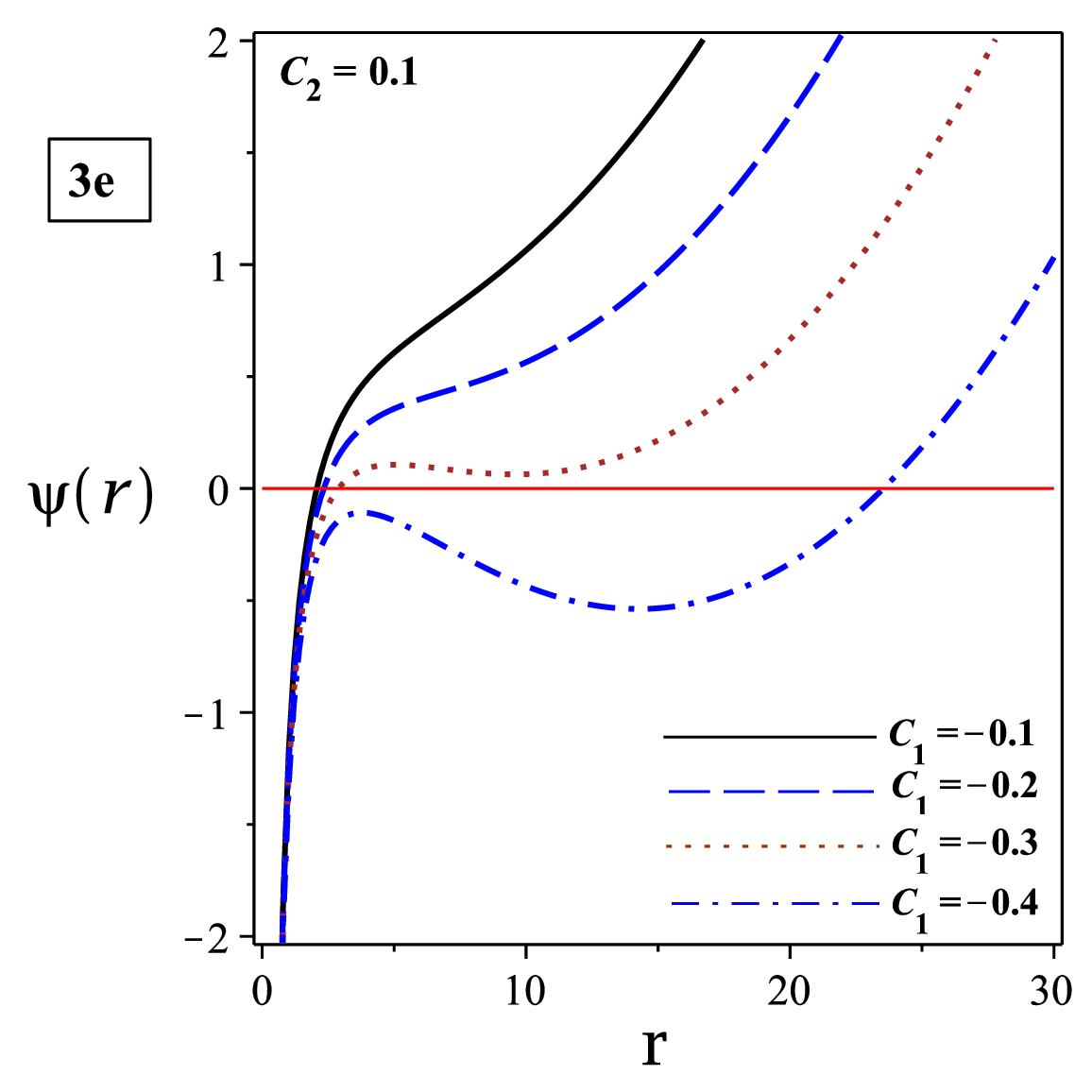} \includegraphics[width=0.3%
\linewidth]{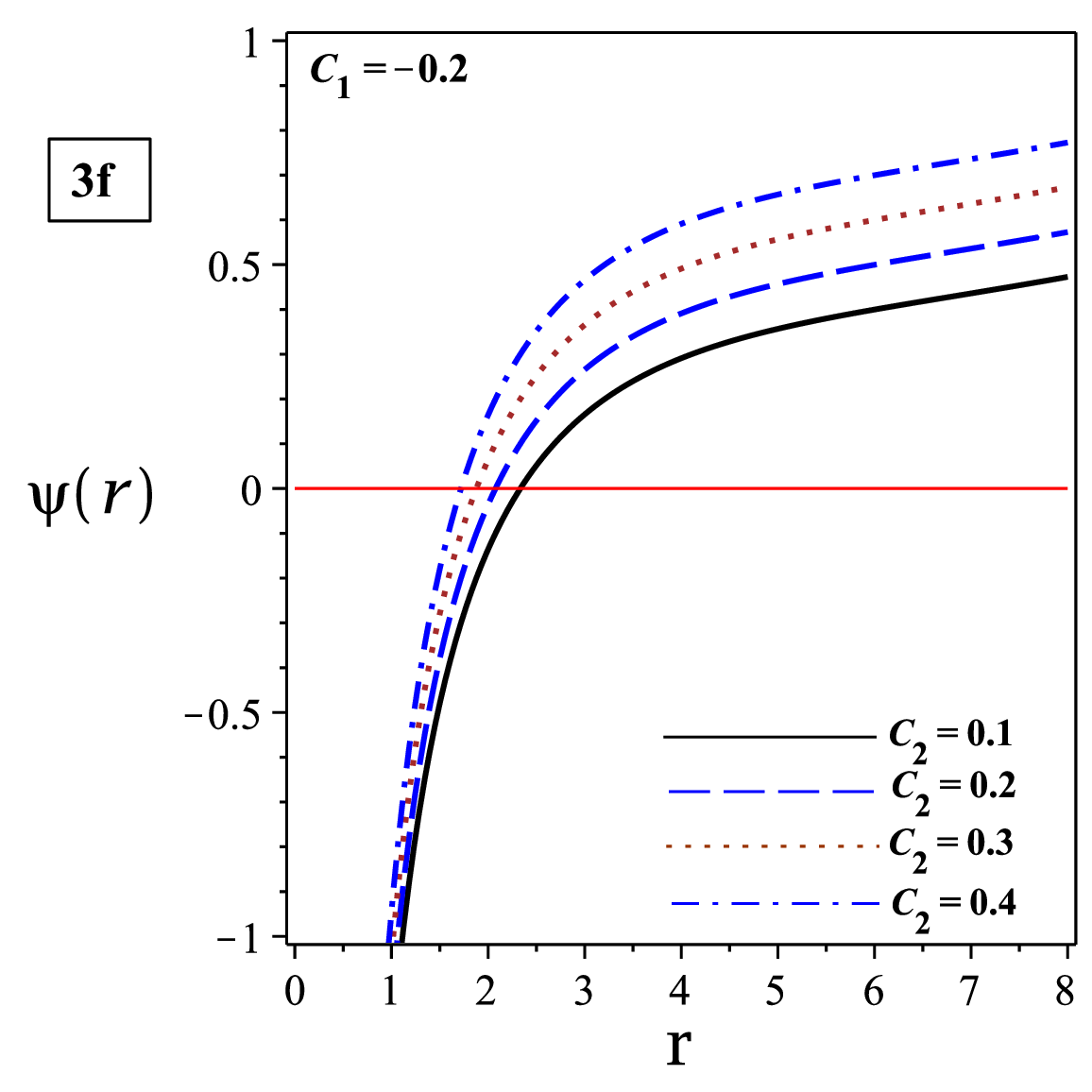} \newline
\caption{The metric function $\protect\psi (r)$ versus $r$ for $q=0.5$, $%
m_{0}=1$, $\protect\eta =-1$, and $\Lambda =-0.02$. Left panels for
different values of $C_{1}$, and right panels for different values of $C_{2}$%
.}
\label{Fig1b}
\end{figure}

To evaluate the effects of massive parameters with different signatures and
values, we plot six panels in Fig. \ref{Fig1b}. Our analysis indicates that:

i) By considering the negative value of $C_{2}$, the radii of phantom AdS
black holes decrease by increasing the positive value of $C_{1}$ (see Fig. %
\ref{Fig1b}a).

ii) For the positive constant value of $C_{1}$, we encounter with large
phantom AdS black holes when the negative value of $C_{2}$ increases (see
Fig. \ref{Fig1b}b).

iii) The radii of phantom AdS black holes decrease when the positive value
of $C_{1}$ increases, provided $C_{2}>0$ (see Fig. \ref{Fig1b}c).

iv) For $C_{1}>0$, by increasing the positive value of $C_{2}$, the radii of
these black holes decreases (see Fig. \ref{Fig1b}d).

v) By increasing negative value of $C_{1}$, we encounter with large black
holes when $C_{2}>0$ (see Fig. \ref{Fig1b}e).

vi) By considering the negative value of $C_{1}$, the radii of phantom AdS
black holes decrease by increasing the positive value of $C_{2}$ (see Fig. %
\ref{Fig1b}f).

\section{Thermodynamics}

\label{sec4}

In this section, we first calculate the conserved and thermodynamic
quantities of the phantom black hole solutions in massive gravity. We extend
phase space to the extended phase space by considering the cosmological
constant as a thermodynamic variable corresponding to pressure and check the
first law of thermodynamics and the Smarr relation. After that, we
investigate the local and global stability of these black holes by employing
heat capacity, and Helmholtz free energy.

\subsection{Thermodynamic Quantities, First Law, and Smarr Relation}

Here, we get the conserved and thermodynamic quantities of these black
holes, which are necessary to study thermodynamic properties. For this
purpose, we express the mass ($m_{0}$) in terms of the radius of the event
horizon $(r_{e})$\ and other quantities of these solutions such as, the
charge $(q)$, the cosmological constant ($\Lambda $), and massive parameters
($C_{1}$, and $C_{2}$) in the following equation which is extracted by
equating $g_{tt}=\psi (r)$\ to zero 
\begin{equation}
m_{0}=\frac{-\Lambda r_{e}^{3}}{6}+\frac{C_{1}r_{e}^{2}}{4}+\frac{\left(
1+C_{2}\right) r_{e}}{2}+\frac{\eta q^{2}}{2r_{e}}.  \label{mm}
\end{equation}

In order to get the Hawking temperature, we obtain the superficial gravity
for these black holes as 
\begin{equation}
\kappa =\left. \frac{g_{tt}^{\prime }}{2\sqrt{-g_{tt}g_{rr}}}=\right\vert
_{r=r_{e}}=\left. \frac{\psi ^{\prime }(r)}{2}\right\vert _{r=r_{e}},
\label{k}
\end{equation}%
by substituting the mass (\ref{mm})\ within the equation (\ref{k}) and using
the obtained metric function, one can calculate the superficial gravity
which is given by 
\begin{equation}
\kappa =\frac{-\Lambda r_{e}}{2}+\frac{C_{1}}{2}+\frac{1+C_{2}}{2r_{e}}-%
\frac{\eta q^{2}}{2r_{e}^{3}},
\end{equation}%
now, we can get the Hawking temperature in the following form 
\begin{equation}
T=\frac{\kappa }{2\pi }=\frac{-\Lambda r_{e}}{4\pi }+\frac{C_{1}}{4\pi }+%
\frac{1+C_{2}}{4\pi r_{e}}-\frac{\eta q^{2}}{4\pi r_{e}^{3}}.
\label{TemPhMassive}
\end{equation}

Another thermodynamic quantity is the electric charge. We can obtain it by
using the Gauss law. So, the electric charge of the black hole per unit
volume is given 
\begin{equation}
Q=\frac{F_{tr}}{4\pi }\int_{0}^{2\pi }\int_{0}^{\pi }\sqrt{g}d\theta
d\varphi =\frac{q}{4\pi },  \label{Q}
\end{equation}%
in the above equation, $F_{tr}=\frac{q}{r^{2}}$, and for case $t=$\ constant
and $r=$constant, the determinant of metric tensor $g$ is given by $%
r^{4}\sin ^{2}\theta $.

We find the electric potential ($U$) at the event horizon with respect to
the reference ($r\rightarrow \infty $) by using the definition $%
U=-\int_{r_{e}}^{+\infty }F_{tr}dr$, where $F_{tr}=\partial
_{t}A_{r}-\partial _{r}A_{t}$. This quantity is%
\begin{equation}
U=\frac{q}{r_{e}}.  \label{elcpoPhMassive}
\end{equation}

One can use the area law to extract the entropy of black holes, which leads
to 
\begin{equation}
S=\frac{\mathcal{A}}{4},  \label{SFR}
\end{equation}%
where $\mathcal{A}$\ is the horizon area. The horizon area per unit volume
is given by%
\begin{equation}
\mathcal{A}=\left. \int_{0}^{2\pi }\int_{0}^{\pi }\sqrt{g_{\theta \theta
}g_{\varphi \varphi }}\right\vert _{r=r_{e}}=\left. r^{2}\right\vert
_{r=r_{e}}=r_{e}^{2},  \label{A}
\end{equation}%
by replacing the horizon area (\ref{A}) within Eq. (\ref{SFR}), we can get
the entropy of phantom black holes per unit volume in massive gravity in the
following form 
\begin{equation}
S=\frac{r_{e}^{2}}{4}.  \label{S}
\end{equation}

In the extended phase space, the cosmological constant acts as a
thermodynamic variable corresponding to pressure 
\begin{equation}
P=\frac{-\Lambda }{8\pi },  \label{P}
\end{equation}%
where this postulate leads to an interpretation of the black hole mass as
enthalpy \cite{Kastor}.

We obtain the total mass of these black holes per unit volume by using
Ashtekar-Magnon-Das (AMD) approach \cite{AMDI,AMDII} which is extracted 
\begin{equation}
M=\frac{m_{0}}{4\pi },  \label{AMDMass}
\end{equation}%
where substituting the mass (\ref{mm}) and $\Lambda =-8\pi P$\ within the
equation (\ref{AMDMass}), we find 
\begin{equation}
M=\frac{r_{e}^{3}P}{3}+\frac{C_{1}r_{e}^{2}}{16\pi }+\frac{\left(
1+C_{2}\right) r_{e}}{8\pi }+\frac{\eta q^{2}}{8\pi r_{e}}.  \label{MM}
\end{equation}

It is straightforward to show that the conserved and thermodynamic
quantities satisfy the first law of thermodynamics in extended phase space
in the following form 
\begin{equation}
dM=TdS+\eta UdQ+VdP+\mathcal{C}_{1}dC_{1}+\mathcal{C}_{2}dC_{2}, \label{First}
\end{equation}%
where the conjugate quantities associated with the intensive parameters $%
S,Q,P,C_{i}$'s are 
\begin{eqnarray}
T &=&\left( \frac{\partial M}{\partial S}\right) _{Q,P,C_{i}},  \notag \\
\eta U &=&\left( \frac{\partial M}{\partial Q}\right) _{S,P,C_{i}},  \notag
\\
V &=&\left( \frac{\partial M}{\partial P}\right) _{S,Q,C_{i}}=\frac{r_{e}^{3}%
}{3},  \notag \\
\mathcal{C}_{1} &=&\left( \frac{\partial M}{\partial C_{1}}\right)
_{S,Q,C_{2}}=\frac{r_{e}^{2}}{16\pi },  \notag \\
\mathcal{C}_{2} &=&\left( \frac{\partial M}{\partial C_{2}}\right)
_{S,Q,C_{1}}=\frac{r_{e}}{8\pi },
\end{eqnarray}%
where $T=\left( \frac{\partial M}{\partial S}\right) _{Q,P,C_{i}}$\ and $%
\eta U=\left( \frac{\partial M}{\partial Q}\right) _{S,P,C_{i}}$\ are in
agreement with those of calculated in Eqs. (\ref{TemPhMassive}) and (\ref%
{elcpoPhMassive}), respectively, provided $\Lambda =-8\pi P$.

It is notable that an active astrophysical black hole is always surrounded by an accretion disk containing various types of baryonic matter, including electric charges. With our first law of thermodynamics, we can describe how energy is exchanged between the black hole and its surroundings. A striking feature of phantom solutions is the change in sign of the work term ($UQ$ in Eq. (\ref{First})), which indicates negative energy from electromagnetic action. When this type of energy is exchanged for work done or received by the black hole, the contribution will be symmetrical to that of a normal solution.

Despite this strong characteristic in ghost solutions, the thermodynamic system can present itself like all asymptotically AdS black hole solutions, i.e. the black hole can start by accreting matter and increasing its entropy, initially in an unstable small black hole phase (small event horizon area). As the black hole accretes more matter, it can move on to a new phase, called a large black hole (large area), where the thermodynamic system is stable. So even if the solution is phantom, but the black hole may be in a thermodynamically stable phase in nature.

The Smarr relation can be derived by a scaling (dimensional) argument as 
\begin{equation}
M=2\left( TS-PV\right) +\eta UQ-\mathcal{C}_{1}C_{1},
\end{equation}%
where $C_{2}$\ does not appear in the Smarr relation since it has scaling
weight $0$ \cite{Hendi2017}. Indeed, the $C_{2}$\ term in the metric
function is a constant term in $4-$dimensional spacetime with no
thermodynamical contribution, so we set $dC_{2}=0$.

\subsection{ Local thermal stability in canonical ensemble}

In the canonical ensemble context, the local stability of a thermodynamic
system is determined by studying the heat capacity. Indeed, the
discontinuities of this quantity mark the possible thermal phase transitions
that the system can undergo. Also, the positivity of the heat capacity
corresponds to thermal stability while the opposite indicates instability.
Moreover, the roots of heat capacity may yield possible changes between
stable/unstable states (or bound points). To get this information, we
calculate the heat capacity of the solutions and investigate the local
stability of the phantom black holes. We also study the effects of
parameters of massive gravity on local stability areas.

To obtain the heat capacity, we first re-write the total mass of the black
hole (\ref{MM}) in terms of the electrical charge (\ref{Q}), the entropy (%
\ref{S}), pressure (\ref{P}) and massive parameters as 
\begin{equation}
M\left( S,Q,P,C_{i}\right) =\frac{\pi \eta Q^{2}}{\sqrt{S}}+\frac{\left(
1+C_{2}\right) \sqrt{S}}{4\pi }+\frac{C_{1}S}{4\pi }+\frac{8PS^{3/2}}{3},
\label{MSQ}
\end{equation}%
we re-write the temperature by using the equation (\ref{MSQ}) 
\begin{equation}
T=\left( \frac{\partial M\left( S,Q,P,C_{i}\right) }{\partial S}\right)
_{Q,P,C_{i}}=4P\sqrt{S}+\frac{C_{1}}{4\pi }+\frac{1+C_{2}}{8\pi \sqrt{S}}-%
\frac{\pi \eta Q^{2}}{2S^{3/2}}.  \label{TM}
\end{equation}

Now, we can obtain the heat capacity which is defined 
\begin{equation}
C_{Q,P,C_{i}}=\frac{T}{\left( \frac{\partial T}{\partial S}\right)
_{Q,P,C_{i}}}=\frac{\left( \frac{\partial M\left( S,Q,P,C_{i}\right) }{%
\partial S}\right) _{Q,P,C_{i}}}{\left( \frac{\partial ^{2}M\left(
S,Q,P,C_{i}\right) }{\partial S^{2}}\right) _{Q,P,C_{i}}},  \label{Heat}
\end{equation}%
by considering Eqs. (\ref{MSQ}) and (\ref{TM}), the heat capacity is given
by 
\begin{equation}
C_{Q,P,C_{i}}=\frac{2\left( 32\pi PS^{2}+2C_{1}S^{3/2}\right) }{12\pi
^{2}\eta Q^{2}-\left( 1-32\pi PS+C_{2}\right) S}+\frac{\left( 1+C_{2}\right)
S-4\pi ^{2}\eta Q^{2}}{6\pi ^{2}\eta Q^{2}-\frac{\left( 1-32\pi
PS+C_{2}\right) S}{2}}.  \label{Heat1}
\end{equation}

By studying the heat capacity of the black hole, we can find two important
points that are related to the physical limitation and phase transition
critical points. Indeed, the root of heat capacity $\left( C_{Q,P,C_{i}}=0%
\text{, or }T=\left( \frac{\partial M\left( S,Q,P,C_{i}\right) }{\partial S}%
\right) _{Q,P,C_{i}}=0\right) $\ is representing a border line between
physical $\left( \left( \frac{\partial M\left( S,Q,P,C_{i}\right) }{\partial
S}\right) _{Q,P,C_{i}}>0\right) $\ and non-physical $\left( \left( \frac{%
\partial M\left( S,Q,P,C_{i}\right) }{\partial S}\right)_{Q,P,C_{i}}<0%
\right) $\ black holes (which is known as the physical limitation point).
Notably, the system at this point has a change in the sign of the heat
capacity. In addition, the divergencies of the heat capacity $\left( \left( 
\frac{\partial ^{2}M\left( S,Q,P,C_{i}\right) }{\partial S^{2}}\right)
_{Q,P,C_{i}}=0\right) $ represent phase transition critical points of black
holes. Therefore, these important points are determined by the following
relations 
\begin{equation}
\left\{ 
\begin{array}{c}
\left( \frac{\partial M\left( S,Q,P,C_{i}\right) }{\partial S}\right)
_{Q,P,C_{i}}=0, ~~~\text{bound point} \\ 
\\ 
\left( \frac{\partial ^{2}M\left( S,Q,P,C_{i}\right) }{\partial S^{2}}%
\right) _{Q,P,C_{i}}=0,~~~\text{phase transition point}%
\end{array}%
\right. .  \label{PhysBound}
\end{equation}

We find two phase transition critical points which are 
\begin{equation}
S_{\text{div}_{\pm }}=\frac{1+C_{2}\pm \sqrt{1+2C_{2}+C_{2}^{2}-1536\pi
^{3}\eta Q^{2}P }}{64P },
\end{equation}%
where for phantom AdS black holes, the positive point is $S_{\text{div}_{+}}=%
\frac{1+C_{2}+\sqrt{1+2C_{2}+C_{2}^{2}-1536\pi ^{3}\eta Q^{2}P }}{64P }$. It
is clear that there is a phase transition critical point, and it depends on $%
P $, $Q$, $\eta $, and $C_{2}$. As a result, one of the parameters of
massive gravity (i.e., $C_{1}$) does not affect the divergence of the heat
capacity. The behavior of the figure (\ref{Fig6}) confirms this point. In
other words, by varying the parameter $C_{1}$, the divergence point of the
heat capacity does not change.

\begin{figure*}[tbph]
\centering
\subfloat[$C_{2}=-0.069$ and $ \Lambda=-0.02
$]{\includegraphics[width=0.34\textwidth]{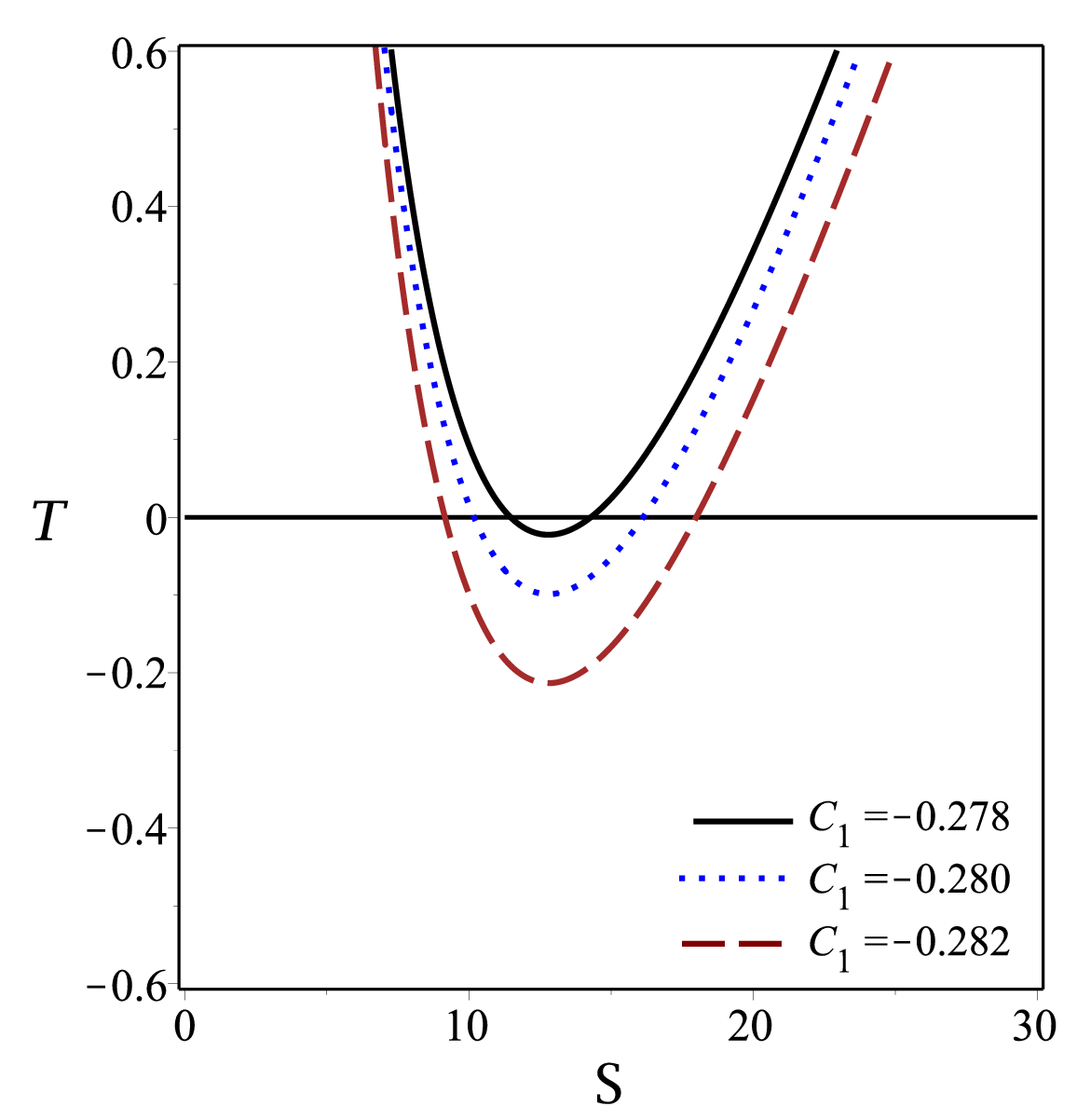}} 
\subfloat[$C_{2}=-0.069$
and $ \Lambda=-0.02 $]{\includegraphics[width=0.355\textwidth]{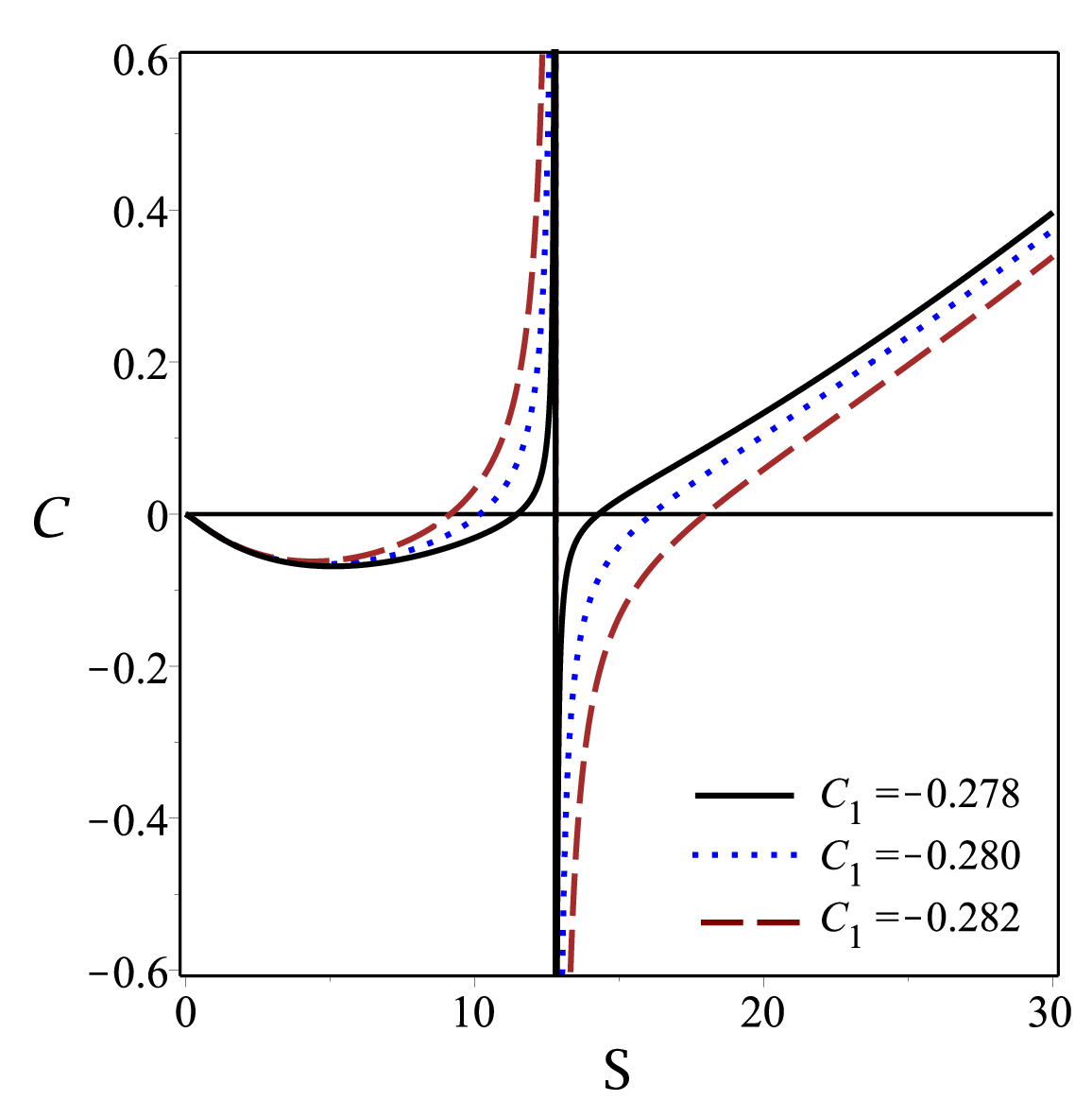}} 
\newline
\caption{The temperature $T$ and heat capacity $C_{Q,P,C_{i}}$ versus $S$
for $Q=0.1$, $\protect\eta=-1$, and different values of negative $C_{1}$.}
\label{Fig4}
\end{figure*}
\begin{figure*}[tbph]
\centering
\subfloat[$C_{1}=0.2$]{\includegraphics[width=0.32\textwidth]{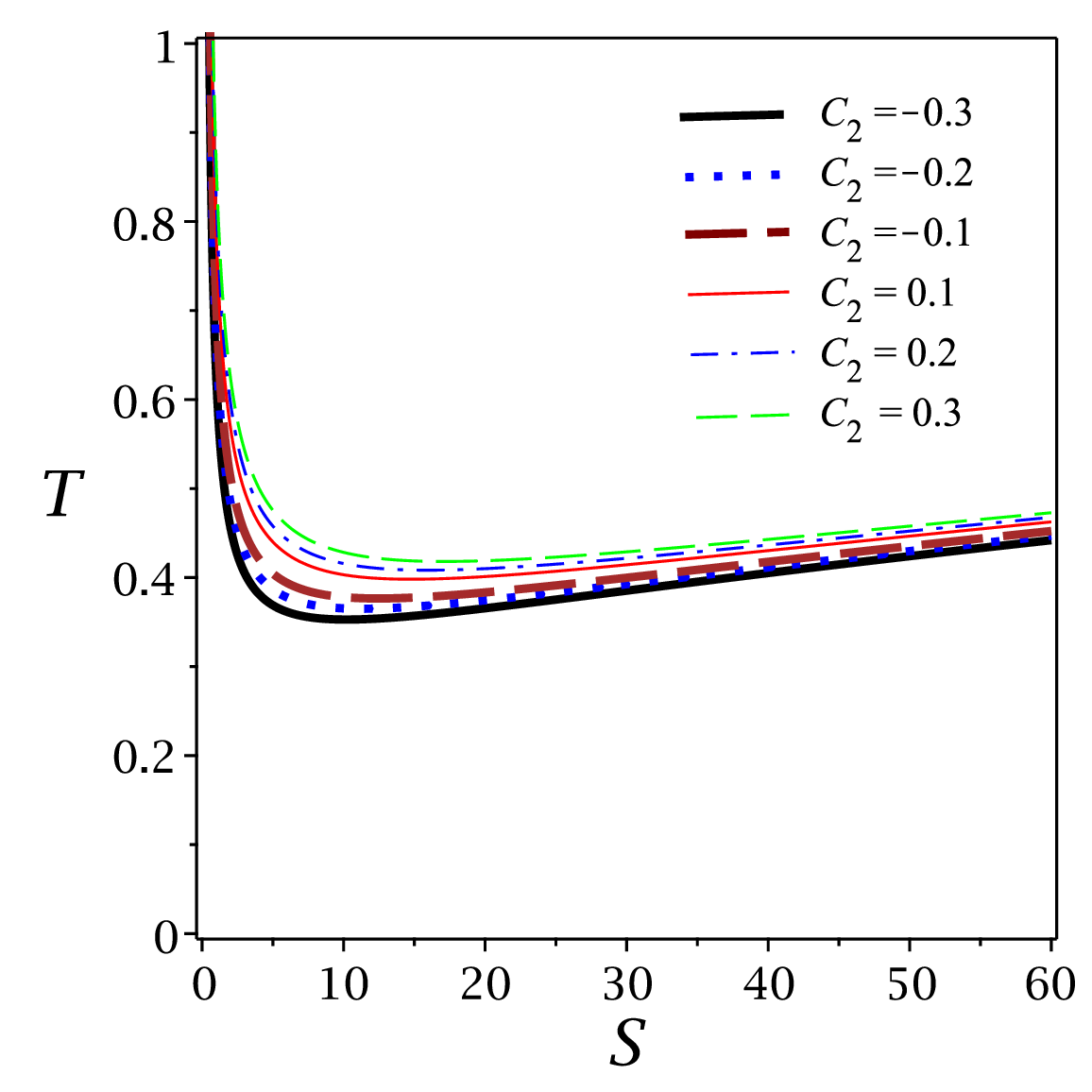}} %
\subfloat[$C_{1}=-0.2$]{\includegraphics[width=0.32\textwidth]{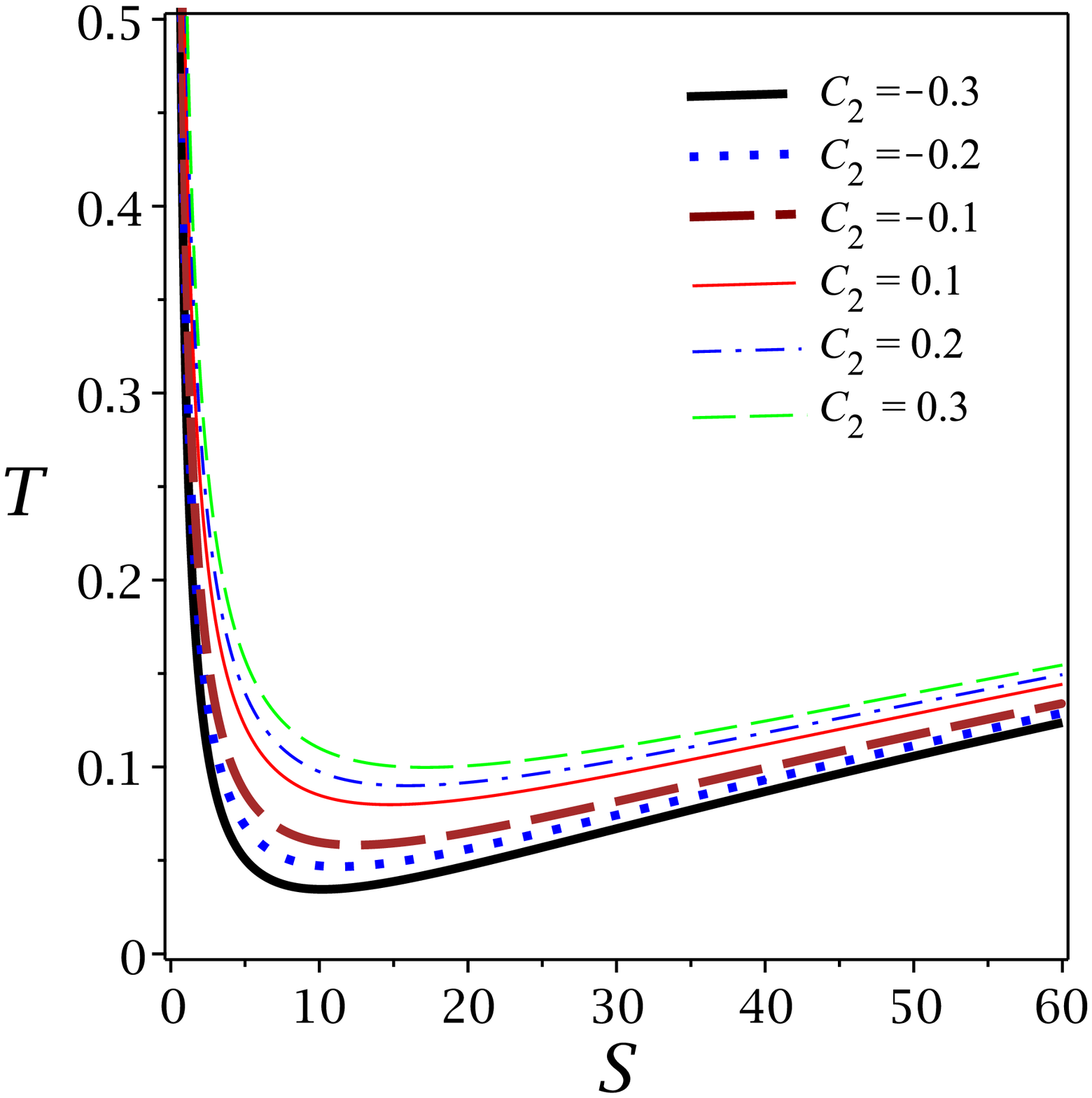}} 
\newline
\subfloat[$C_{1}=0.2$]{\includegraphics[width=0.32\textwidth]{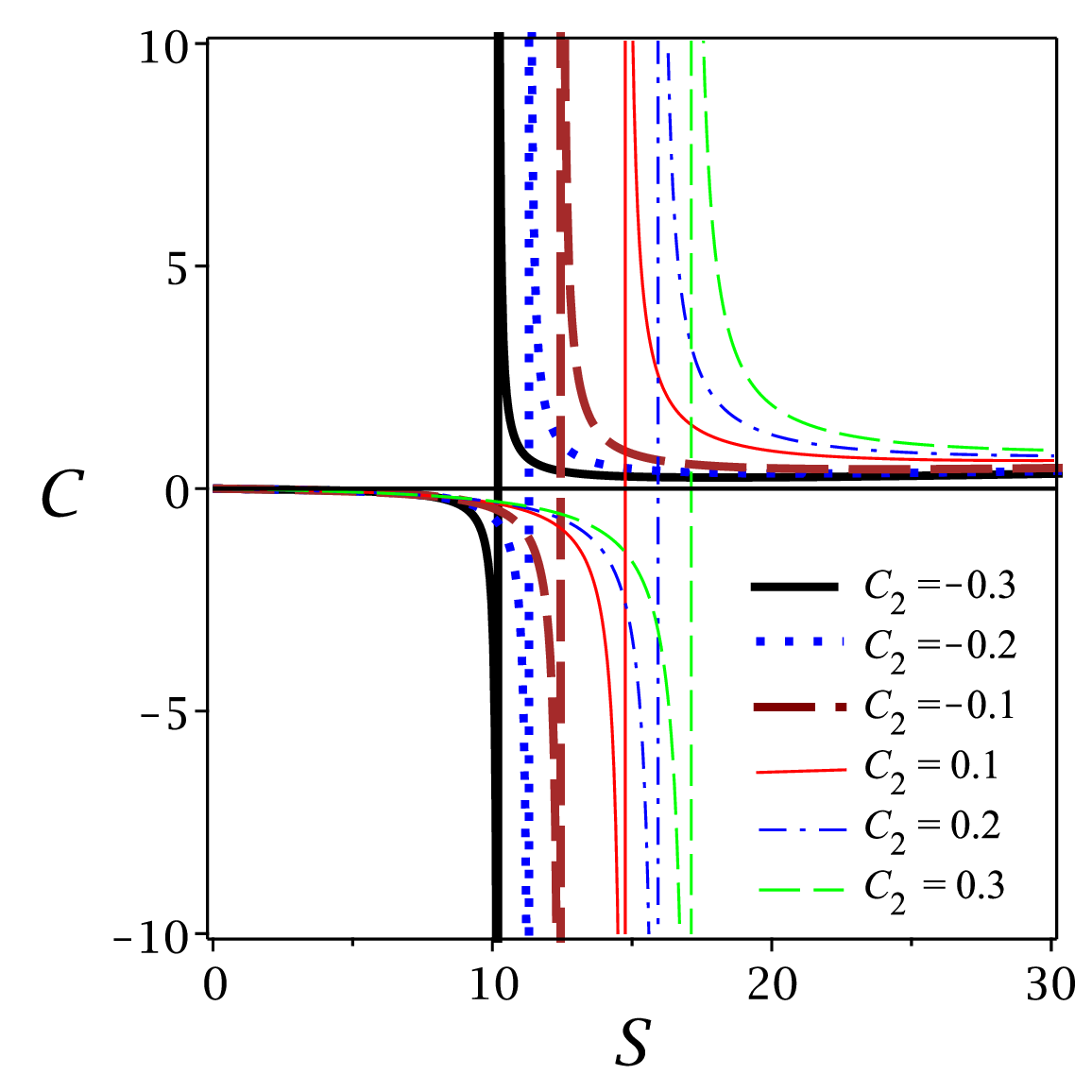}} %
\subfloat[$C_{1}=-0.2$]{\includegraphics[width=0.32\textwidth]{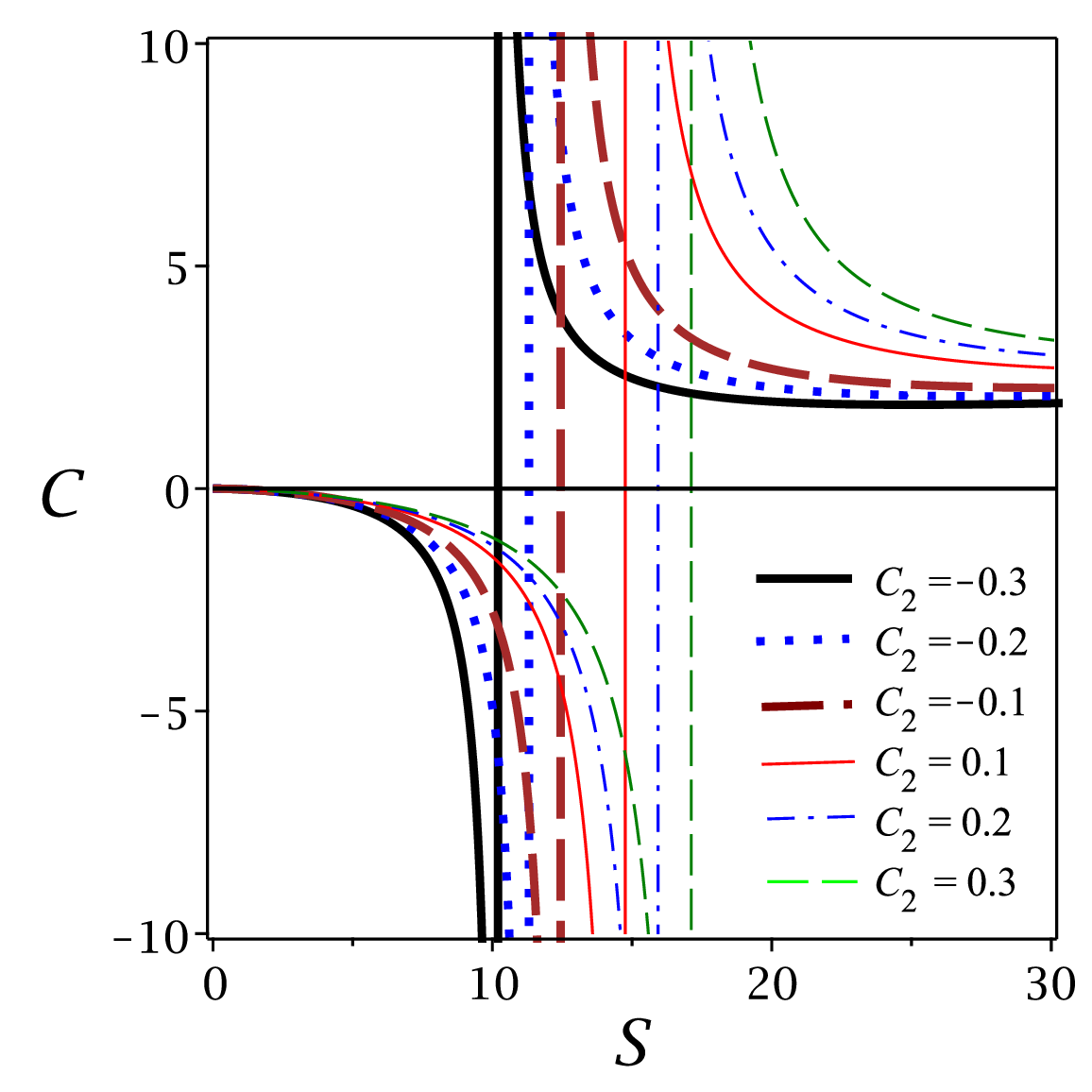}} 
\newline
\caption{The temperature $T$ and heat capacity $C_{Q,P,C_{i}}$ versus $S$
for $Q=0.1$, $\protect\eta=-1$, $\Lambda=-0.02$, and different values of
negative $C_{2}$. }
\label{Fig5}
\end{figure*}
\begin{figure*}[tbph]
\centering
\subfloat[$C_{2}=0.1$]{\includegraphics[width=0.32\textwidth]{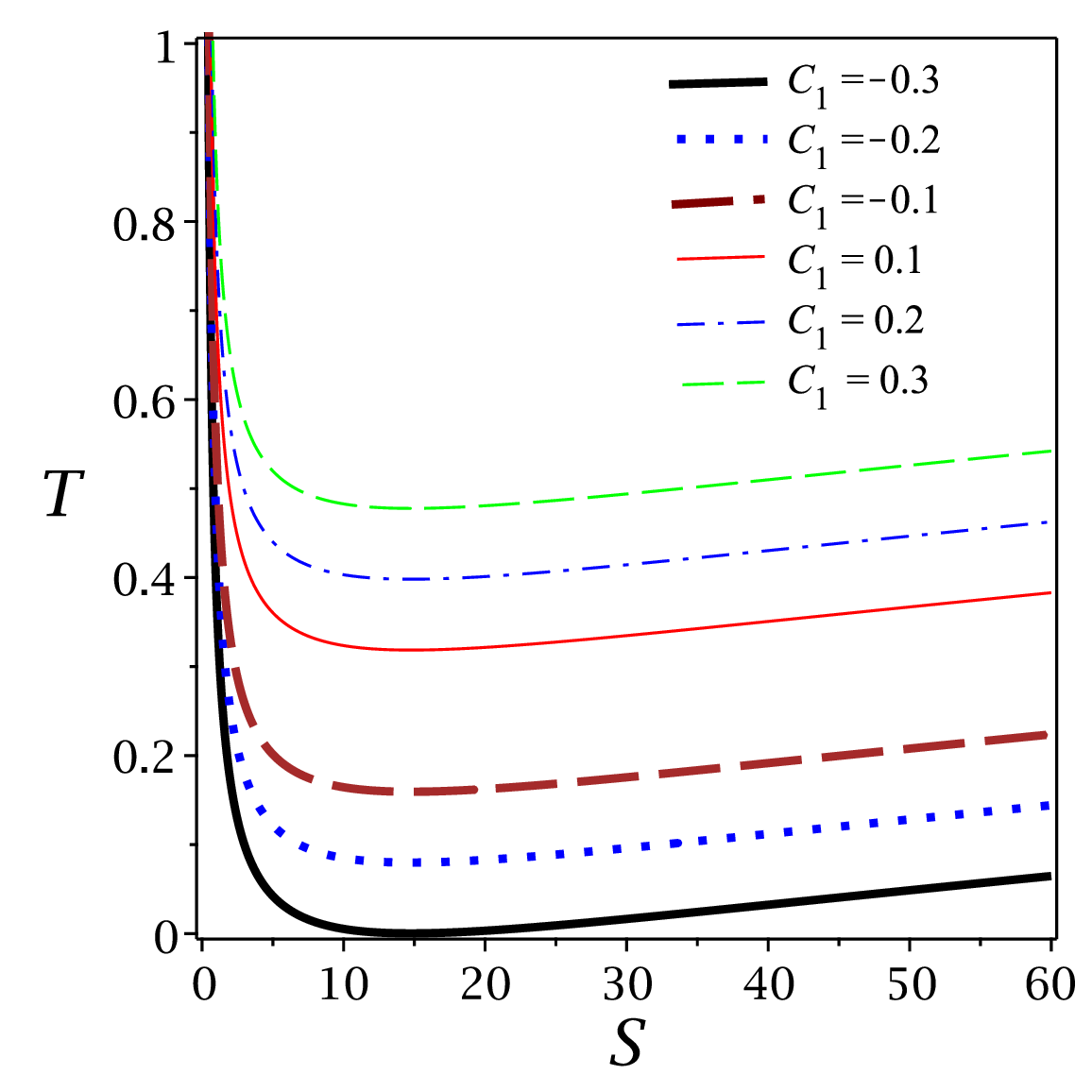}} %
\subfloat[$C_{2}=-0.1$]{\includegraphics[width=0.32\textwidth]{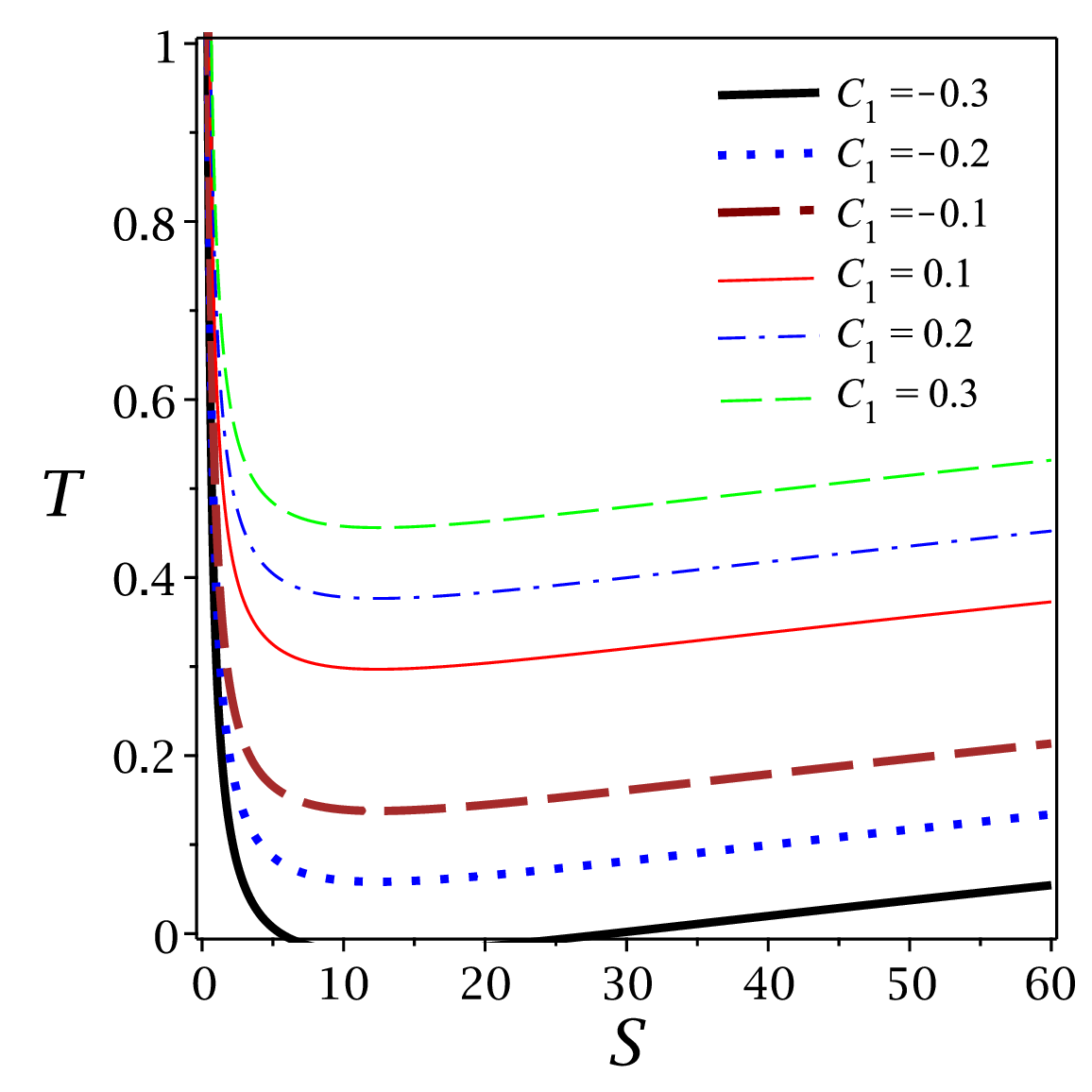}} 
\newline
\subfloat[$C_{2}=0.1$]{\includegraphics[width=0.32\textwidth]{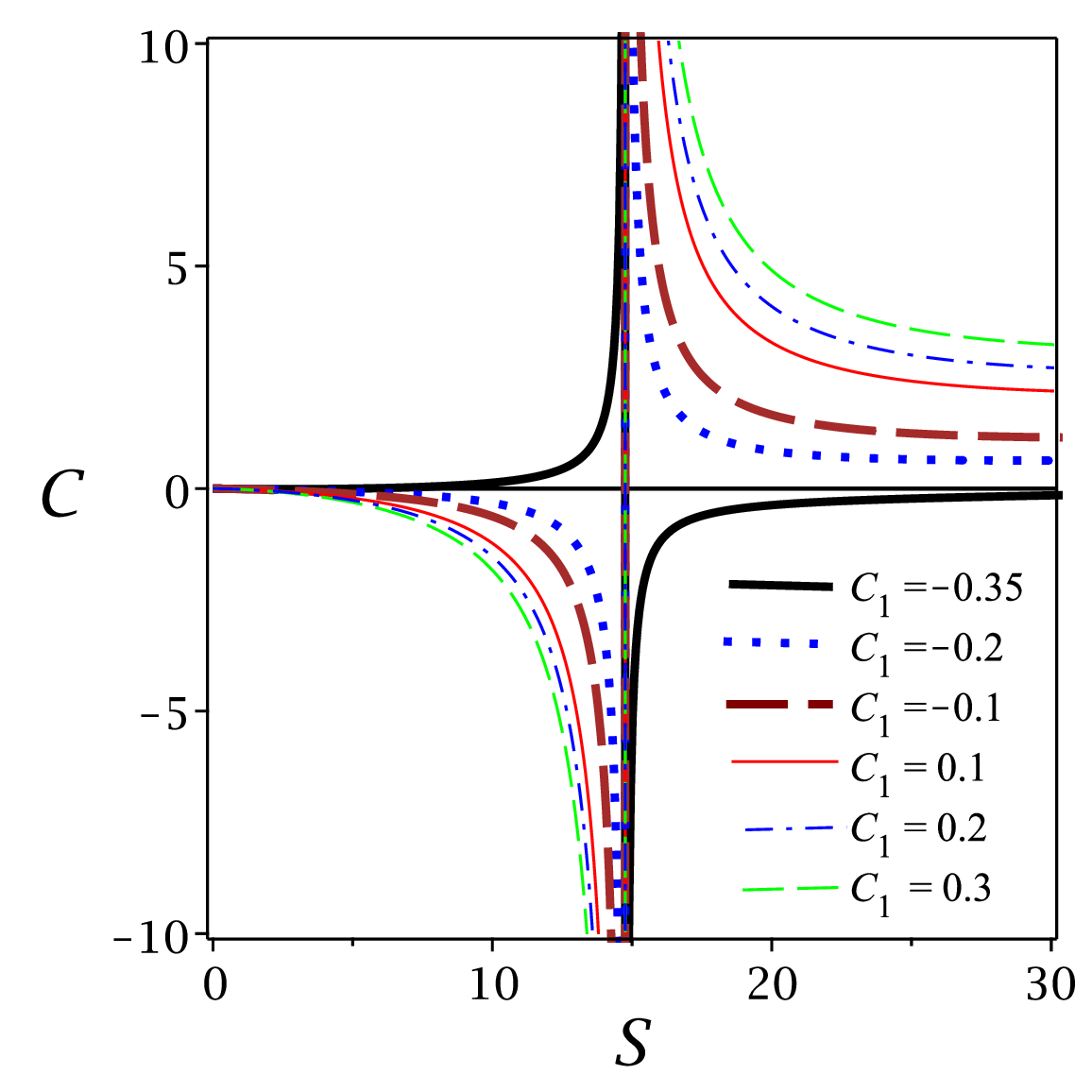}} %
\subfloat[$C_{2}=-0.1$]{\includegraphics[width=0.32\textwidth]{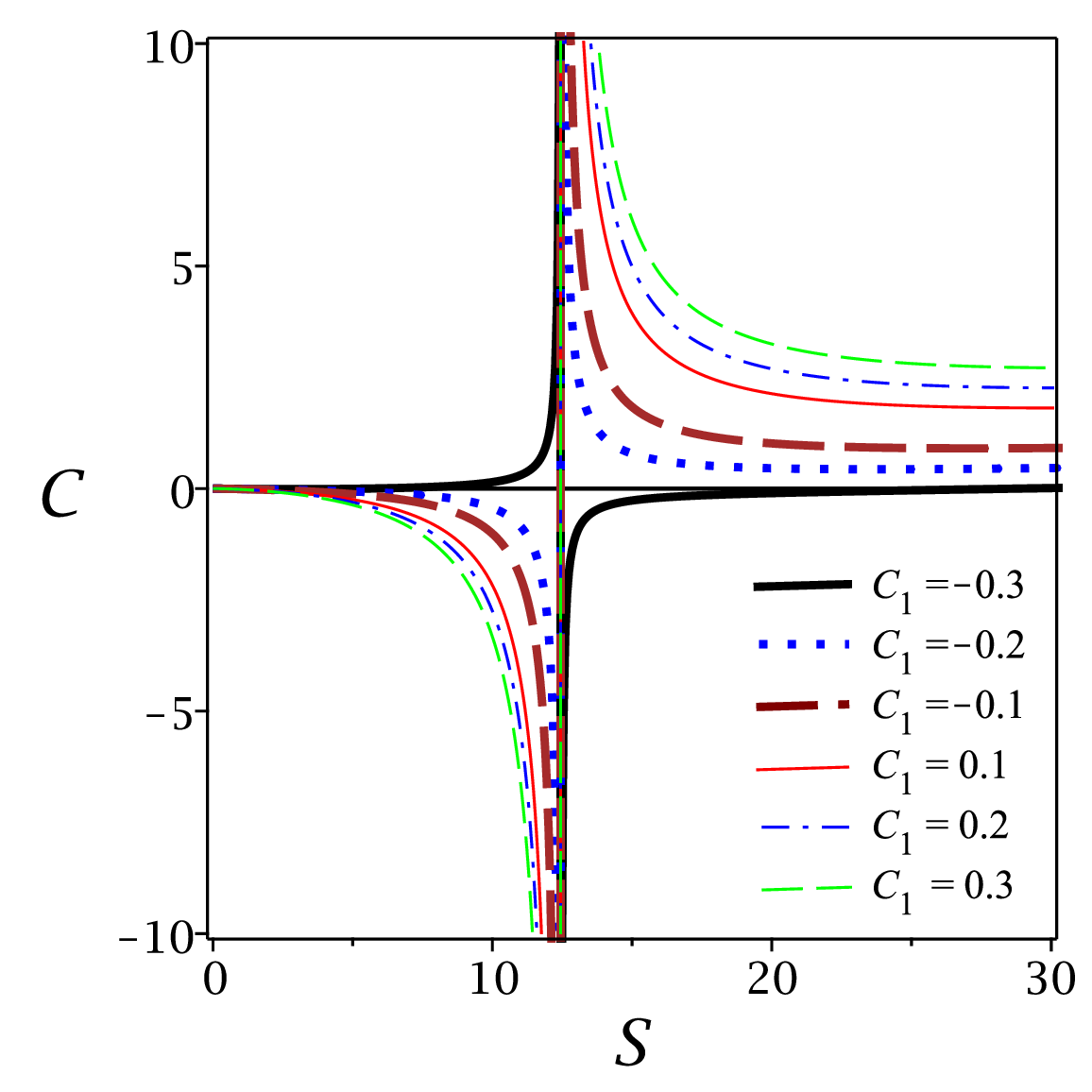}} 
\newline
\caption{The temperature $T$ and heat capacity $C_{Q,P,C_{i}}$ versus $S$
for $Q=0.1$, $\protect\eta=-1$, $\Lambda=-0.02$, and different values of
negative $C_{1}$.}
\label{Fig6}
\end{figure*}

It seems that the obtained heat capacity in Eq. (\ref{Heat1}), may have a
few physical limitation points. For this purpose, and also to understand the
effects of various parameters on these points, we plot the heat capacity and
the temperature in Figs. (\ref{Fig4}), (\ref{Fig5}), and (\ref{Fig6}). We
can find different behavior for the physical limitation points in these
figures, which depend on the parameters of our system. For example, by
adjusting the parameters, we can find two physical limitation points in Fig. %
\ref{Fig4}(a), or no point in the top panels of Fig. \ref{Fig5}.

To study the behavior of the temperature (\ref{TM}), we expand it\ for small
and large values of entropy in the following forms 
\begin{eqnarray}
\underset{S\rightarrow 0}{\lim }T &\propto &\frac{-\pi \eta Q^{2}}{2S^{3/2}},
\\
&&  \notag \\
\underset{S\rightarrow \infty }{\lim }T &\propto &4P\sqrt{S},
\end{eqnarray}%
where confirms that the temperature is always positive for small and large
phantom AdS black holes because $\eta =-1$. It is worthwhile to mention that
for medium black holes, the temperature is dependent on parameters $C_{1} $
and $C_{2}$ $\left( T\propto \frac{1+C_{2}+2C_{1}\sqrt{S}}{8\pi \sqrt{S}}%
\right) $. This relation determines that the temperature can be positive or
negative. The temperature of medium phantom AdS black holes is negative
(positive) when $C_{1}<\frac{-\left( 1+C_{2}\right) }{2\sqrt{S}}$ $\left(
C_{1}>\frac{-\left( 1+C_{2}\right) }{2\sqrt{S}}\right) $. Therefore, we find
two different behaviors for the temperature, which are:

i) by considering $C_{1}<\frac{-\left( 1+C_{2}\right) }{2\sqrt{S}}$, the
phantom AdS black holes with medium size cannot be physical objects because
the temperature is negative. However, small and large black holes are
physical objects (see Fig. \ref{Fig4}(a) and black continuous line of Fig. %
\ref{Fig6}(b)).

ii) according to this fact that the temperature is always positive when $%
C_{1}>\frac{-\left( 1+C_{2}\right) }{2\sqrt{S}}$, so the phantom AdS black
holes with different sizes are physical objects (see Figs. (\ref{Fig5}) and (%
\ref{Fig6})).

Our results indicate that only for specific values of the parameters $C_{1}$
and $C_{2}$, small (large) phantom AdS black holes are thermally stable
(unstable) (see black continuous lines of Figs. \ref{Fig6}(c) and \ref{Fig6}%
(d)); otherwise, they are thermally unstable (stable) (see down panels of
Figs. (\ref{Fig5}) and (\ref{Fig6})).

Another result is related to the effect of parameter $C_{2}$ on the local
stability area of these black holes. From the down panels of Fig. \ref{Fig5}%
, one can find that the local stability area decreases by increasing $C_{2}$
from $-0.3$ to $0.3$.

\subsection{Global thermal stability in grand canonical ensemble}

The idea of studying the black hole's global stability was first suggested
by Hawking and Page \cite{Hawking:1983}. According to their suggestion, the
black hole's global stability can be examined in the grand canonical
ensemble by calculating the Gibbs free energy, such that a black hole is
globally stable provided that its Gibbs free energy is positive \cite%
{Dehghani2019,Dehghani2020}. In this subsection, we want to study the global
stability of phantom AdS black holes in the context of massive gravity by
using the Gibbs free energy approaches.

In the case under consideration, the Gibbs free energy can be defined via
the following relation 
\begin{equation}
G=M\left( S,Q,P,C_{i}\right) -TS-QU,
\end{equation}%
where, we can get the free energy by using Eqs. (\ref{MSQ}) and (\ref{TM}),
which leads to 
\begin{equation}
G=\frac{-32\pi P S^{2}+3\left( 1+C_{2}\right) S+12\pi ^{2}Q^{2}(3\eta-4)}{%
24\pi \sqrt{S}},  \label{F}
\end{equation}%
and we obtain the roots of free energy by solving $F=0$ that are 
\begin{equation}
\left\{ 
\begin{array}{c}
S_{G_{1}}=\frac{3\left( 1+C_{2}\right) -\sqrt{9\left( 1+C_{2}\right)
^{2}+512P\pi ^{3}Q^{2}\left( 9\eta -12\right) }}{64\pi P } \\ 
\\ 
S_{G_{2}}=\frac{3\left( 1+C_{2}\right) +\sqrt{9\left( 1+C_{2}\right)
^{2}+512P\pi ^{3}Q^{2}\left( 9\eta -12\right) }}{64\pi P }%
\end{array}%
\right. .
\end{equation}

\begin{figure*}[tbph]
\centering
\subfloat[$C_{1}=0.2$]{\includegraphics[width=0.32\textwidth]{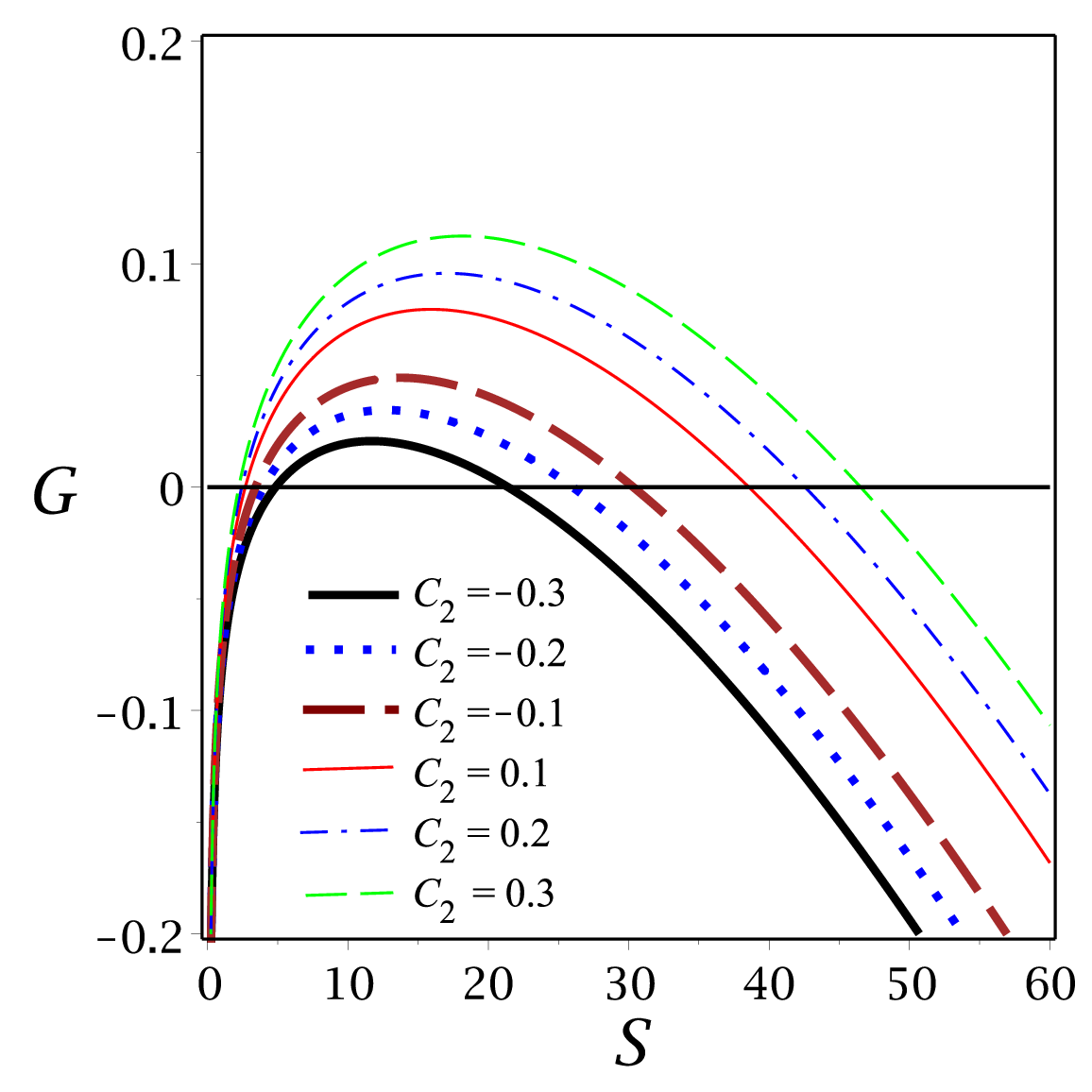}} %
\subfloat[$C_{1}=-0.2$]{\includegraphics[width=0.32\textwidth]{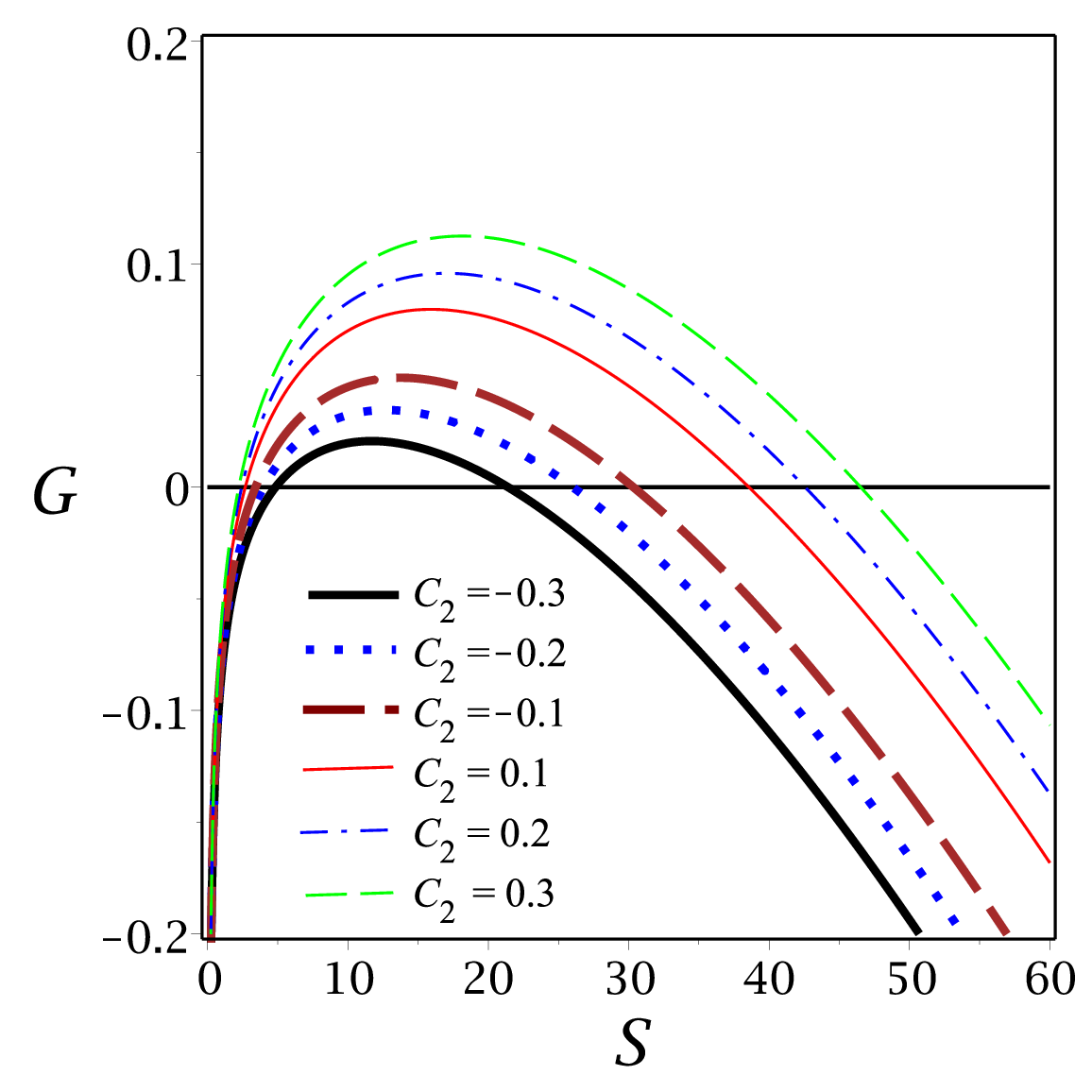}} 
\newline
\caption{The Gibbs free energy ($G$) versus $S$ for $Q=0.1 $, $\protect\eta%
=-1 $, and $\Lambda=-0.02 $. }
\label{Fig7}
\end{figure*}

To study the behavior of the Gibbs free energy (Eq. (\ref{F})), we expand it
versus $S$ in the following form 
\begin{equation}
G=\frac{\pi Q^{2}\left(3\eta-4\right)}{2\sqrt{S}}+\frac{\left(
1+C_{2}\right) \sqrt{S}}{8\pi }-\frac{4P S^{3/2}}{3},  \label{FS}
\end{equation}%
the free energy is only dependent on one of the parameters of massive
gravity, i.e., $C_{2}$. In addition, the Gibbs free energy for small and
large values of entropy depends on 
\begin{eqnarray}
\underset{S\rightarrow 0}{\lim }G &\propto &\frac{\pi
Q^{2}\left(3\eta-4\right)}{2\sqrt{S}},  \label{FsmallS} \\
&&  \notag \\
\underset{S\rightarrow \infty }{\lim }G &\propto &-\frac{4P S^{3/2}}{3},
\label{FlargeS}
\end{eqnarray}%
where indicates $G$ is always negative for very small and large phantom AdS
black holes. So there are two global instability areas. To study the
behavior of free energy carefully, we plot it versus $S$ in Fig. \ref{Fig7}\
for different values of $C_{2}$.

Briefly, our results reveal that there are three (un)stable areas (two
unstable areas and one stable area):

i) the global instability areas are located in the range $S<$ $S_{G_{1}}$
because $G<0$.

ii) in the range $S_{G_{1}}<S<$ $S_{G_{2}}$, the Gibbs free energy is
positive and the phantom AdS black holes satisfy the global stability
condition.

iii) the Gibbs free energy is negative in the range $S>$ $S_{G_{2}}$, so the
large phantom AdS black holes are unstable.

Notably, the massive parameter $C_{2}$ changes the global stability area in
such a way that the global stability area increases by increasing $C_{2}$
from $-0.3$ to $0.3$ (see Fig. \ref{Fig7}).

As a result, very small and large phantom AdS black holes are
physical objects all the time; whereas medium black holes can be
physical/nonphysical depending on the values of parameters. As was shown in
Fig. 4, medium black holes are nonphysical only for very specific values of
massive parameters. From the local stability point of view, small (large)
black holes are unstable (stable) except for very specific values of massive
parameters for which small (large) black holes are stable (unstable). From a
global stability perspective, very small and large phantom AdS black holes
are globally unstable all the time, whereas medium black holes are globally
stable.

Before ending this section, we would like to examine the effect of graviton
mass on thermodynamic quantities. Using definition $C_{1}=m_{g}^{2}cc_{1}$,
and $C_{2}=m_{g}^{2}c^{2}c_{2}$, one can explore the effect of graviton mass
on the temperature, heat capacity, and free energy. For fixed $c$, $c_{1}$,
and $c_{2}$, an increase in $C_{1}$ and $C_{2}$ is equivalent to an increase
in the graviton mass. From Fig. \ref{Fig5}(b), it can be seen that the
temperature increases by increasing $C_{2}$ from $0.1$ to $0.3$, meaning
that increasing graviton mass leads to increasing the temperature. Also,
Fig. \ref{Fig5}(d) shows that divergence shifts to the larger entropy with
an increase in the graviton mass. It reveals the fact that the phase
transition takes place in larger entropy in the presence of more massive
gravitons. Regarding the influence of the graviton mass on the free energy,
it is clear from Fig. \ref{Fig6} that the graviton mass increases the free
energy. In fact, increasing $m_{g}$ makes increasing the region of global
stability.

\section{Optical Properties}

\label{sec5}

In this section, we perform an in-depth study of the optical features of
phantom AdS black holes in massive gravity, given by the solution (\ref%
{f(r)ENMax}), such as the shadow, energy emission rate, and deflection
angle. Taking into account these optical quantities, we explore how the
parameters of the model affect these quantities.

\subsection{Photon Sphere and Shadow}

Here, we intend to analyze the motion of a free photon in the black hole
background (\ref{f(r)ENMax}). We first obtain the radius of the photon
sphere and the shadow of the corresponding black hole, and then find
admissible regions of the black hole parameters for which an acceptable
optical result can be observed.

Lagrangian governing the motion of the photon is \cite{Wei104027}

\begin{eqnarray}
2\mathcal{L} &=&g_{\mu \nu }\dot{x}^{\mu }\dot{x}^{\nu },  \notag \\
&&  \label{Lagrangian} \\
\dot{x}^{\mu } &=&\frac{dx^{\mu }}{d\lambda },  \notag
\end{eqnarray}%
where $\lambda $ is the affine parameter along the geodesics. Since the
black hole solution (\ref{f(r)ENMax}) is spherically symmetric, we consider
photons orbiting around the black hole on the equatorial hyperplane with $%
\theta=\pi /2$, without a significant loss of generality. So Eq. (\ref%
{Lagrangian}) takes the following form

\begin{equation}
2\mathcal{L}=-\psi (r)\dot{t}^{2}+\frac{\dot{r}^{2}}{\psi (r)}+r^{2}\dot{\phi%
}^{2}.  \label{EqLagrangian}
\end{equation}

From the Lagrangian, we can obtain the equations of motion as 
\begin{eqnarray}
\dot{t} &=&-\frac{p_{t}}{\psi (r)},  \notag \\
&&  \notag \\
\dot{r} &=&p_{r}\psi (r),  \label{Eqmotion} \\
&&  \notag \\
\dot{\varphi} &=&\frac{p_{\varphi }}{r^{2}},  \notag
\end{eqnarray}%
where $p$ is the generalized momentum defined by $p_{\mu }=\partial \mathcal{%
L}/\partial \dot{x}^{\mu }=g_{\mu \nu }\dot{x}^{\nu }$.

According to Eqs. (\ref{metric}) and (\ref{f(r)ENMax}), the coefficients are
independent of "$t $" and "$\varphi $". So, one can consider $p_{t}\equiv -E$
and $p_{\varphi}\equiv L $ as constants of motion, where $E $ and $L$ are
the energy and angular momentum of the photon, respectively. Using the
equations of motion and two conserved quantities, one can rewrite the null
geodesic equation as follows 
\begin{equation}
\dot{r}^{2}+V_{\mathrm{eff}}(r)=0,  \label{EqVef1}
\end{equation}%
where the effective potential $V_{\mathrm{eff}}$ is 
\begin{equation}
V_{\mathrm{eff}}(r)=\psi (r)\left[ \frac{L^{2}}{r^{2}}-\frac{E^{2}}{\psi (r)}%
\right] .  \label{Eqpotential}
\end{equation}

To have the spherical geodesics, we apply two conditions 
\begin{eqnarray}
V_{\mathrm{eff}}(r)|_{r=r_{ph}} &=&0,  \notag \\
&&  \label{Eqcondition} \\
\frac{\partial V_{\mathrm{eff}}(r)}{\partial r}\Big\vert_{r=r_{ph}} &=&0, 
\notag
\end{eqnarray}%
where $r_{ph}$ is the radius of photon sphere. With the use of Eq. (\ref%
{Eqpotential}), solving $V_{eff}^{\prime }(r_{ph})=0$ leads to the following
relation 
\begin{equation}
C_{1}r_{ph}^{3}+4(C_{2}+1)r_{ph}^{2}-12m_{0}r_{ph}-8q^{2}=0,  \label{Eqcubic}
\end{equation}%
as we see, Eq. (\ref{Eqcubic}) is cubic in $r_{ph}$ and its discriminant $%
\Delta $ can be calculated as 
\begin{equation}
\Delta =\frac{P^{3}}{27}+\frac{\mathcal{Q}^{2}}{4},  \label{EqDEl}
\end{equation}%
in which $P$ and $Q$ are given by 
\begin{eqnarray}
P &=&-\frac{12m_{0}}{C_{1}}-\frac{16(C_{2}+1)^{2}}{3C_{1}^{2}}, \\
&&  \notag \\
\mathcal{Q} &=&-\frac{8q^{2}}{C_{1}}+\frac{16m_{0}(C_{2}+1)}{C_{1}^{2}}+%
\frac{128(C_{2}+1)^{3}}{27C_{1}^{3}}.  \label{spq}
\end{eqnarray}

\begin{figure*}[tbph]
\centering
\subfloat[$q=0.5$]{\includegraphics[width=0.32\textwidth]{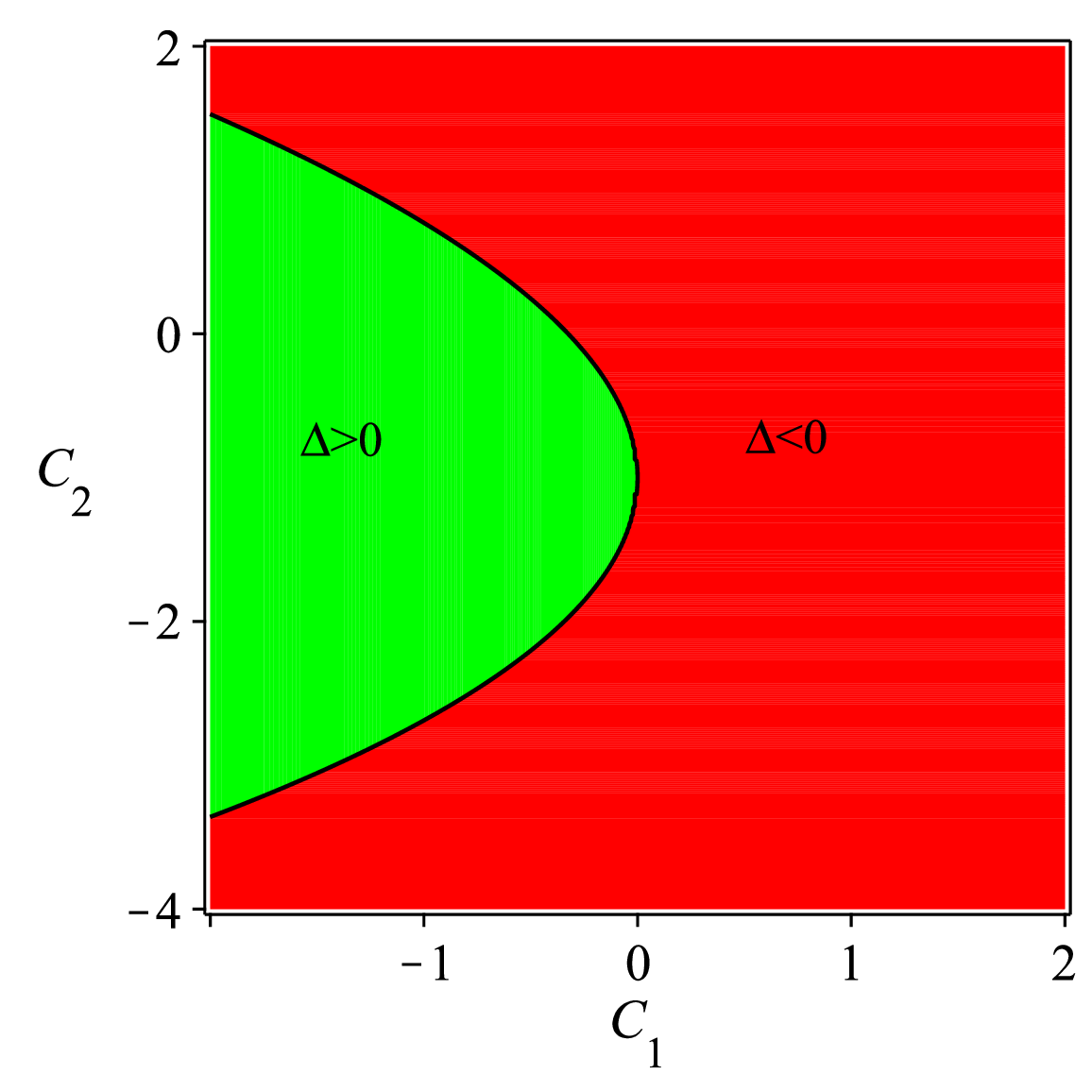}} %
\subfloat[$q=0.8$]{\includegraphics[width=0.32\textwidth]{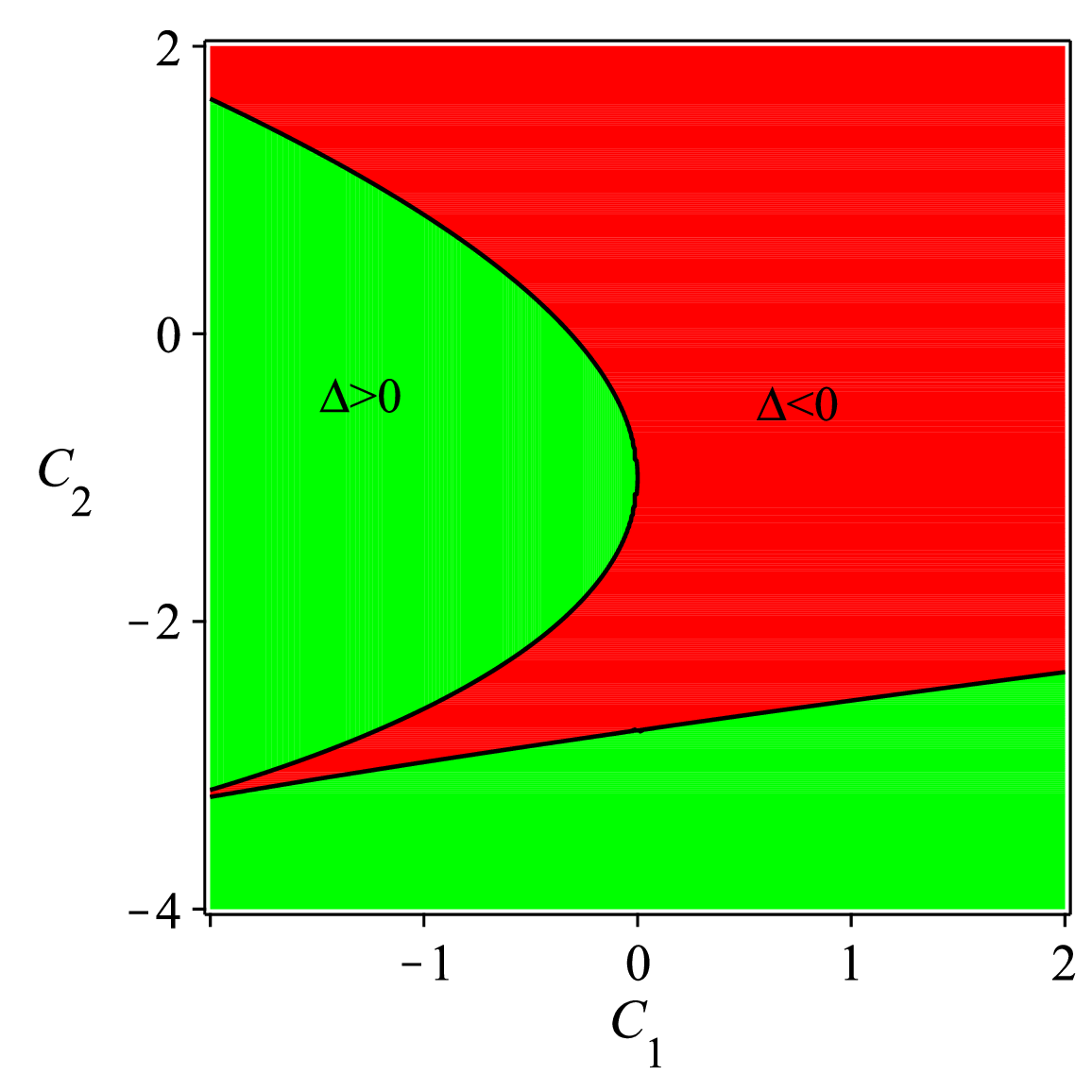}} %
\subfloat[$q=0.9$]{\includegraphics[width=0.32\textwidth]{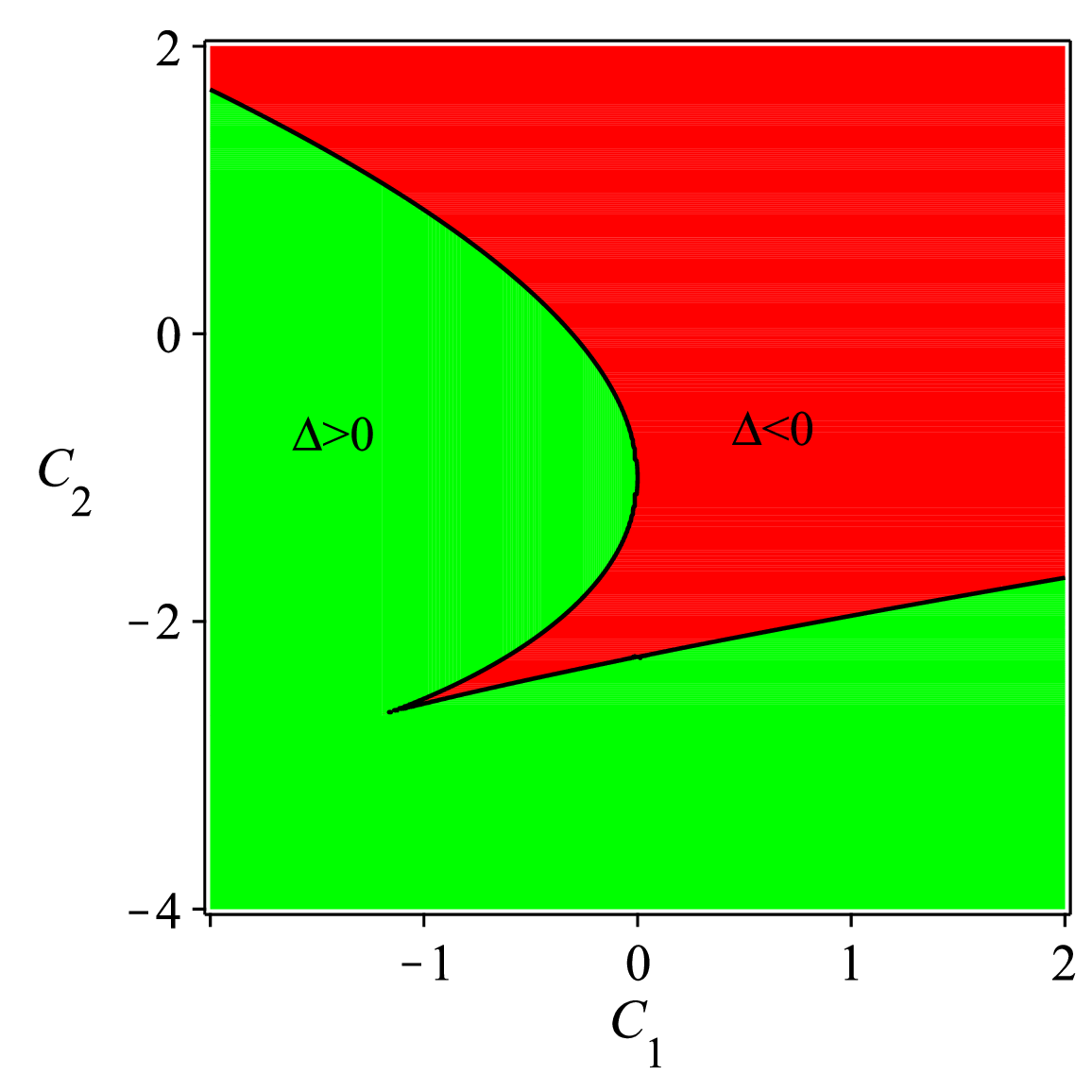}} 
\newline
\caption{Positive or negative regions of discriminant $\Delta$ for $\protect%
\eta =-1$ and $m_{0}=1$. }
\label{FigJ1}
\end{figure*}

Depending on the values of the black hole parameters, $\Delta $ can be
positive or negative. Fig. \ref{FigJ1} displays positive/negative regions of 
$\Delta $, giving us a more precise picture of this function. For the case
of positive $\Delta $, we will have only one real solution defined as 
\begin{equation}
r_{ph}=\left(-\frac{q}{2}+\sqrt{\Delta} \right)^{\frac{1}{3}}+\left(-\frac{q%
}{2}-\sqrt{\Delta} \right)^{\frac{1}{3}}-\frac{2}{3}\frac{(1+C_{2})}{C_{1}}.
\label{Eqroot1}
\end{equation}

For $\Delta <0$, there are three real solutions which are determined as 
\begin{eqnarray}
r_{ph}^{(1)} &=&\frac{2\sqrt{-P}\sin \left[ \frac{\sin ^{-1}\left( \frac{3%
\sqrt{3}\mathcal{Q}}{2(\sqrt{-P})^{3}}\right) }{3}\right] }{\sqrt{3}}-\frac{%
2(1+C_{2})}{3C_{1}}, \\
&&  \notag \\
r_{ph}^{(2)} &=&-\frac{2\sqrt{-P}\sin \left[ \frac{\sin ^{-1}\left( \frac{3%
\sqrt{3}\mathcal{Q}}{2(\sqrt{-P})^{3}}\right) }{3}+\frac{\pi }{3}\right] }{%
\sqrt{3}}-\frac{2(1+C_{2})}{3C_{1}}, \\
&&  \notag \\
r_{ph}^{(3)} &=&\frac{2\sqrt{-P}\cos \left[ \frac{\sin ^{-1}\left( \frac{3%
\sqrt{3}\mathcal{Q}}{2(\sqrt{-P})^{3}}\right) }{3}+\frac{\pi }{6}\right] }{%
\sqrt{3}}-\frac{2(1+C_{2})}{3C_{1}}.
\end{eqnarray}

According to our analysis, $r^{(3)}_{ph} $ given by Eq. (63) is the largest
positive root. From the first condition of Eq.(\ref{Eqcondition}), the
shadow radius can be obtained as \cite{Perlick104031} 
\begin{equation}
r_{sh}=\frac{L_{p}}{E}=\frac{r_{ph}}{\sqrt{\psi(r_{ph})}}.  \label{Eqrsh}
\end{equation}

The apparent shape of a shadow can be obtained by a stereo-graphic
projection in terms of the celestial coordinates $x$ and $y$ which are
defined as \cite{Vazquez489} 
\begin{eqnarray}
x &=&\lim_{r_{0}\rightarrow \infty }\left( -r_{0}^{2}\sin \theta _{0}\frac{%
d\varphi }{dr}\Big|_{(r_{0},\theta _{0})}\right) ,  \notag \\
&&  \label{celestial:1a} \\
y &=&\lim_{r_{0}\rightarrow \infty }\left( r_{0}^{2}\frac{d\theta }{dr}\Big|%
_{(r_{0},\theta _{0})}\right) ,  \notag
\end{eqnarray}%
where $r_{0}$ is the distance between the observer and the black hole, and $%
\theta_{0}$ is the inclination angle.

\begin{table*}[htb!]
\caption{The event horizon ($r_{e}$), photon sphere radius ($r_{ph}$) and
shadow radius ($r_{sh}$) for the variation of $q$, $C_{1}$, $C_{2} $ and $%
\Lambda$ for $\protect\eta =-1$ and $m_{0}=1$.}
\label{table1}\centering
\begin{tabular}{||c|c|c|c|c||}
\hline
{\footnotesize $q$ \hspace{0.3cm}} & \hspace{0.3cm}$0.1$ \hspace{0.3cm} & 
\hspace{0.3cm} $0.4$\hspace{0.3cm} & \hspace{0.3cm} $0.7$\hspace{0.3cm} & 
\hspace{0.3cm}$1.2$\hspace{0.3cm} \\ \hline
$r_{e}$ ($C_{1}=0.5$, $C_{2}=0.2 $ and $\Lambda =-0.02 $) & $1.3055$ & $%
1.3632$ & $1.4735$ & $1.7198$ \\ \hline
$r_{ph}$ ($C_{1}=0.5$, $C_{2}=0.2 $ and $\Lambda =-0.02 $) & $2.0642 $ & $%
2.1450$ & $2.3044$ & $2.7194$ \\ \hline
$r_{sh}$ ($C_{1}=0.5$, $C_{2}=0.2 $ and $\Lambda =-0.02 $) & $2.3474 $ & $%
2.3985$ & $2.4975$ & $2.6627$ \\ \hline
$r_{ph}>r_{e}$ & \checkmark & \checkmark & \checkmark & \checkmark \\ \hline
$r_{sh}>r_{ph}$ & \checkmark & \checkmark & \checkmark & $\times$ \\ 
\hline\hline
&  &  &  &  \\ 
{\footnotesize $C_{1}$ \hspace{0.3cm}} & \hspace{0.3cm} $-0.4$ \hspace{0.3cm}
& \hspace{0.3cm} $-0.01$\hspace{0.3cm} & \hspace{0.3cm} $0.1$\hspace{0.3cm}
& \hspace{0.3cm} $0.82$\hspace{0.3cm} \\ \hline
$r_{e}$ ($q=0.5$, $C_{2}=0.2 $ and $\Lambda =-0.02 $) & $22.6371$ & $1.7669$
& $1.6536$ & $1.2691$ \\ \hline
$r_{ph}$ ($q=0.5$, $C_{2}=0.2 $ and $\Lambda =-0.02 $) & $8.3236 $ & $%
45.3448 $ & $2.5311$ & $2.0092$ \\ \hline
$r_{sh}$ ($q=0.5$, $C_{2}=0.2 $ and $\Lambda =-0.02 $) & $-16.7572 $ & $%
12.7763$ & $3.4441$ & $2.0058$ \\ \hline
$r_{ph}>r_{e}$ & $\times$ & \checkmark & \checkmark & \checkmark \\ \hline
$r_{sh}>r_{ph}$ & $\times$ & $\times$ & \checkmark & $\times$ \\ \hline\hline
&  &  &  &  \\ 
{\footnotesize $C_{2}$ \hspace{0.3cm}} & \hspace{0.3cm}$-0.4$ \hspace{0.3cm}
& \hspace{0.3cm} $0.1$\hspace{0.3cm} & \hspace{0.3cm} $0.5$\hspace{0.3cm} & 
\hspace{0.3cm}$1.1$\hspace{0.3cm} \\ \hline
$r_{e}$ ( $C_{1}=q=0.5 $ and $\Lambda =-0.02 $) & $1.9255 $ & $1.4658$ & $%
1.1653$ & $0.9628$ \\ \hline
$r_{ph}$ ($C_{1}=q=0.5 $ and $\Lambda =-0.02 $) & $3.1697 $ & $2.3147$ & $%
1.7967$ & $1.4658$ \\ \hline
$r_{sh}$ ($C_{1}=q=0.5 $ and $\Lambda =-0.02 $) & $3.5360 $ & $2.5819$ & $%
1.9151$ & $1.4637$ \\ \hline
$r_{ph}>r_{e}$ & \checkmark & \checkmark & \checkmark & \checkmark \\ \hline
$r_{sh}>r_{ph}$ & \checkmark & \checkmark & \checkmark & $\times$ \\ 
\hline\hline
&  &  &  &  \\ 
{\footnotesize $\Lambda$ \hspace{0.3cm}} & \hspace{0.3cm}$-0.01$ \hspace{%
0.3cm} & \hspace{0.3cm} $-0.05$\hspace{0.3cm} & \hspace{0.3cm} $-0.09$%
\hspace{0.3cm} & \hspace{0.3cm}$-0.14$\hspace{0.3cm} \\ \hline
$r_{e}$ ($C_{1}=q=0.5 $ and $C_{2}=0.2 $) & $1.3997 $ & $1.3824$ & $1.3663$
& $1.3476$ \\ \hline
$r_{ph}$ ($C_{1}=q=0.5 $ and $C_{2}=0.2 $) & $2.1904 $ & $2.1904$ & $2.1904$
& $2.1904$ \\ \hline
$r_{sh}$ ($C_{1}=q=0.5 $ and $C_{2}=0.2 $) & $2.4513 $ & $2.3587$ & $2.2758$
& $2.1835$ \\ \hline
$r_{ph}>r_{e}$ & \checkmark & \checkmark & \checkmark & \checkmark \\ \hline
$r_{sh}>r_{ph}$ & \checkmark & \checkmark & \checkmark & $\times$ \\ 
\hline\hline
\end{tabular}%
\end{table*}

\begin{figure}[tbph]
\centering 
\subfloat[$q=C_{2}=0.2$ and
$\Lambda=-0.02$]{\includegraphics[width=0.3\textwidth]{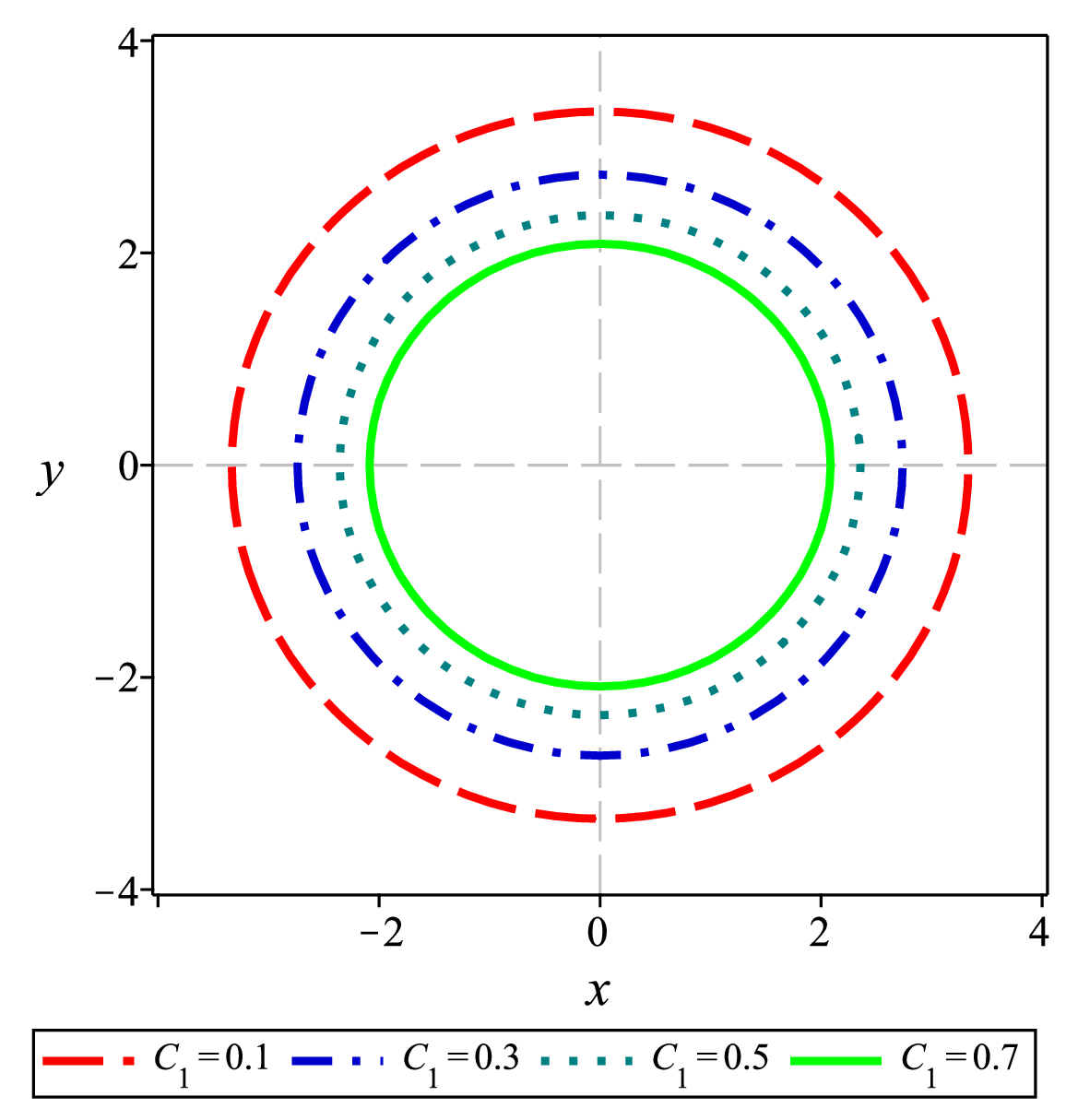}} 
\subfloat[$C_{1}=0.5$, $q=0.2$ and
$\Lambda=-0.02$]{\includegraphics[width=0.3\textwidth]{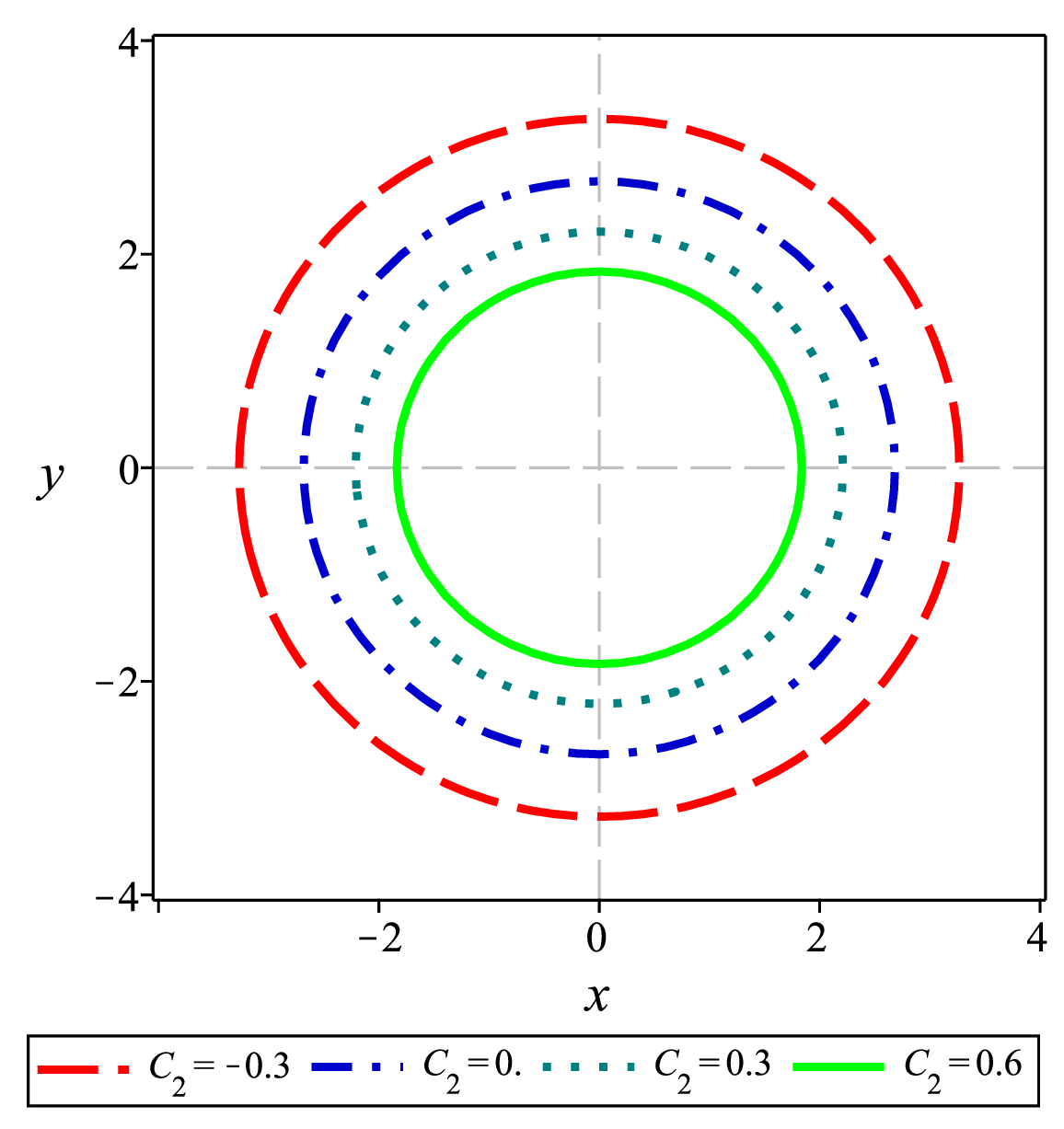}} \newline
\subfloat[$C_{1}=0.5$, $C_{2}=0.2$ and
$\Lambda=-0.02$]{\includegraphics[width=0.3\textwidth]{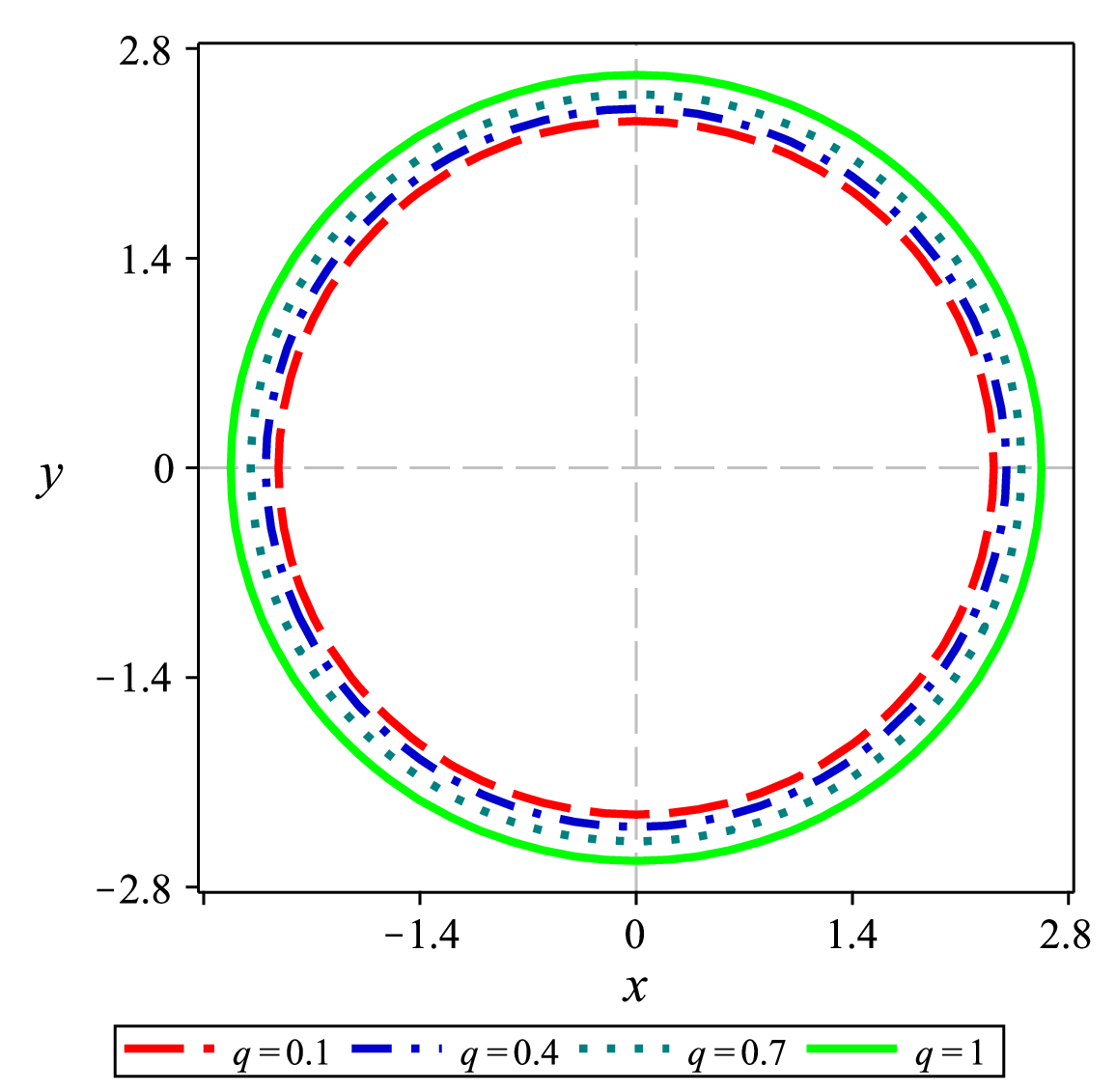}} 
\subfloat[$C_{1}=0.5$ and
$q=C_{2}=0.2$]{\includegraphics[width=0.3\textwidth]{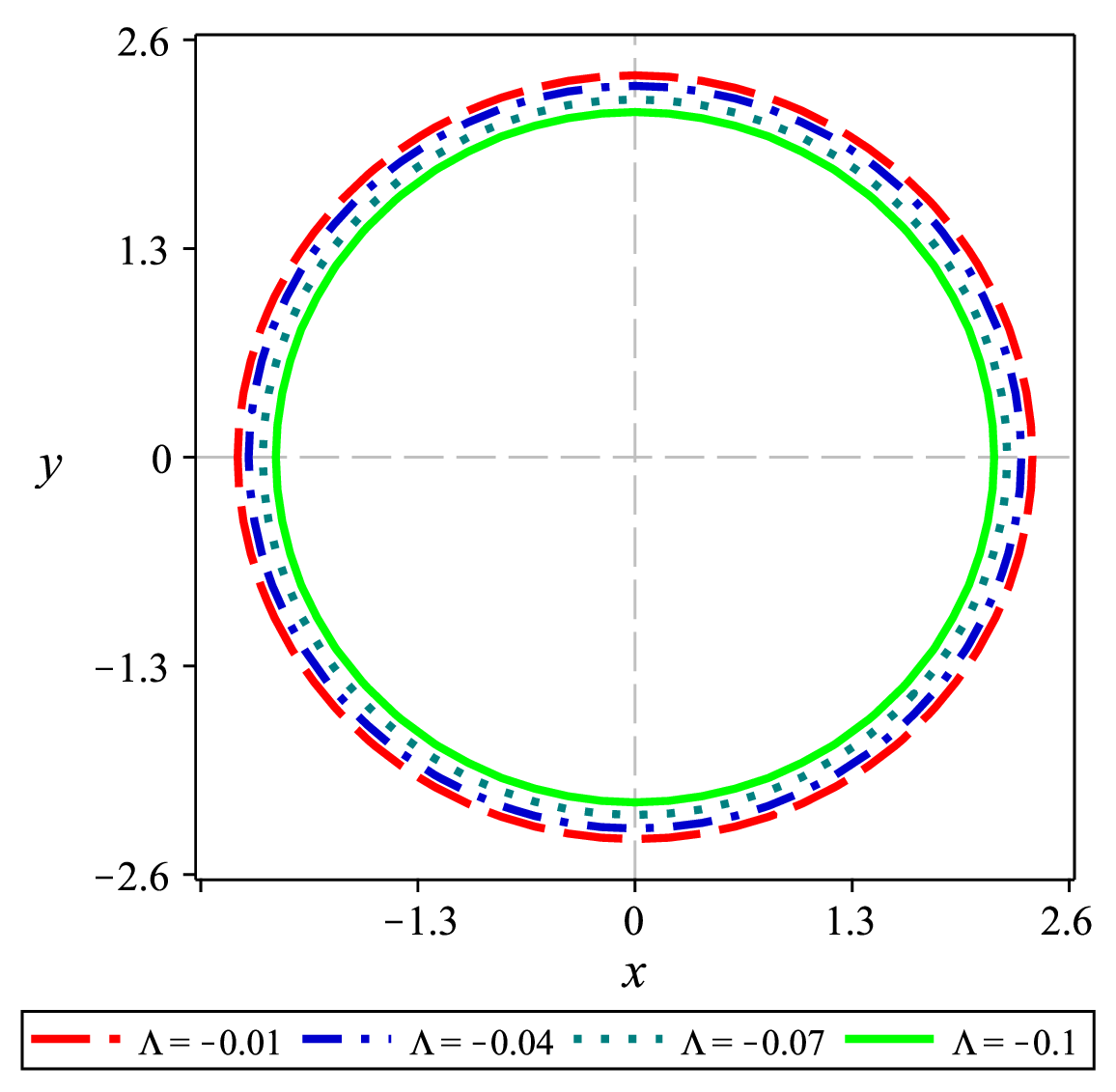}} \newline
\caption{The black hole shadow in the celestial plane $(x-y)$ for $\protect%
\eta =-1$ and $m_{0}=1$.}
\label{FigJ2}
\end{figure}

Now with $r_{ph}$ and $r_{sh}$ in hand, we can investigate allowed regions
of the parameters to have acceptable optical behavior. To do so, we need to
investigate the condition $r_{e}<r_{ph}<r_{sh}$, where $r_{e} $ is the
radius related to the event horizon. For clarity, we list several values of
the event horizon, the photon sphere radius, and the shadow radius in table %
\ref{table1}. As we see, for large values of the electric charge, the
cosmological constant, and parameters $C_{1}$ and $C_{2} $, the shadow size
is smaller than the photon sphere radius, which is not acceptable
physically. As a remarkable point regarding the parameter $C_{1}$, we notice
that an acceptable optical result can be observed just for positive values
of this parameter. From this table, one can also see how the event horizon,
photon sphere radius, and shadow size change by varying the parameters of
the model. Studying the effect of electric charge on these quantities, we
find that the increase of this parameter leads to the increase of all three
quantities. Regarding the effect of the cosmological constant and parameters 
$C_{1}$ and $C_{2}$, our analysis shows that these parameters have
decreasing contributions to the event horizon, the photon sphere radius, and
the shadow radius. To better understand the effect of these parameters on
the shadow size, we have plotted Fig. \ref{FigJ2}. From this figure, it is
clear that the influence of parameters $C_{1}$ and $C_{2}$ on the size of
the black hole shadow is significant compared to the electric charge and the
cosmological constant. As was already mentioned, for fixed $c$, 
$c_{1}$, and $c_{2}$, the change of $C_{1}$ and $C_{2}$ shows the variation
of the massive graviton. Therefore, Fig. \ref{FigJ2}(a) represents the
increasing contribution of the graviton mass on the shadow size. %

\subsection{Energy emission}

\begin{figure}[tbph]
\centering 
\subfloat[$q=C_{2}=0.2$ and $\Lambda=-0.02 $] {\includegraphics[width=0.3\textwidth]{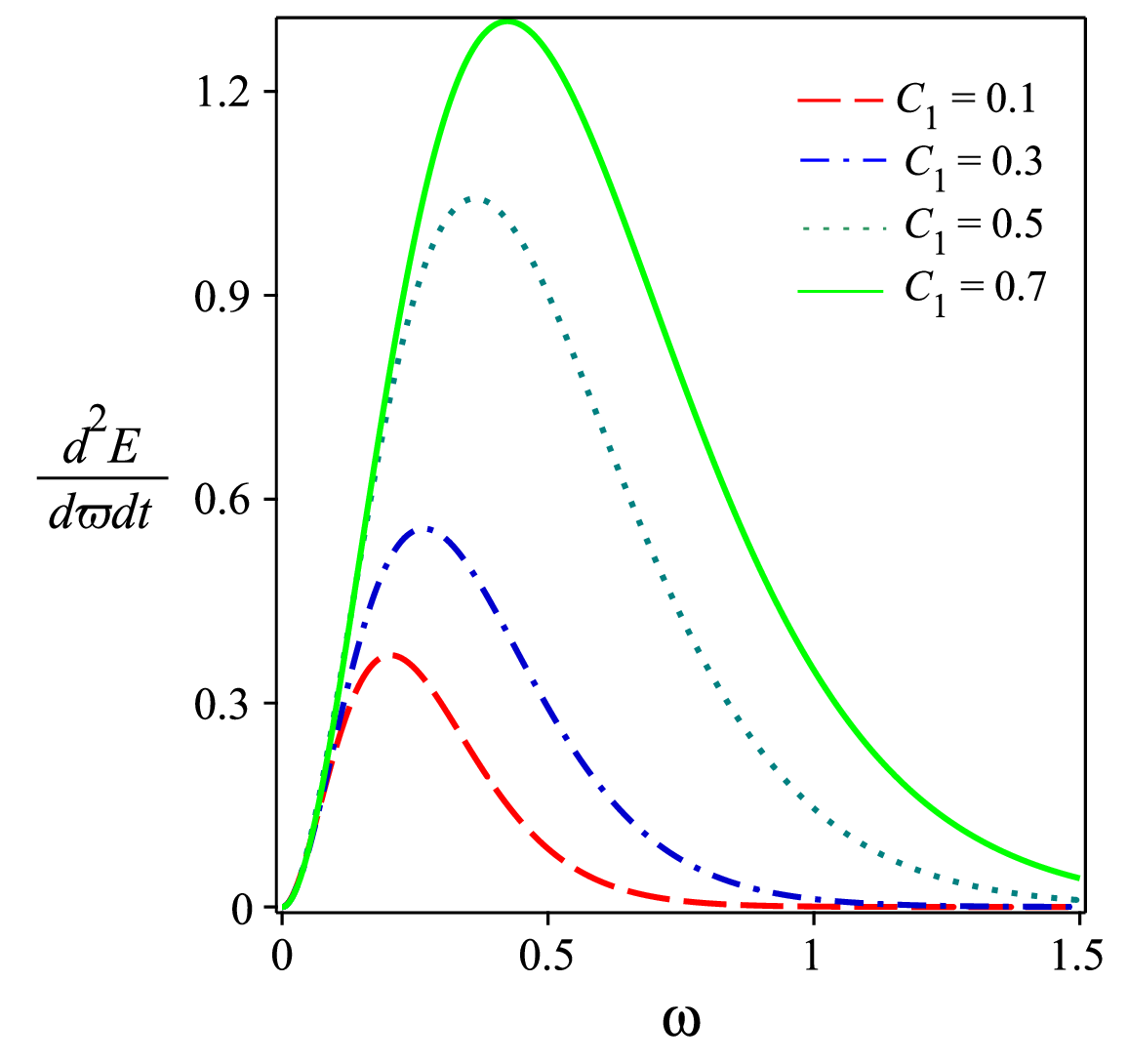}} 
\subfloat[$C_{1}=0.5$, $q=0.2$ and $\Lambda=-0.02 $] {\includegraphics[width=0.3\textwidth]{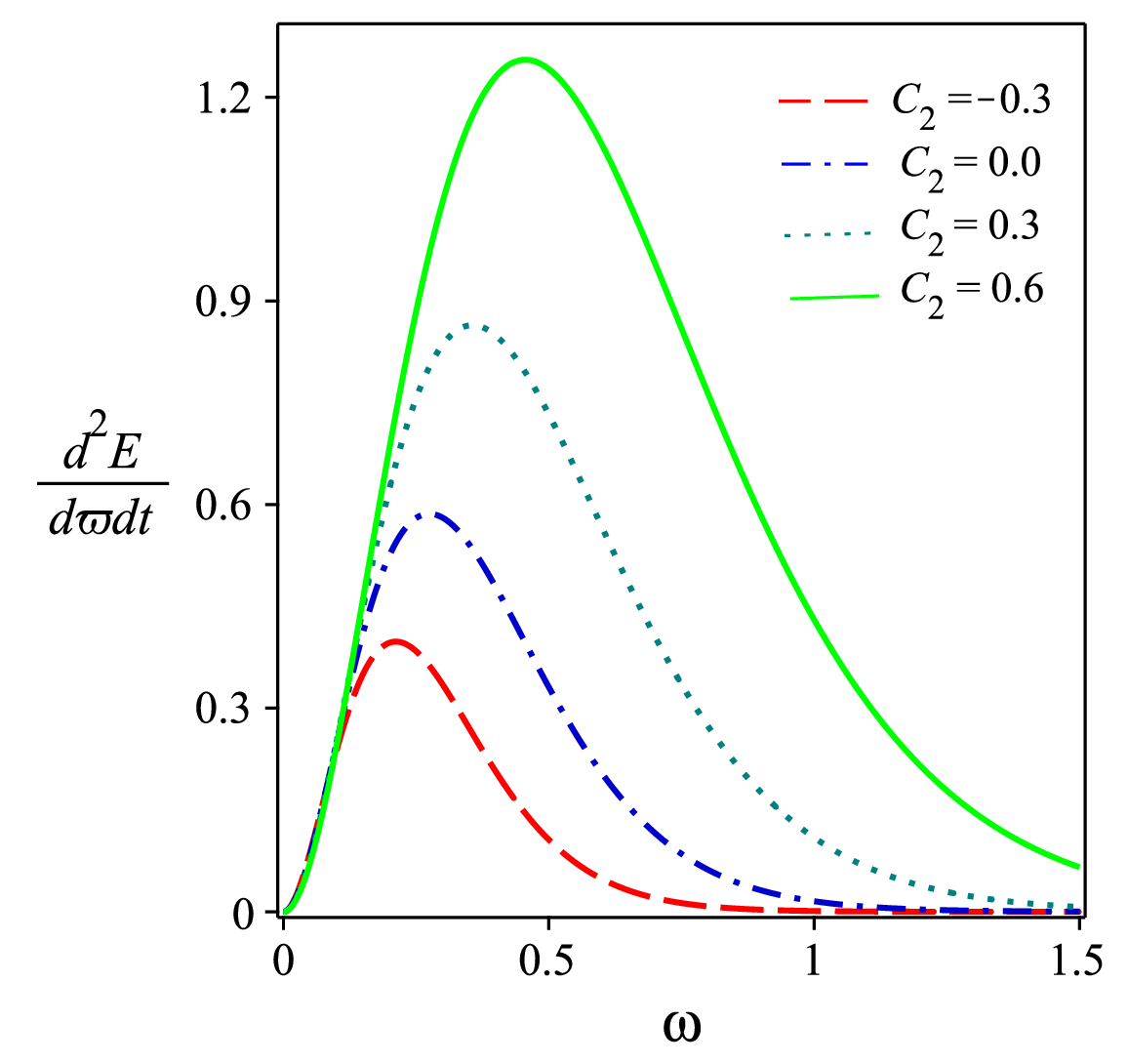}}
\newline
\subfloat[$C_{1}=0.5$, $ C_{2}=0.2$ and $\Lambda=-0.02 $] {\includegraphics[width=0.3\textwidth]{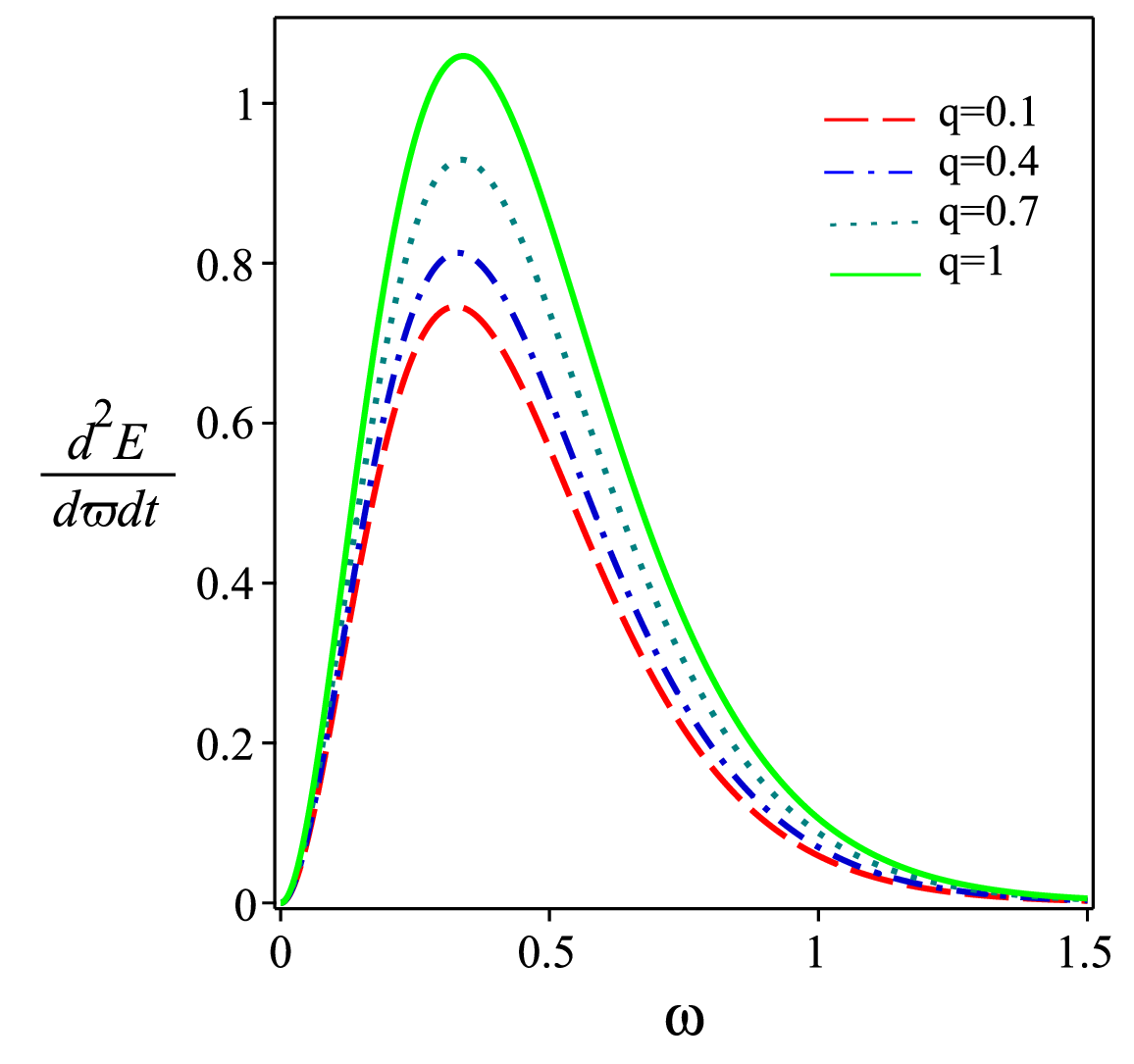}} 
\subfloat[$C_{1}=0.5$ and $ q=C_{2}=0.2$] {\includegraphics[width=0.3\textwidth]{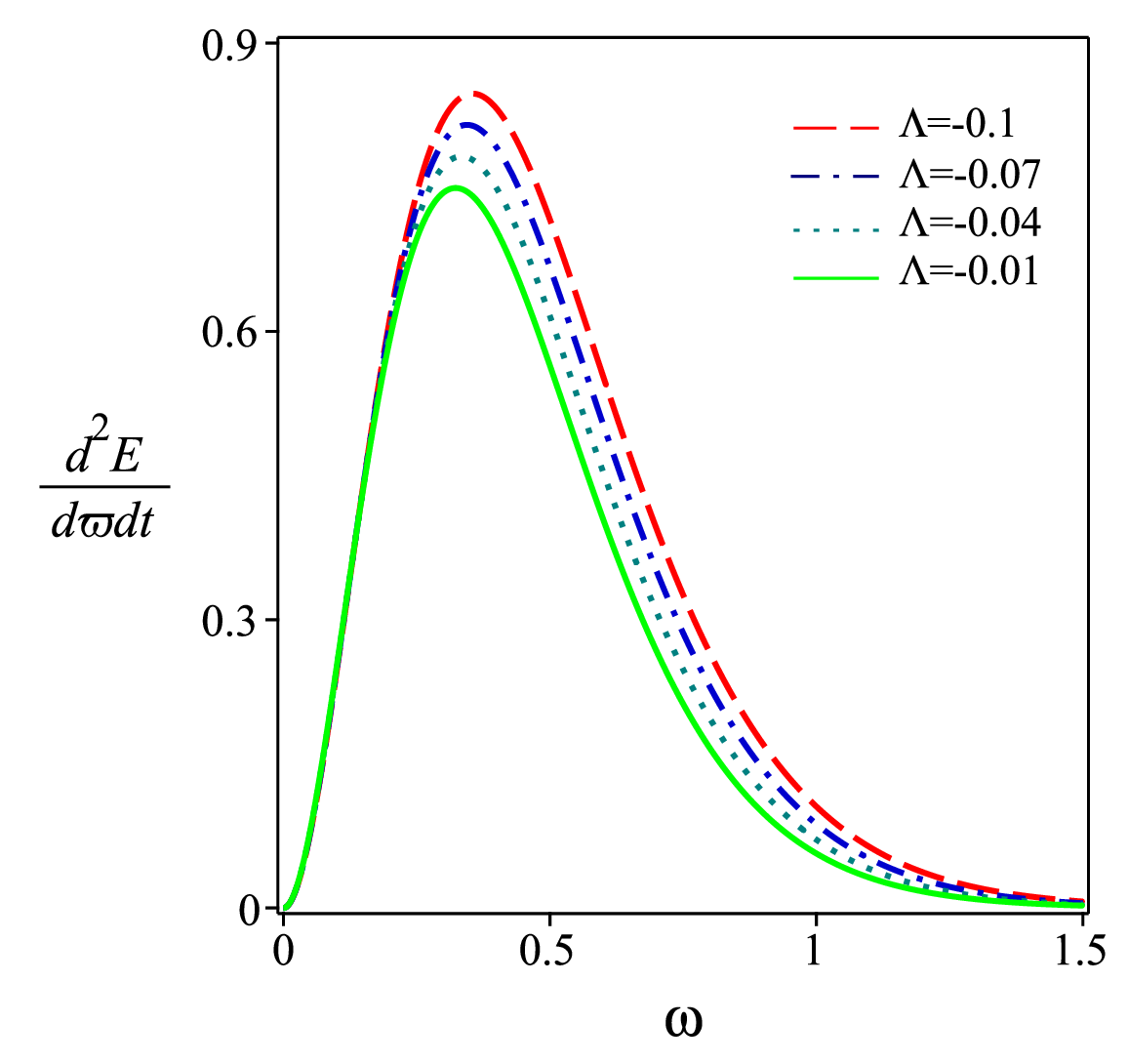}} \newline
\caption{The energy emission rate versus $\protect\omega $ for the
corresponding black hole with $\protect\eta =-1$, $m_{0}=1$ and different
values of $C_{1}$, $C_{2} $, $q $ and $\Lambda$.}
\label{FigJ3}
\end{figure}

Having the black hole shadow, it is possible to study the emission of
particles around the black hole. It has been known that the black hole
shadow corresponds to its high energy absorption cross-section for the
observer located at infinity \cite{Wei063}. In fact, the absorption
cross-section oscillates around a limiting constant value $\sigma _{lim}$
for a spherically symmetric black hole. Since the shadow measures the
optical appearance of a black hole, it is approximately equal to the area of
the photon sphere ($\sigma_{lim}\approx \pi r_{sh}^{2}$). For $4-$%
dimensional spacetime, the energy emission rate is defined as 
\begin{equation}
\frac{d^{2}\mathcal{E}(\omega )}{dtd\omega }=\frac{2\pi ^{3}\omega
^{3}r_{sh}^{2}}{e^{\frac{\omega }{T}}-1},  \label{Eqemission}
\end{equation}%
in which $\omega $ and $T$ are, respectively, the emission frequency and
Hawking temperature. Inserting Eq. (\ref{TemPhMassive}) into Eq. (\ref%
{Eqemission}), the energy emission rate can be obtained for the present
black hole solution. In Figure \ref{FigJ3}, these energetic aspects are
plotted as a function of the emission frequency for different values of the
electric charge, cosmological constant, and parameters $C_{1}$ and $C_{2}$.
As it is transparent, there exists a peak of the energy emission rate, which
decreases and shifts to low frequencies as the parameters $C_{1}$, $C_{2}$, $%
q$, and $|\Lambda |$ decrease. From this figure, all four parameters have an
increasing effect on the emission rate. This shows that as the influence of
these parameters becomes stronger, the evaporation process becomes faster. As a result, the black hole has a shorter lifetime in the
presence of the more massive gravitons and strong electric fields, or a high
curvature background. 

\subsection{Deflection Angle}

Gravitational lensing is a subject of wide interest that has a tremendous
impact on the distribution of matters and the constituents of the Universe.
The gravitational lensing effect states that a light beam is distorted when
it passes through a huge object, which is one of the most important
predictions of general relativity. The deflection of light by gravitational
fields has been studied with great interest in astrophysics as well as in
theoretical physics. Determining the mass of galaxies and clusters \cite%
{Henk177,Brouwer481}, as well as discovering dark energy and dark matter 
\cite{Vanderveld103518} is an important application of gravitational lensing
in astronomy and cosmology. In this subsection, we employ the Gauss-Bonnet
theorem and study the light deflection around the black hole solution (\ref%
{f(r)ENMax}). We exploit the optical geometry of the black hole solution and
find the Gaussian curvature in weak gravitational lensing.

Using weak field approximation, the Gauss-Bonnet theorem in optical metric
can be expressed as follows \cite{Gibbons2350,Werner3047,Jusufi8403} 
\begin{equation}
\int \int_{\mathcal{D}_{R}}\mathcal{K}dS+\oint_{\partial \mathcal{D}%
_{R}}kdt+\sum_{i}\theta _{i}=2\pi \mathcal{X}(\mathcal{D}_{R}),
\end{equation}%
where $D_{R}$ is a region consisting of the source of light waves, the
observer, and the focal point of the lens. $\mathcal{K}$ and $k$ are,
respectively, the Gaussian curvature of $D_{R}$ and the geodesic curvature
of $\partial \mathcal{D}_{R}$. Also, $dS$ is the area element and $%
\theta_{i} $ refers to the exterior angle at the $i-th$ vertex.

As the light beam approaches from infinity up to a large distance, and since
we are working with weak field limitations, the light beam is almost
straight. So, we employ the straight-line approximation $r=\frac{b}{%
sin(\varphi)} $, where $b$ is the impact parameter. With this, the
asymptotic deflection angle $\alpha $ can be calculated as \cite%
{Gibbons2350,Jusufi8403,Kumaran2210} 
\begin{equation}
\alpha=-\int_{0} ^{\pi} \int_\frac{b}{\sin\phi} ^{\infty} \mathcal{K}dS.
\label{HF13}
\end{equation}

To calculate the deflection angle, we need to determine the optical path,
which is obtained by the null geodesic condition $ds^{2} = 0$. Since we have
considered the motion of photon in the equatorial plane ($\theta=\pi/2 $),
the optical path metric can be obtained as

\begin{equation}
dt^{2}=\tilde{g}_{ij}dx^{i}dx^{j}=\frac{1}{\psi ^{2}(r)}dr^{2}+\frac{r^{2}}{%
\psi (r)}d\varphi ^{2}.  \label{dt}
\end{equation}

To obtain the optical Gaussian curvature $\mathcal{K}$ from the optical
metric Equation (\ref{dt}), we use the expression $\mathcal{K}=R/2$ \cite%
{Upadhyay2210}, in which $R$ is the Ricci scalar calculated using the
optical metric. Here, the infinitesimal surface element can be computed as 
\cite{Upadhyay2210} 
\begin{equation}
dS=\sqrt{det\left( \tilde{g}\right) }drd\varphi =\frac{r}{ \psi (r)^{3/2}}%
drd\varphi .  \label{dt1}
\end{equation}

Using Equation (\ref{HF13}) and the value of optical Gaussian curvature, the
deflection angle $\alpha $ of the corresponding black hole is calculated as 
\begin{eqnarray}
\alpha &\simeq &\frac{3\sqrt{3}\pi \left( 9C_{1}^{3}+48C_{1}\Lambda
(C_{2}+1)-128m_{0}\Lambda ^{2}\right) }{256b^{2}\left( -\Lambda \right) ^{%
\frac{5}{2}}}+\frac{\sqrt{3}\left( 48\Lambda
(C_{2}+1)^{2}+9C_{1}^{2}(C_{2}+1)\right) }{24b^{3}\left( -\Lambda \right) ^{%
\frac{5}{2}}}+\frac{1701\sqrt{3}\pi m_{0}q^{4}}{512b^{10}\left( -\Lambda
\right) ^{\frac{5}{2}}}  \notag \\
&&+\frac{45\sqrt{3}\pi \left( \left( 9C_{1}^{2}-32\Lambda (C_{2}+1)\right)
m_{0}-8\Lambda C_{1}q^{2}\right) }{512b^{4}\left( -\Lambda \right) ^{\frac{5%
}{2}}}-\frac{\sqrt{3}\left( 96\Lambda m_{0}C_{1}+64\Lambda ^{2}q^{2}\right) 
}{24b^{3}\left( -\Lambda \right) ^{\frac{5}{2}}}+\frac{1536\sqrt{3}q^{6}}{%
847b^{11}\left( -\Lambda \right) ^{\frac{5}{2}}}  \notag \\
&&+\frac{128\sqrt{3}q^{2}\left( 3m_{0}^{2}-(C_{2}+1)q^{2}\right) }{%
21b^{9}\left( -\Lambda \right) ^{\frac{5}{2}}}+\frac{2\sqrt{3}\Lambda \left(
C_{1}^{2}+(C_{2}+1)\Lambda \right) }{3b\left( -\Lambda \right) ^{\frac{5}{2}}%
}+\frac{3\sqrt{3}\left( \left( 16\Lambda (C_{2}+1)+9C_{1}^{2}\right)
q^{2}\right) }{10b^{5}\left( -\Lambda \right) ^{\frac{5}{2}}}  \notag \\
&&+\frac{3\sqrt{3}\left( 32\Lambda m_{0}^{2}+24C_{1}m_{0}(C_{2}+1)\right) }{%
10b^{5}\left( -\Lambda \right) ^{\frac{5}{2}}}+\frac{15\sqrt{3}\pi \left(
21C_{1}q^{2}(C_{2}+1)-27C_{1}m_{0}^{2}\right) }{128b^{6}\left( -\Lambda
\right) ^{\frac{5}{2}}}  \notag \\
&&+\frac{15\sqrt{3}\pi \left( \left( 28\Lambda q^{2}+12(C_{2}+1)^{2}\right)
m_{0}\right) }{128b^{6}\left( -\Lambda \right) ^{\frac{5}{2}}}-\frac{24\sqrt{%
3}\left( 126m_{0}^{2}(C_{2}+1)+153m_{0}C_{1}q^{2}\right) }{245b^{7}\left(
-\Lambda \right) ^{\frac{5}{2}}}  \notag \\
&&+\frac{24\sqrt{3}\left( 2q^{2}\left( 14\Lambda
q^{2}+27(C_{2}+1)^{2}\right) \right) }{245b^{7}\left( -\Lambda \right) ^{%
\frac{5}{2}}}-\frac{945\sqrt{3}\pi \left(
3C_{1}q^{4}-6m_{0}^{3}+14m_{0}q^{2}(C_{2}+1)\right) }{2048b^{8}\left(
-\Lambda \right) ^{\frac{5}{2}}},  \notag
\end{eqnarray}

\begin{figure}[tbph]
\centering 
\subfloat[$ q=C_{2}=0.2$ and $ \Lambda=-0.08 $]{
        \includegraphics[width=0.3\textwidth]{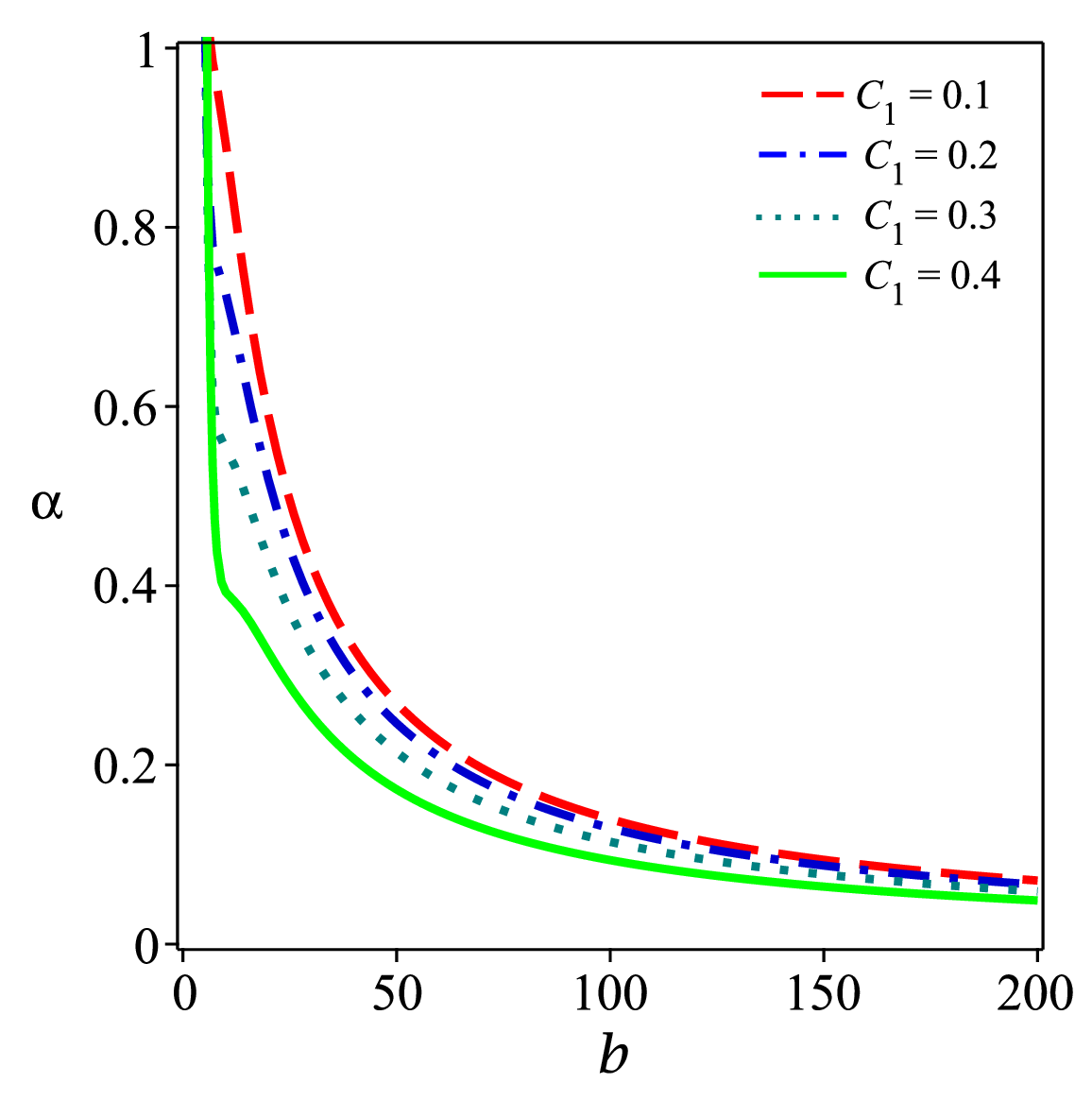}} 
\subfloat[$ q=C_{1}=0.2$ and $ \Lambda=-0.08 $]{
        \includegraphics[width=0.3\textwidth]{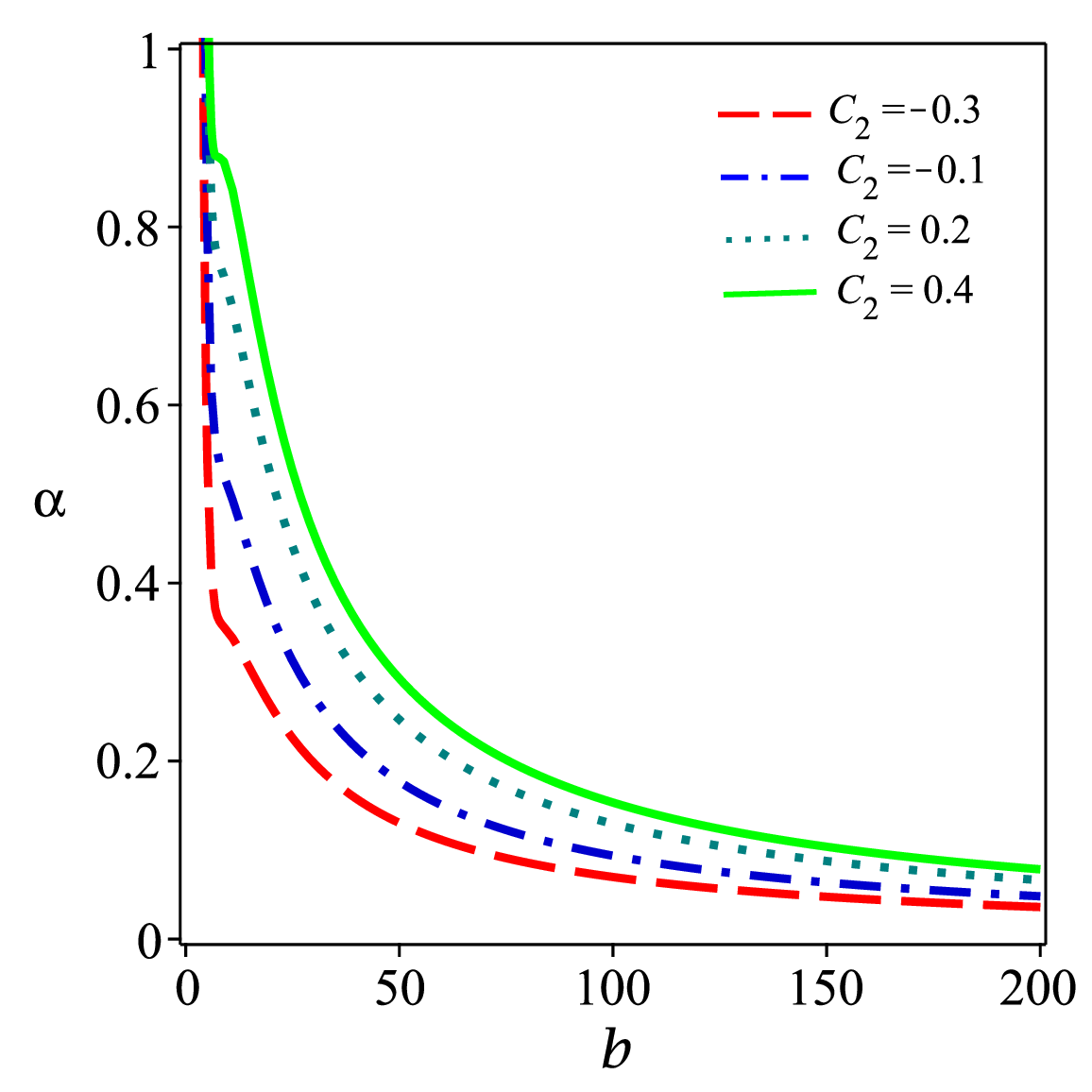}} \newline
\subfloat[ $ C_{1}=C_{2}=0.2$ and $ \Lambda=-0.08 $]{
        \includegraphics[width=0.3\textwidth]{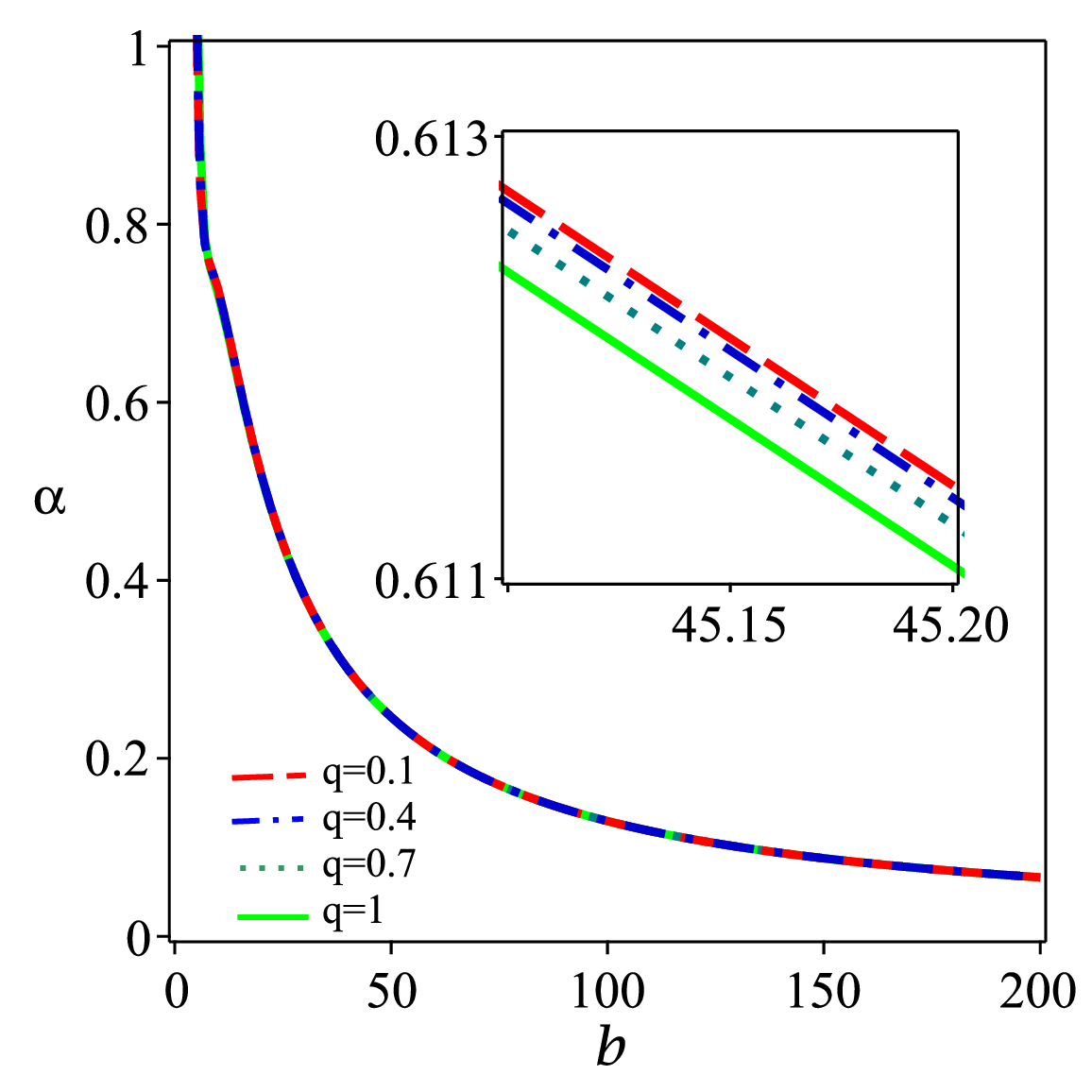}} 
\subfloat[  $ C_{1}=C_{2}=q=0.2$]{
        \includegraphics[width=0.3\textwidth]{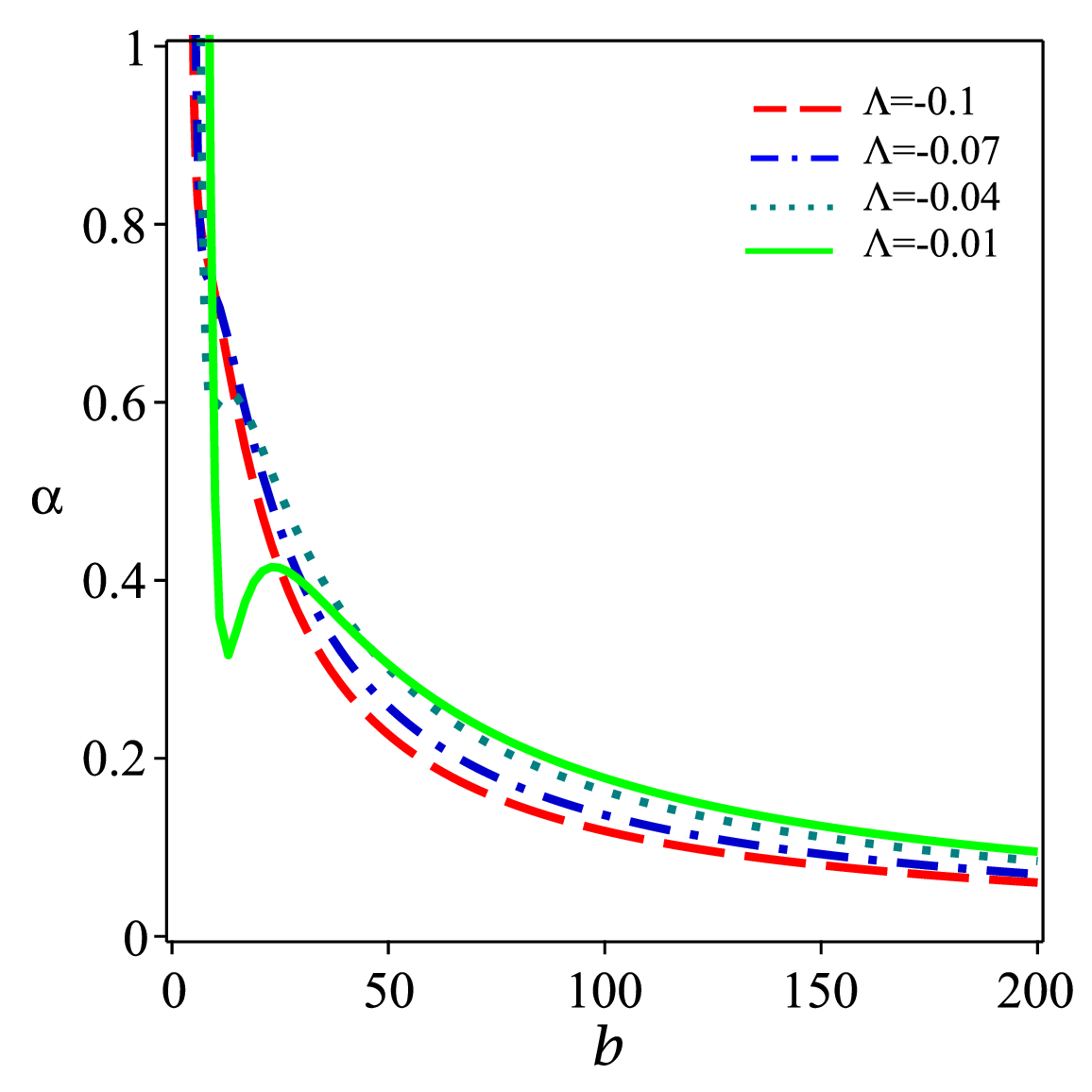}} \newline
\caption{The behavior of $\protect\alpha$ with respect to the impact
parameter $b $ for $\protect\eta =-1$, $m_{0}=1$ and different values of $%
C_{1}$, $C_{2} $, $q$ and $\Lambda$.}
\label{FigJ4}
\end{figure}

The behavior of $\alpha$ concerning the impact parameter, $b$, is
illustrated in Figure \ref{FigJ4}. As we see, for small (large) values of
the parameter $C_{1} $ ($\vert \Lambda \vert $), the deflection angle is a
decreasing function of $b$. Otherwise, $\alpha $ behavior is slightly
different in such a way that the deflection angle first decreases to a
minimum value with the increase of $b$, then increases and reaches a maximum
value with the further increase of the impact parameter, after that, it
gradually decreases as $b $ goes to infinity. Fig. \ref{FigJ4} also shows
how $\alpha$ changes by varying black hole parameters. According to Fig. \ref%
{FigJ4}(a), the parameter $C_{1}$ has a decreasing contribution to the
deflection angle. For fixed $c$, $c_{1}$, and $c_{2}$, this is
equivalent to the decreasing effect of the graviton mass on the deflection
angle. In contrast to $C_{1}$, the parameter $C_{2}$ has an
increasing effect on $\alpha$ (see Fig. \ref{FigJ4}(b)). Regarding the
electric charge effect, we see from Fig. \ref{FigJ4}(c), that its effect is
similar to that of the parameter $C_{1} $. But as it is clear, its effect is
insignificant compared to other parameters, indicating that variation of the
electric charge does not have a remarkable effect on the deflection angle.
Analyzing Fig. \ref{FigJ4}(d) we find that for intermediate values of the
impact parameter, $\alpha$ is an increasing function of the cosmological
constant, whereas, for large values of $b$, it is a decreasing function of $%
\Lambda$. This reveals the fact that photons with large angular momentum
deviate much more from their straight path in a low-curvature background.

\section{Quasinormal Modes}

\label{sec6}

From a theoretical point of view, perturbations of a black hole's spacetime
can be performed in two ways: by perturbing the black hole metric (the
background) or by adding fields to the black hole's spacetime. When black
holes are perturbed, they tend to relax toward equilibrium. Such a process
is done through the emission of QNMs. QNMs are complex frequencies, $%
\omega=\omega _{R}-i\omega_{I}$, which encode important information related
to the stability of the black hole under small perturbations. The sign of
the imaginary part determines if the mode is stable or unstable. For
positive $\omega _{I}$ the mode is unstable, whereas negativity represents
stable modes. For a stable mode, the real part provides the frequency of the
oscillation, while the inverse of $|Im(\omega)| $ determines the damping
time $t_{D}^{-1}=|Im(\omega )|$ \cite{Blazquez2020}. There are some types of
perturbation such as scalar, Dirac, vector, or tensor. In this paper, we are
interested in investigating scalar perturbations by a massive field around a
black hole.

\textbf{Massive Scalar Perturbations:} Here, we will study the QN
frequencies of a massive scalar field. For a scalar field $\Psi $ of mass $%
\mu$ in the background of the metric $g_{\mu \nu }$, the equation of motion
is given by the following Klein-Gordon equation 
\begin{equation}
(\nabla ^{\nu }\nabla _{\nu }-\mu ^{2})\Psi =0,  \label{EqKG}
\end{equation}%
where $\nabla _{\nu }$ is the covariant derivative. The above equation can
be written explicitly as follows: 
\begin{equation}
\frac{1}{\sqrt{-g}}\partial _{\nu }(\sqrt{-g}g^{\mu \nu }\partial _{\mu
}\Psi )-\mu ^{2}\Psi =0.
\end{equation}

Using the method of separation of variables, the scalar field can be written
as 
\begin{equation}
\Psi (t,r,\theta,\phi )=\frac{u_{L}(r)}{r}e^{-i\omega t }
Y_{Lm}(\theta,\phi),  \label{separable}
\end{equation}
which $Y_{lm}(\theta,\phi) $ denotes the spherical harmonics and the
function $u_{l}(r) $ satisfies an ordinary second order linear differential
equation as follows 
\begin{equation}
\frac{d^{2}u_l}{dx^{2}}+ \left[ \omega^{2} - V(r) \right]u_l=0,
\end{equation}
where $x=\int \frac{dr}{\psi (r)} $ represents the tortoise coordinates. The
corresponding effective potential barrier is given by 
\begin{equation}
V_{l}(r)=\psi (r)\left( \mu^{2}+\frac{l(l+1)}{r^{2}}+ \frac{\psi^{\prime} (r)%
}{r}\right) ,  \label{EqEV}
\end{equation}
where $l$ is the multipole number.

QNMs were investigated with different methods over the last decades. Here,
we employ the semi-analytical WKB approximation, which is based on the
matching of the WKB expansion of the modes at the event horizon and spatial
infinity with the Taylor expansion of the effective potential near the peak
of the potential barrier. This method was first proposed in the 1980s \cite%
{SIyer,QNM2} and then extended to 6th order \cite{QNM3}, and to 13th order 
\cite{Konoplya1}. It should be noted that increasing the WKB order does not
always lead to a better approximation for the frequency, so we employ the
3rd order expansion for our study, which is given by the following formula 
\begin{equation}
\omega ^{2}=[V_{0}+(-2V_{0}^{^{\prime \prime }})^{\frac{1}{2}}\Lambda ]-i\nu
(-2V_{0}^{^{\prime \prime }})^{\frac{1}{2}}(1+\Omega ),  \label{Eqomega}
\end{equation}%
where 
\begin{eqnarray}
\Lambda (n) &=&\frac{\left( \frac{V_{0}^{(4)}}{V_{0}^{^{\prime \prime }}}%
\right) \left( \frac{1}{4}+\nu ^{2}\right) }{8\left( -2V_{0}^{^{\prime
\prime }}\right) ^{1/2}}-\frac{\left( \frac{V_{0}^{^{\prime \prime \prime }}%
}{V_{0}^{^{\prime \prime }}}\right) ^{2}\left( 7+60\nu ^{2}\right) }{%
288\left( -2V_{0}^{^{\prime \prime }}\right) ^{1/2}}\ ,  \label{4.14} \\
&&  \notag \\
\Omega (n) &=&\frac{5\left( \frac{V_{0}^{^{\prime \prime \prime }}}{%
V_{0}^{^{\prime \prime }}}\right) ^{4}(77+188\nu ^{2})}{6912\left(
-2V_{0}^{^{\prime \prime }}\right) ^{1/2}}-\frac{\left( \frac{%
V_{0}^{^{\prime \prime \prime 2}}V_{0}^{(4)}}{V_{0}^{^{\prime \prime 3}}}%
\right) (51+100\nu ^{2})}{384\left( -2V_{0}^{^{\prime \prime }}\right) ^{1/2}%
}+\frac{\left( \frac{V_{0}^{(4)}}{V_{0}^{^{\prime \prime }}}\right)
^{2}(67+68\nu ^{2})}{2304\left( -2V_{0}^{^{\prime \prime }}\right) ^{1/2}} 
\notag \\
&&+\frac{\left( \frac{V_{0}^{^{\prime \prime \prime }}V_{0}^{(5)}}{%
V_{0}^{^{\prime \prime 2}}}\right) (19+28\nu ^{2})}{288\left(
-2V_{0}^{^{\prime \prime }}\right) ^{1/2}}-\frac{\left( \frac{V_{0}^{(6)}}{%
V_{0}^{^{\prime \prime }}}\right) (5+4\nu ^{2})}{288\left( -2V_{0}^{^{\prime
\prime }}\right) ^{1/2}}.  \label{4.15}
\end{eqnarray}

\begin{figure}[tbph]
\centering 
\subfloat[$q=0.2$]{\includegraphics[width=0.3\textwidth]{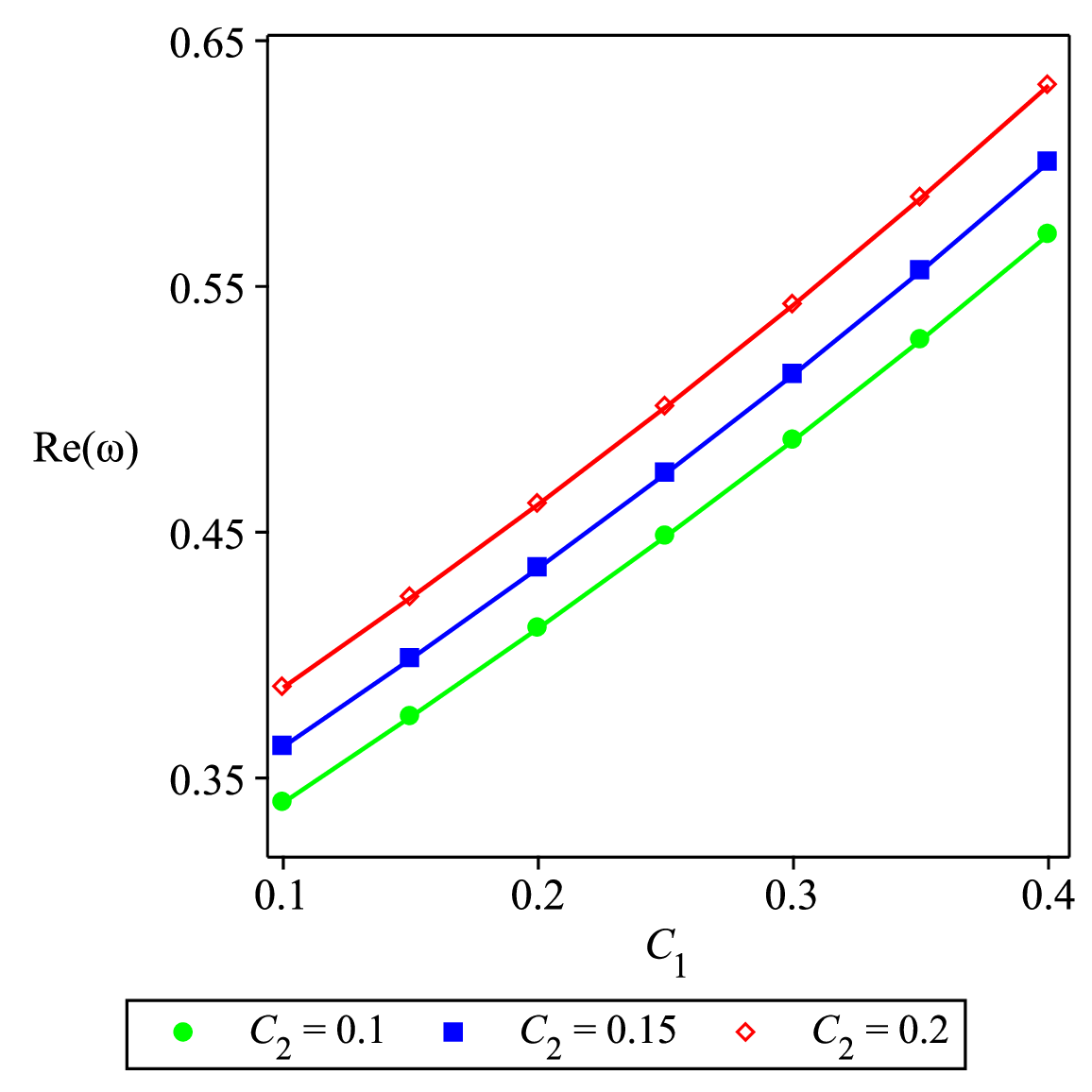}} %
\subfloat[$q=0.2$]{\includegraphics[width=0.3\textwidth]{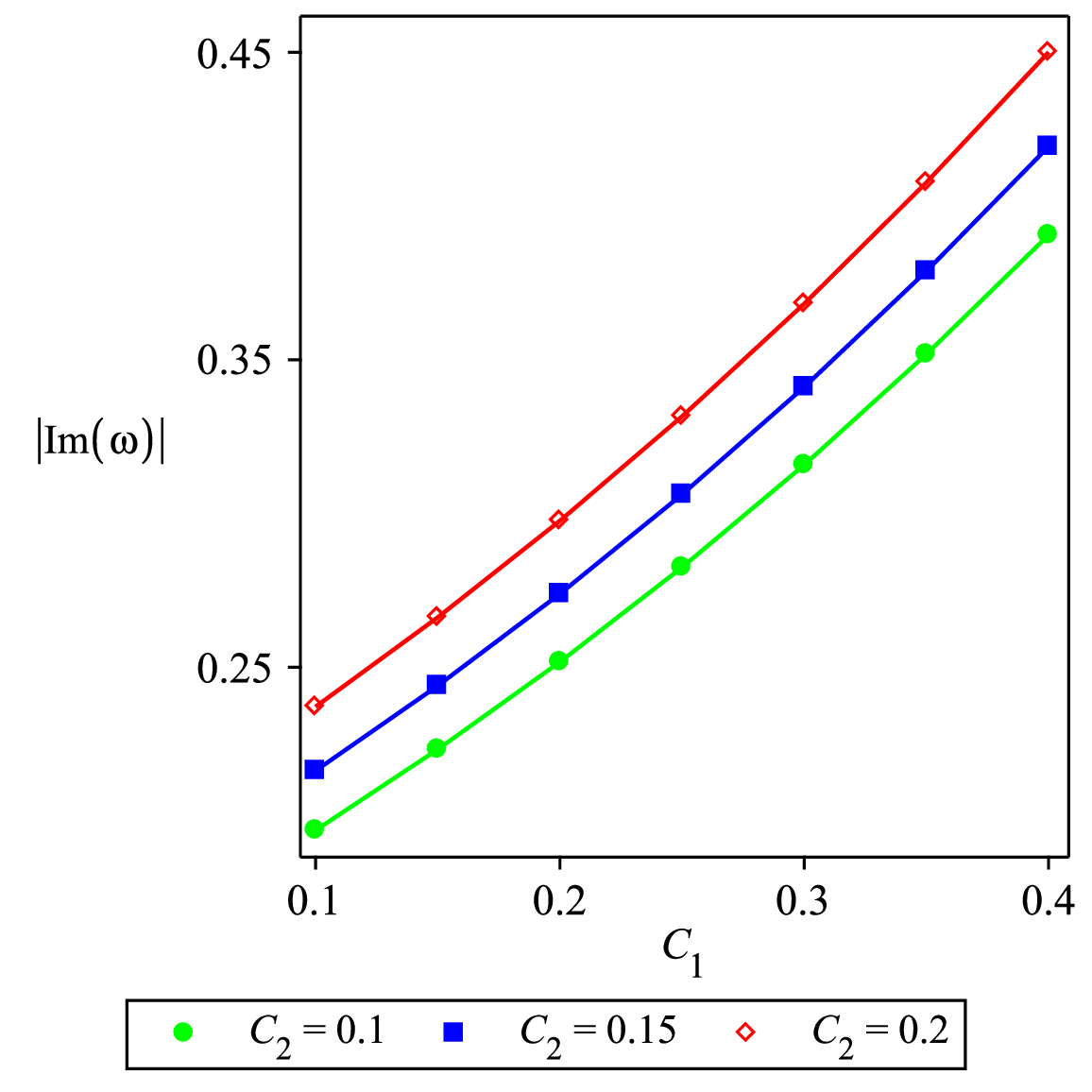}} \newline
\subfloat[$C_{2}=0.2$]{\includegraphics[width=0.3\textwidth]{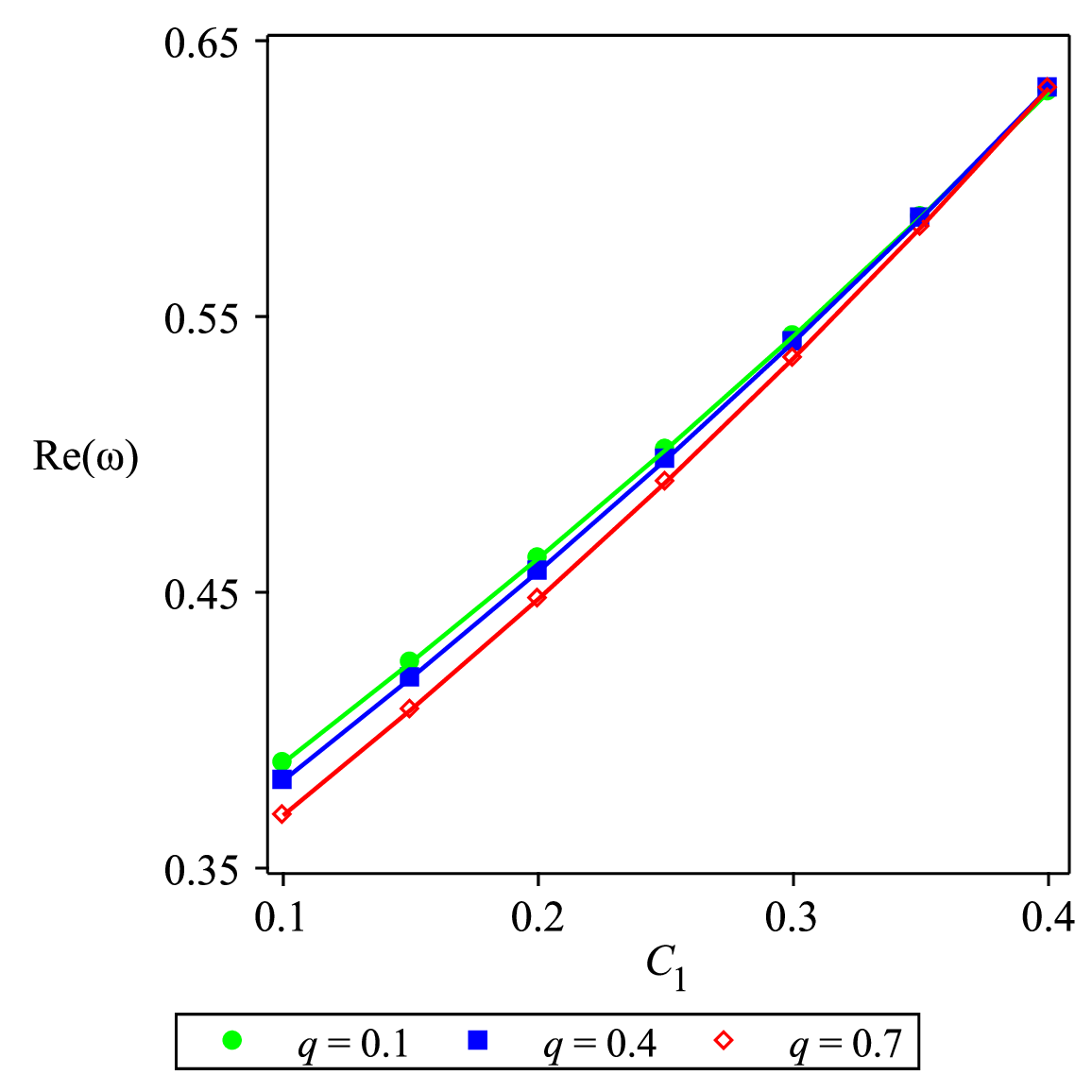}} %
\subfloat[$C_{2}=0.2$]{\includegraphics[width=0.3\textwidth]{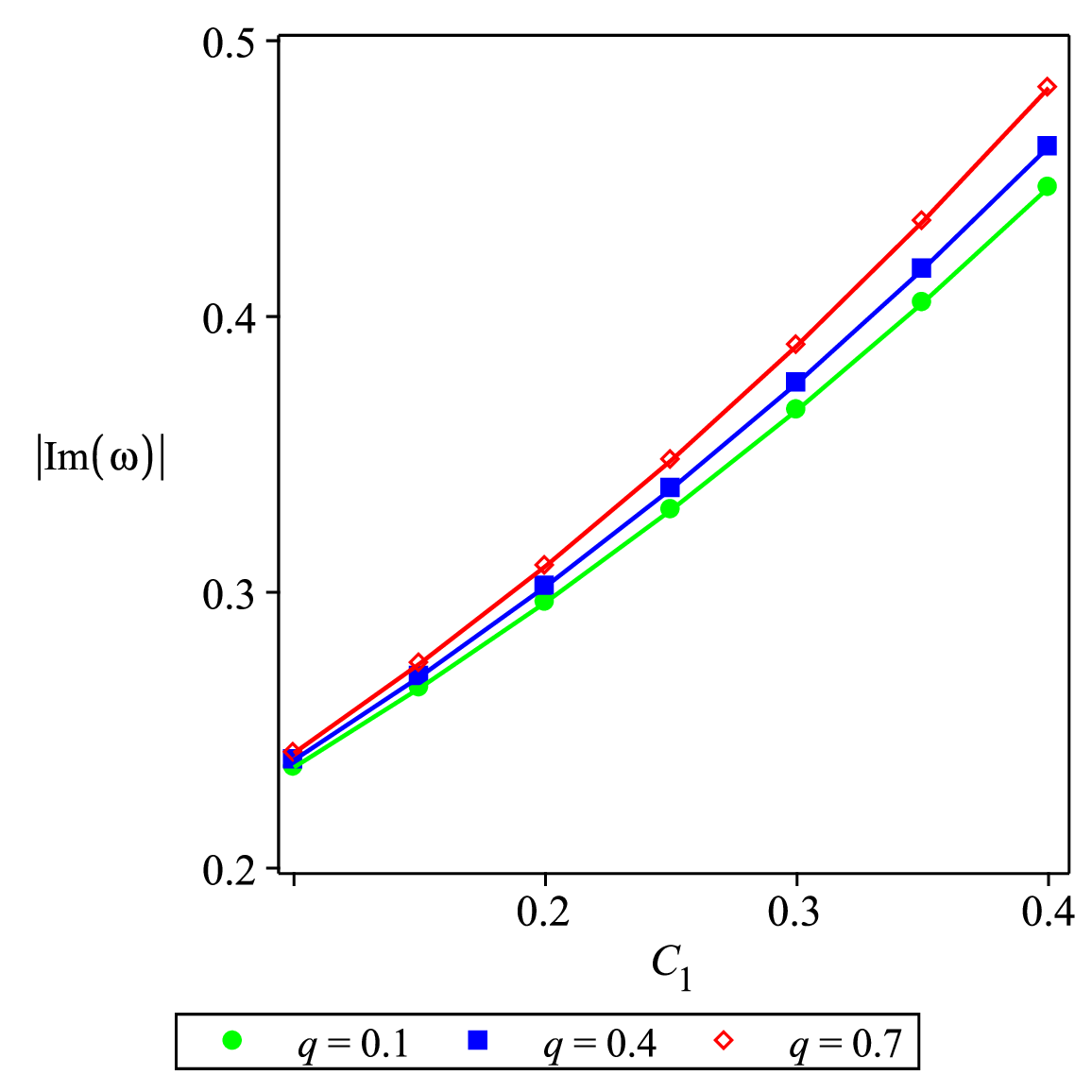}} 
\newline
\caption{The behavior of $Re(\protect\omega) $ and $\vert Im(\protect\omega%
)\vert$ with the parameter $C_{1} $ for $\protect\eta =-1$, $m_{0}=1$ and
different values of $C_{2} $ and $q$.}
\label{FigJ5}
\end{figure}

In the relations above, $\nu =n+\frac{1}{2} $ and $n$ is the overtone
number. $V_{0}^{(j)} $ denotes the $j$-order derivative of the effective
potential on the point maximum $x_{0} $. It is worth pointing out that the
WKB formula does not give reliable frequencies for $n\geq l $, whereas it
leads to quite accurate values for $n<l $. Now, using Eq. (\ref{EqEV}) and
Eq. (\ref{Eqomega}), we can calculate the QNMs for the corresponding black
hole, which is shown in Fig. \ref{FigJ5} with different values of black hole
parameters. We see that as the parameter $C_{1} $ increases, both the real
and absolute value of imaginary parts of the QN frequency grows
monotonically. This shows that when this parameter gets more, the scalar
perturbations oscillate more rapidly, and due to the increasing imaginary
part, they decay faster. For fixed $c$, $c_{1}$, and $c_{2}$,
one can also say that the scalar perturbations oscillate with more energy
and decay faster in the presence of the more massive gravitons. Figs. \ref{FigJ5}(a) and \ref{FigJ5}(b) show how $Re (\omega) $ and $\vert
Im(\omega)\vert$ change under varying the $C_{2} $ parameter. The curves
shift upwards by increasing $C_{2} $, meaning that both real and imaginary
parts of the QNMs are an increasing function of this parameter. Figs. \ref%
{FigJ5}(c) and \ref{FigJ5}(d) give a simple illustration of how the
dependence of the QNMs on the electric charge. As we see, $q$ decreases
(increases) the real (absolute value of the imaginary) part of $\omega $.
This implies that the scalar field perturbations around the black hole
oscillate with less energy and decay faster in a powerful electric field.

\section{Conclusions}

\label{sec7}

In this paper, we have obtained AdS phantom solutions in the dRGT-like
massive gravity. Investigating some geometrical properties, we have shown
that such solutions can be interpreted as black hole solutions. Then, we
studied the thermodynamic structure of solutions in the extended phase space
and examined the validity of the first law of thermodynamics. Calculating
heat capacity, we investigated the local stability of the system and showed
how the parameters of the model affect the region of the stability.
According to our analysis, large (small) phantom AdS black holes are in a
local stable (unstable) state all the time. We also examined the global
stability of the system through the use of Helmholtz-free energy and found
that large and small phantom AdS black holes are globally stable, whereas
medium black holes cannot satisfy the global stability condition.

In the next step, we studied the optical features of black holes and noticed
that some constraints should be imposed on the range of parameters of the
model under consideration to have acceptable optical behavior. Studying the
effect of the parameters on the size of the shadow and the radius of the
photon sphere showed that the electric charge has an increasing effect on
these two optical quantities, while the parameters $C_{1} $ and $C_{2} $
have a decreasing contribution to them. Regarding the cosmological constant
effect, although the photon sphere radius is not affected by this parameter,
the shadow size shrinks by increasing $\vert \Lambda \vert$.

We continued our investigation by studying the energy emission rate and the
impact of black hole parameters on the radiation process. Our analysis
showed that all parameters have an increasing contribution to the emission
rate, namely, the evaporation process gets faster as the effects of these
parameters get stronger. This reveals the fact that the lifetime of the
black hole would be shorter with the increase of the parameters of the model.

After that, we performed an in-depth analysis of the gravitational lensing
of light around such black holes. Depending on the choice of the black hole
parameters, we observed different behaviors. For large (small) values of the
cosmological constant (parameter $C_{1}$), the deflection angle $\alpha$ is
a decreasing function of the impact parameter $b$. While for small (large)
values of $\vert \Lambda \vert$ ($C_{1}$ parameter), $\alpha$ first
decreases as $b$ increases, then increases and reaches a local maximum as $b$
increases further, after that, it gradually decreases as $b$ goes to
infinity. Regarding the influence of parameters on $\alpha$, we found that
the parameter $C_{1}$, electric charge, and the absolute value of the
cosmological constant have a decreasing contribution to $\alpha$, whereas
the parameter $C_{2}$ has an increasing effect. This shows that photons are
more deflected in a weak electric field or when the effect of massive
parameter $C_{2}$ ($C_{1}$) gets strong (weak).

Finally, we presented a study of the quasinormal modes of scalar
perturbation. We found that both massive parameters $C_{1}$ and $C_{2}$
increase the real part and absolute value of the imaginary part of the
quasinormal frequencies. This means that the quasinormal modes oscillate
with more energy and decay faster around the black holes in the presence of
massive gravity. Studying the effect of the electric charge showed that the
real (imaginary) part of the quasinormal modes decreases (increases) with
the growth of this parameter, indicating that the scalar field perturbations
in the presence of electric charge oscillate more slowly and decay faster as
compared to neutral black holes.

\begin{acknowledgements}

We are grateful to the anonymous referees for the insightful comments and
suggestions, which have allowed us to improve this paper significantly. B. E. P thanks the University of Mazandaran. M. E. R  thanks Conselho Nacional de Desenvolvimento Cient\'ifico e Tecnol\'ogico - CNPq, Brazil, for partial financial support.

\end{acknowledgements}

\end{document}